\documentclass[aps,pra,preprint,10pt,superscriptaddress]{revtex4-2}
\usepackage{amsmath,amsfonts,amssymb}
\usepackage{bm}
\usepackage{xcolor}

\usepackage{xpatch}
\makeatletter
\xpatchcmd{\@ssect@ltx}{\@xsect}{\protected@edef\@currentlabelname{#8}\@xsect}{}{}
\xpatchcmd{\@sect@ltx}{\@xsect}{\protected@edef\@currentlabelname{#8}\@xsect}{}{}
\makeatother

\usepackage{hyperref}
\hypersetup{
	colorlinks=true,
	linkcolor=black,
	citecolor=black,
	filecolor=black,      
	urlcolor=black,
}

\usepackage{graphicx}
\usepackage{grffile}
\usepackage{array}
\usepackage{tabularx}
\usepackage{setspace}

\begin{document}
 
\title{Modulation of metastable ensemble dynamics explains the inverted-U relationship between tone discriminability and arousal in auditory cortex}

\author{Lia Papadopoulos}
\affiliation{Institute of Neuroscience,  University of Oregon, Eugene, OR 97403, USA}

\author{Suhyun Jo}
\affiliation{Institute of Neuroscience,  University of Oregon, Eugene, OR 97403, USA}

\author{Kevin Zumwalt}
\affiliation{Institute of Neuroscience,  University of Oregon, Eugene, OR 97403, USA}

\author{Michael Wehr}
\affiliation{Institute of Neuroscience,  University of Oregon, Eugene, OR 97403, USA}
\affiliation{Department of Psychology, University of Oregon, Eugene, OR 97403, USA}

\author{Santiago Jaramillo}
\affiliation{Institute of Neuroscience,  University of Oregon, Eugene, OR 97403, USA}
\affiliation{Department of Biology, University of Oregon, Eugene, OR 97403, USA}

\author{David A. McCormick}
\affiliation{Institute of Neuroscience,  University of Oregon, Eugene, OR 97403, USA}
\affiliation{Department of Biology, University of Oregon, Eugene, OR 97403, USA}

\author{Luca Mazzucato}
\altaffiliation{Lead Contact}
\altaffiliation{Correspondence: lmazzuca@uoregon.edu}
\affiliation{Institute of Neuroscience,  University of Oregon, Eugene, OR 97403, USA}
\affiliation{Department of Biology, University of Oregon, Eugene, OR 97403, USA}
\affiliation{Department of Mathematics, University of Oregon, Eugene, OR 97403, USA}
\affiliation{Department of Physics, University of Oregon, Eugene, OR 97403, USA}

\maketitle

\begin{center}
\textbf{SUMMARY}
\end{center}
~\\
Past work has reported inverted-U relationships between arousal and auditory task performance, but the underlying neural network mechanisms remain unclear. To make progress, we recorded auditory cortex activity from behaving mice during passive tone presentation and simultaneously monitored pupil-indexed arousal. In these experiments, neural discriminability of tones was maximized at intermediate arousal, revealing a neural correlate of the inverted-U. We explained this arousal-dependent sound processing using a spiking model with clusters. In the model, stimulus discriminability peaked as the network transitioned from a multi-attractor phase exhibiting slow switching between metastable cluster activations (low arousal) to a single-attractor phase with uniform activity (high arousal). This transition also qualitatively captured arousal-induced reductions of neural variability observed in the data. Altogether, this study elucidates computational principles to explain interactions between arousal, neural discriminability, and variability, and suggests that transitions in the dynamical regime of cortical networks could underlie nonlinear modulations of sensory processing.

\clearpage
\newpage

\section*{Introduction}
\label{s:intro}

Variations in brain state -- such as levels of arousal -- significantly impact sensory responses and information processing \cite{mcginley2015waking,mccormick2020neuromodulation,lee2012neuromodulation,flavell2022emergence,harris2011cortical,busse2017sensation,aston2005integrative,yerkes1908relation}. During wakefulness, increases in arousal and arousal-related neuromodulator activity are associated with increases in pupil diameter \cite{mathot2018pupillometry,reimer2016pupil,collins2023cholinergic}, enabling non-invasive monitoring of arousal state in behaving animals. Using pupillometry, several studies have reported an intriguing ``inverted-U" relationship between baseline pupil diameter and task performance, indicating optimal performance at intermediate arousal \cite{mcginley2015cortical, hulsey2023decision, murphy2011pupillometry, waschke2019local, beerendonk2024disinhibitory, schriver2018pupil,ebitz2014pupil}. 

The inverted-U relationship between pupil-indexed arousal and performance has been demonstrated particularly clearly in the context of auditory processing, with examples observed during sound detection and discrimination tasks in mice \cite{mcginley2015cortical, hulsey2023decision} and humans \cite{murphy2011pupillometry, waschke2019local, beerendonk2024disinhibitory}. Although some work has revealed neural substrates of optimal sound detection \cite{mcginley2015cortical}, network-level dynamical principles mediating the inverted-U relationship, especially for sound discrimination, remain unclear. Here, we combined electrophysiology and circuit modeling to shed light on potential network mechanisms underlying optimal, arousal-dependent performance states for auditory discrimination.

Given that neural signatures of inverted-U arousal-performance relationships have been observed in auditory cortex (ACtx) without task engagement \cite{mcginley2015cortical}, we examined how arousal impacted neural discriminability of pure tones during passive stimulus presentation. To this end, we used Neuropixels probes to record sound-evoked activity from ACtx of behaving mice, while simultaneously monitoring arousal levels with pupillometry. We found that tone frequency was most accurately decoded from ACtx activity during intermediate pupil dilation, in line with an inverted-U relationship. This finding extends previous results on neural correlates of optimal sound detection in ACtx \cite{mcginley2015waking} to population-based neural discriminability of auditory stimuli.

To illuminate potential network mechanisms governing this neural manifestation of the inverted-U relationship, we modeled ACtx as a recurrently-connected network of excitatory and inhibitory spiking neurons. We compared two canonical models for the network architecture -- uniform or clustered. In the latter model, neurons were organized into strongly-connected modules representing functional neural assemblies \cite{amit1997model}. Unlike uniformly-connected networks, clustered networks exhibit metastable activity patterns, characterized by spontaneous, transient activations of different neural assemblies \cite{deco2012neural,litwin2012slow,mazzucato2015dynamics, mazzucato2016stimuli, mazzucato2019expectation,wyrick2021state, brinkman2022metastable,la2019cortical}.

We hypothesized that non-monotonic variations of stimulus discriminability might emerge from modulations of the metastable assembly dynamics present in clustered networks. To investigate this, we presented model networks with sensory stimulation and an arousal modulation. Motivated by experimental studies, an increase in arousal was modeled as a suppression of recurrent excitatory synaptic transmission \cite{gil1997differential,favero2012state,levy2006nicotinic,hsieh2000differential,eggermann2009cholinergic, kimura1997distinct,dinh2009norepinephrine,ohshima2017alpha2a,kobayashi2009presynaptic,kobayashi2000selective,hasselmo1997noradrenergic} and an increase in external drive (representing increased thalamic activation) \cite{poulet2012thalamic,mcginley2015cortical,nestvogel2022visual,petty2021effects,urbain2015whisking,aydin2018locomotion,erisken2014effects,molnar2021cell}. Under this implementation, the clustered model reproduced the inverted-U relationship between arousal and stimulus discriminability, while the uniform model failed to display such an effect.

In the clustered model, we show that the inverted-U relationship emerges via a transition from a multi-attractor phase, characterized by slow switching between different highly-active neural assemblies (low arousal), to a single-attractor phase with uniform activity (high arousal). Optimal stimulus discriminability occurred between these two regimes, where assembly activation dynamics were present but flexibly modulated. The clustered model additionally predicted a reduction of neural variability with increasing arousal, and we found evidence of that trend in the experimental data. As a whole, our results suggest that arousal-induced transitions in the collective dynamical regime of a cortical circuit may explain certain aspects of arousal-dependent stimulus processing and neural variability in auditory cortex. Although our study does not rule out alternative explanations for these phenomena, it provides insight into one possible computational mechanism that can be further tested and built upon in future work. 

\begin{figure*}[ht!]
    \centering
    \includegraphics[width=\textwidth]{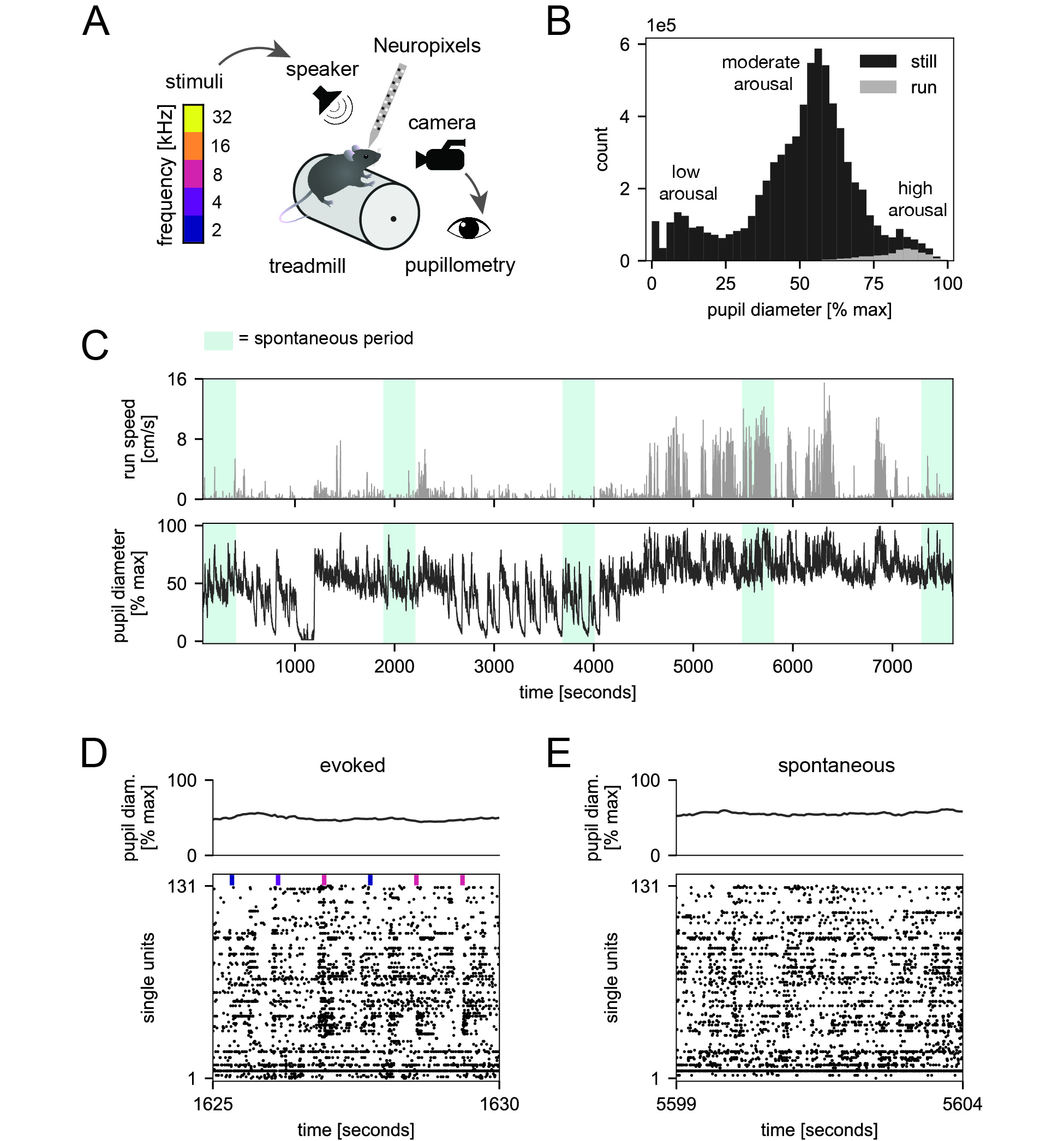}
    \caption[Neuropixels recordings from mouse ACtx during a range of arousal states.]{\textbf{Neuropixels recordings from mouse ACtx during a range of arousal states.} \textbf{(A)} Schematic of experimental setup (\nameref{s:Methods}). \textbf{(B)} Pupil diameter distributions from an example session during stillness (dark gray) or running (light gray). \textbf{(C)} Running speed and pupil diameter traces from an example session. \textbf{(D)} Pupil diameter trace and population raster across 5 seconds of evoked activity; vertical ticks indicate stimulus onset times. \textbf{(E)} Same as \textbf{(D)} but for spontaneous activity.}
    \label{f:experimental_setup}
\end{figure*}

\begin{figure*}[b!]
    \includegraphics[width=\textwidth]{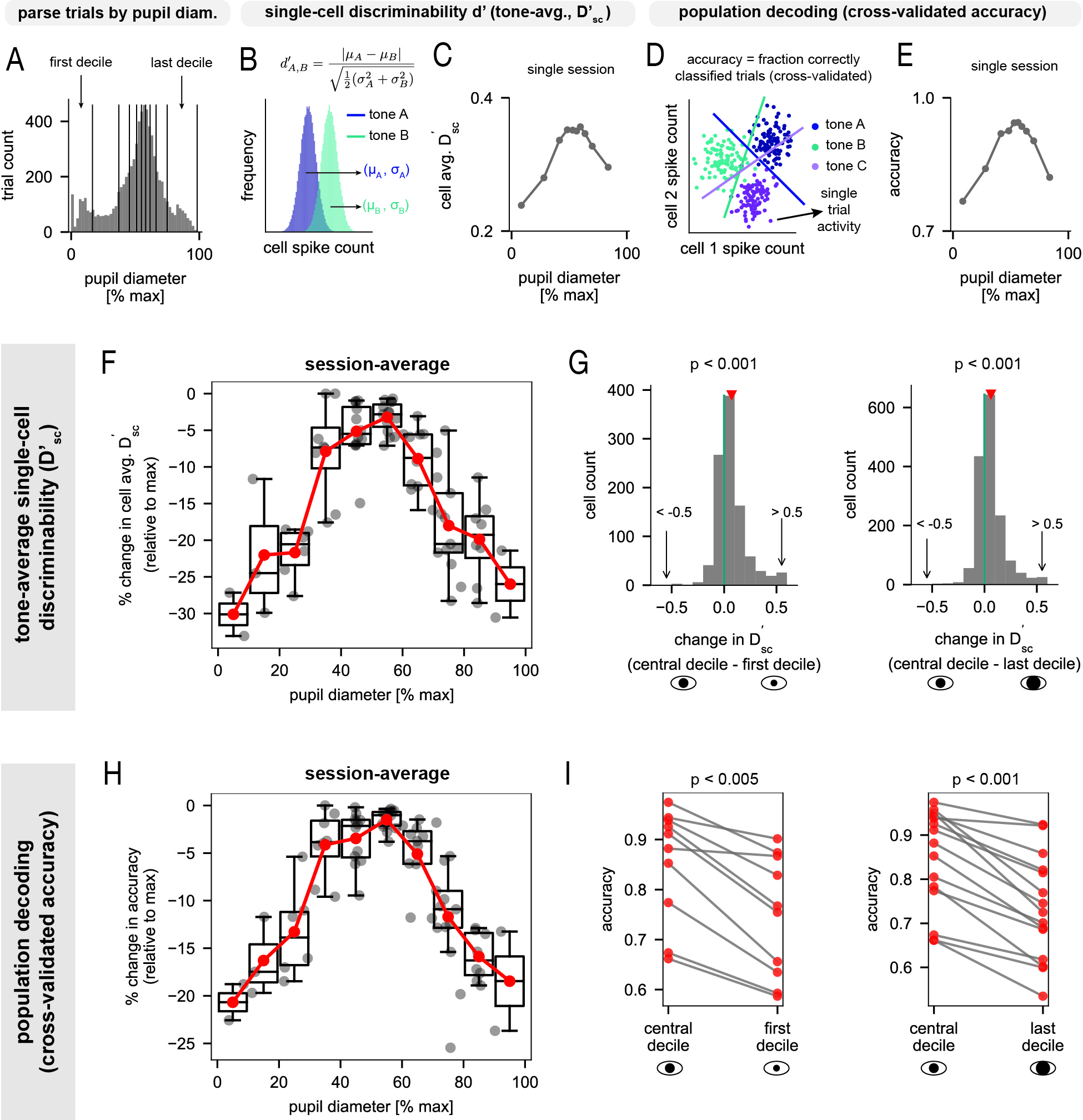}
\end{figure*}

\begin{figure*}[t!]
    \caption[Neural discriminability of pure tones is optimal at intermediate arousal in ACtx.]{\textbf{Neural discriminability of pure tones is optimal at intermediate arousal in ACtx.} \textbf{(A)} Histogram of pre-stimulus pupil diameter from an example session; black lines indicate deciles. \textbf{(B)} Schematic of single-cell discriminability index ($d'$) for one pair of stimuli. An overall measure, $D'_{\mathrm{sc}}$, was computed by averaging $d'$ across all tone pairs (\nameref{s:Methods}). \textbf{(C)} Cell-averaged $D'_{\mathrm{sc}}$ in each pupil decile from \textbf{(A)}. \textbf{(D)} Schematic of linear population decoding analysis (\nameref{s:Methods}). \textbf{(E)} Cross-validated decoding accuracy in each pupil decile from \textbf{(A)}. \textbf{(F)} Percent change in cell-averaged $D'_{\mathrm{sc}}$ (relative to session-maximum) \textit{vs.} pupil diameter (group data from $n=15$ sessions). In each pupil diameter bin, we show single-session data (gray) with the corresponding boxplot and session-average (red). \textbf{(G)} \textit{Left:} Difference in $D'_{\mathrm{sc}}$ between the most central and first pupil decile of a session ($n=1002$ units from 10 sessions with average pupil diameter of first decile $\leq 33 \%$ max dilation; red triangle indicates mean difference; $p < 0.001$, Wilcoxon signed-rank test). \textit{Right:} Distribution of the difference in $D'_{\mathrm{sc}}$ between the most central and last pupil decile of a session ($n=1552$ units from 15 sessions with average pupil diameter of last decile $\geq 67 \%$ max dilation; red triangle indicates mean difference; $p < 0.001$, Wilcoxon signed-rank test). \textbf{(H)} Same as \textbf{(F)} but for cross-validated decoding accuracy. \textbf{(I)} \textit{Left:} Accuracy in the most central and first pupil decile of a session (data from $n=10$ sessions with average pupil diameter of first decile $\leq 33 \%$ max dilation; $p < 0.005$, Wilcoxon signed-rank test). \textit{Right:} Accuracy in the most central and last pupil decile of a session (data from $n=15$ sessions with average pupil diameter of last decile $\geq 67 \%$ max dilation; $p < 0.001$, Wilcoxon signed-rank test). See also Fig.~\ref{f:decoding_dprime_supp_data}.}
\label{f:decoding_allTrials}
\end{figure*}

\section*{Results}

We used Neuropixels probes to record from populations of auditory cortical neurons in head-fixed mice (primarily targeting A1), while simultaneously monitoring locomotion speed and pupil diameter (Fig.~\ref{f:experimental_setup}A; \nameref{s:Methods}). The pupil diameter in each session was normalized by its maximum value and was used as a proxy for arousal level (Fig.~\ref{f:experimental_setup}B,C). A full spectrum of arousal states was expressed in several sessions, and the middle-to-high pupil range was expressed in the remaining recordings. Single-unit activity was measured during passive sound presentation (Fig.~\ref{f:experimental_setup}D, ``evoked" periods) and in the absence of auditory stimuli (Fig.~\ref{f:experimental_setup}E, ``spontaneous" periods). During evoked periods, mice were presented with brief pure tones (25 ms, 2-32 kHz).

\subsection*{Neural discriminability of pure tones is optimal at intermediate arousal in ACtx}

To test whether arousal impacts neural discriminability of pure tones in ACtx, we split the trials in each session by pupil diameter (see Fig.~\ref{f:decoding_allTrials}A for an example). Within each pupil-based partition, we then computed a single-cell measure of neural stimulus discriminability ($D'_{\mathrm{sc}}$; Fig.~\ref{f:decoding_allTrials}B,C,  \nameref{s:Methods}). On average across sessions, the cell-averaged $D'_{\mathrm{sc}}$ followed an inverted-U relationship with normalized pupil diameter (Fig.~\ref{f:decoding_allTrials}F). At the population level, intermediate pupil diameters were associated with significant increases in $D'_{\mathrm{sc}}$ relative to small or large diameters (Fig.~\ref{f:decoding_allTrials}G), and in individual sessions, the cell-averaged $D'_{\mathrm{sc}}$ was highest at moderate pupil diameters (Fig.~\ref{f:decoding_dprime_supp_data}A). 

Auditory information is also encoded in the collective activity of neuronal ensembles \cite{averbeck2006neural,panzeri2022structures,bathellier2012discrete,harris2012cell,pachitariu2015state,christison2017contribution,ince2013neural}. To understand how arousal might affect the ability of downstream areas to read out sound information from ACtx ensemble activity, we trained an ideal-observer linear classifier to decode tone frequency from single-trial population spike-counts (Fig.~\ref{f:decoding_allTrials}D, \nameref{s:Methods}). Training and testing was done separately for each pupil-based partition in a session, and results were summarized with the cross-validated decoding accuracy (Fig.~\ref{f:decoding_allTrials}E).

On average across sessions, decoding performance exhibited an inverted-U relationship with pupil diameter (Fig.~\ref{f:decoding_allTrials}H), and there was a statistically significant increase in accuracy in states of moderate pupil dilation relative to more constricted or highly-dilated pupil states (Fig.~\ref{f:decoding_allTrials}I). Moreover, the best performance in all sessions was achieved at intermediate pupil diameters, and the worst performance at relatively small or large diameters (Fig.~\ref{f:decoding_dprime_supp_data}B). The session-averaged decoding performance still followed an inverted-U after excluding locomotion trials (Fig.~\ref{f:decoding_dprime_supp_data}C,D), though the right-hand-side decline was less pronounced. This may be due to the fact that average pupil diameters were smaller without movement data (Fig.~\ref{f:decoding_dprime_supp_data}E). The decoding results were also similar when pupil diameter was normalized by the global maximum across all sessions (Fig.~\ref{f:decoding_dprime_supp_data}F,G). Altogether, our findings reveal neural signatures of an inverted-U relationship between arousal level and neural sound discriminability at single-cell and population levels.

\begin{figure*}
    \centering
    \includegraphics[width=\textwidth]{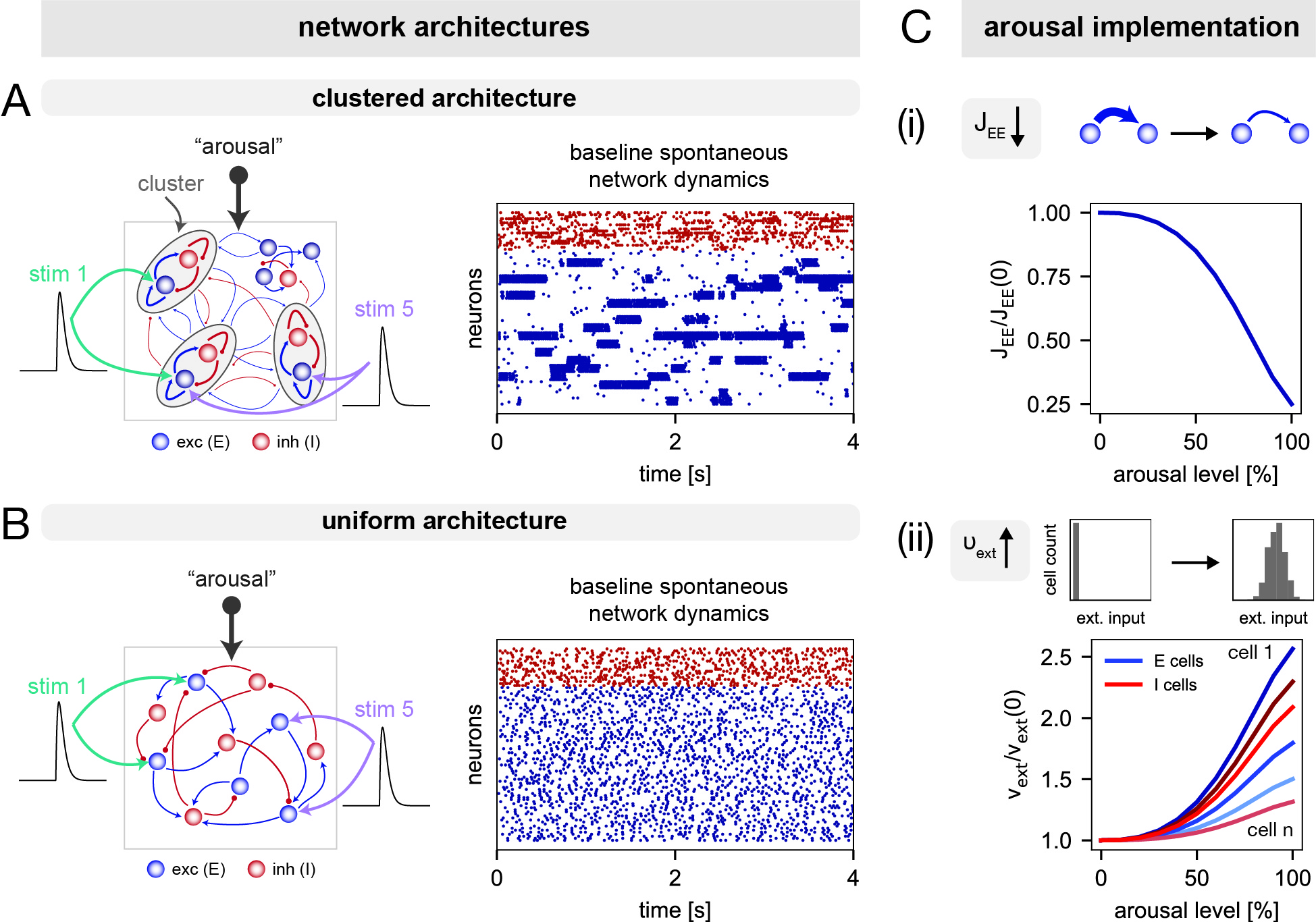}
    \caption[Network modeling of arousal-dependent stimulus processing in ACtx.]{\textbf{Network modeling of arousal-dependent stimulus processing in ACtx.} \textbf{(A, B)} ACtx was modeled as a recurrent network of spiking neurons arranged in either a clustered (\textbf{A Left}) or uniform (\textbf{B Left}) architecture. Both networks were subjected to an arousal signal and sensory stimulation. Raster plots show spontaneous baseline activity of a subset of neurons from a clustered (\textbf{A Right}) or uniform (\textbf{B Right}) network. \textbf{(C)} An increase in arousal was modeled as a  simultaneous modulation of two parameters. \textbf{(C-i)} Synaptic strength $J_{\mathrm{EE}}$ (relative to baseline value) \textit{vs.} arousal. \textbf{(C-ii)} External input $\nu_{\mathrm{ext}}$ (relative to baseline value) \textit{vs.} arousal; different curves correspond to different example cells. See also Table~\ref{t:model_params} and Figs.~\ref{f:rate_corr_data_model}, \ref{f:sd_nu_ext_e_arousalModel}. See \nameref{s:Methods} for model details.}
    \label{f:model_overview}
\end{figure*}

\subsection*{Network modeling of arousal-dependent stimulus processing in ACtx}

What circuit mechanisms can explain the inverted-U relationship between tone discriminability and arousal in ACtx? Because the inverted-U relationship is nonlinear, we speculated that it could stem from a modulation of collective dynamics akin to a phase transition. To investigate this, we modeled ACtx as a recurrently-connected network of excitatory (E) and inhibitory (I) integrate-and-fire neurons (Fig.~\ref{f:model_overview}A,B; \nameref{s:Methods}). This type of spiking model strikes a balance between biological plausibility and tractability, and also allows for analysis of spiking variability (Fig.~\ref{f:fano_factor_main}).

The collective activity of a cortical network depends on the architecture of synaptic couplings. Here, we considered a model with ``clustered" architecture (Fig.~\ref{f:model_overview}A) and compared it to a control with ``uniform" architecture (Fig.~\ref{f:model_overview}B). By comparing these alternative models, we aimed to elucidate potential dynamical principles underlying the experimental observations. Following previous work, neurons in the clustered model were arranged into clusters with strong internal synaptic coupling and relatively weak coupling to other clusters \cite{litwin2012slow,wyrick2021state,mazzucato2015dynamics,amit1997model,deco2012neural}. Neurons in the uniform model were instead connected randomly with homogeneous coupling strengths. In both models, auditory stimuli were implemented as external excitatory inputs (Fig.~\ref{f:model_overview}A,B). In the clustered model, stimulus input targeted cells in a randomly-chosen subset of the clusters, and in the uniform model, stimulus input targeted a random subset of all excitatory cells. To match the experiments, we modeled five stimuli and allowed for overlap in the cell subgroups targeted by different stimuli, in line with the fact that auditory neurons can respond to multiple tones.

As shown in past studies, the clustered model exhibits metastable attractor dynamics \cite{litwin2012slow,brinkman2022metastable,mazzucato2015dynamics,mazzucato2019expectation,wyrick2021state,la2019cortical, deco2012neural, amit1997model}. These dynamics occur in a regime of strong intracluster coupling, where the network’s state space contains a multiplicity of attractors corresponding to different configurations of active clusters (Fig.~\ref{f:mft_supp}A,B) \cite{litwin2012slow,brinkman2022metastable,mazzucato2015dynamics}. In such a regime, random fluctuations can cause transitions between attractors; this generates metastable dynamics wherein individual clusters switch between high and low firing modes over time, and collective network activity moves between different collective states characterized by different groups of simultaneously active clusters (Fig.~\ref{f:model_overview}A Right). Throughout the text, we use the term  ``multistability" to denote the property of the state space having multiple attractors, and the term ``metastable" to denote the itinerant dynamics describing spontaneous transitions between attractors. We hypothesized that modulations of metastable cluster dynamics could provide a mechanism for the inverted-U relationship. In contrast to the clustered model, the uniform model has a single attractor with asynchronous activity (Fig.~\ref{f:model_overview}B Right).

The second aspect of the model is the arousal implementation. Experimental studies indicate that arousal modulates sensory processing through various neuromodulatory systems [in particular, via the actions of acetylcholine (Ach) and norepinephrine (NE)] and thalamocortical mechanisms \cite{zagha2014neural,berridge2003locus,slater2022neuromodulatory,semba1991cholinergic,mccormick2020neuromodulation,mcginley2015waking,sara2009locus,mccormick1989cholinergic,mccormick1997sleep,poulet2019cortical,lee2012neuromodulation}. Given that these pathways can induce a variety of effects on cortical circuits \cite{yang2021cholinergic,picciotto2012acetylcholine, colangelo2019cellular,o2012norepinephrine,berridge2003locus,slater2022neuromodulatory,nadim2014neuromodulation,mccormick1997sleep,mccormick1989cholinergic,salgado2016layer,poulet2019cortical,lee2012neuromodulation}, there are several possibilities for how to model the impact of arousal. Here, we considered one implementation involving a simultaneous modulation of two parameters. In particular, an increase in arousal was modeled as (i) a global decrease in the strength of recurrent E-to-E synapses, and (ii) an increase in the external excitatory drive to E and I cells; to improve biological plausibility, we also introduced cell-to-cell variability in the external drive modulation (Fig.~\ref{f:model_overview}C).

The first component of the arousal model is motivated by experiments showing that increases in Ach and NE can have suppressive effects on intracortical excitatory synaptic transmission \cite{gil1997differential,favero2012state,levy2006nicotinic,hsieh2000differential,eggermann2009cholinergic,kimura1997distinct,hasselmo1992cholinergic,favero2012state,dinh2009norepinephrine,ohshima2017alpha2a,kobayashi2009presynaptic,kobayashi2000selective,hasselmo1997noradrenergic}. We reasoned that such a modulation of synaptic efficacy would strongly impact assembly dynamics in the clustered model -- which rely on recurrent excitation -- and thus impact stimulus processing. The second component of the arousal model is motivated by studies showing that active states are associated with increases in thalamic activity, which provides a major source of excitatory drive to sensory cortices \cite{poulet2012thalamic,mcginley2015cortical,nestvogel2022visual,petty2021effects,urbain2015whisking,aydin2018locomotion,erisken2014effects,molnar2021cell}. Importantly, this arousal implementation qualitatively captured the heterogeneity of arousal-related firing rate modulations observed in the data (Fig.~\ref{f:rate_corr_data_model}).

In what follows, we examine how collective network dynamics are affected by the above arousal modulation. We refer to the induced changes in network dynamics as the ``computational" or ``network mechanism", in contrast to the ``biological implementation" of arousal, which refers to the specific neurophysiological pathways and modulations of network parameters used to model arousal. Though here we examined one biological implementation of arousal supported by empirical evidence, alternative implementations can produce similar network mechanisms (see Fig.~\ref{f:sd_nu_ext_e_arousalModel} and Discussion). 

\begin{figure*}[b!]
    \centering
    \includegraphics[width=\textwidth]{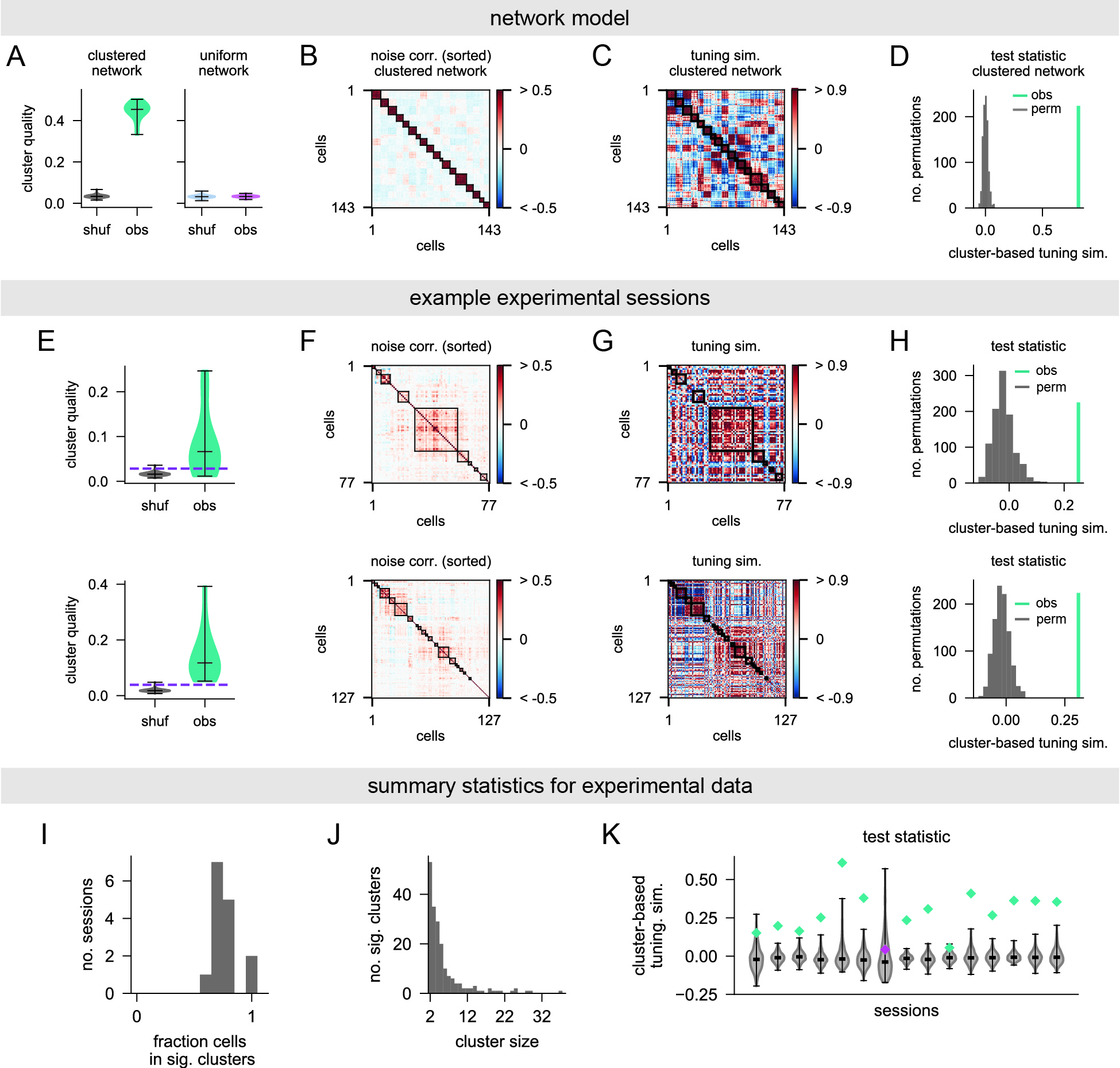}
\end{figure*}

\begin{figure*}[t!]
    \centering
    \caption[Evidence of functionally-organized correlation-based clusters in ACtx.]{\textbf{Evidence of functionally-organized correlation-based clusters in ACtx.} \textbf{(A-D)} Clustering analysis in the network models. \textbf{(A)} Violin plots of cluster quality obtained from hierarchical clustering of observed (``obs”) and trial-shuffled (``shuf”) noise-correlation matrices. In the example clustered network, all observed clusters are significant relative to trial-shuffled data ($p<0.05$, Bonferroni-corrected); in the example uniform network, no clusters are significant. \textbf{(B)} Noise correlation matrix of a random subsample of cells from a clustered network, sorted according to detected clusters. \textbf{(C)} Tuning similarity matrix for the same cells and sorting in \textbf{(B)}. \textbf{(D)} Test-statistic for cluster-based tuning similarity, computed from the data in \textbf{(B,C)}. The observed value exceeds the distribution obtained by permuting cluster labels across cells ($p<0.001$). \textbf{(E-H)} Clustering analysis in two experimental sessions (1 session/row). \textbf{(E)} Same as \textbf{(A)}, but for the experimental sessions. Clusters in the observed distribution above the dashed line are significant relative to trial-shuffled data ($p<0.05$, Bonferroni-corrected). \textbf{(F)} Noise correlation matrices sorted according to detected clusters; significant clusters are outlined in black. \textbf{(G)} Tuning similarity matrices for the same cells and sorting in \textbf{(F)}. \textbf{(H)} Same as \textbf{(D)}, but for the experimental sessions. The observed values exceed the distributions obtained by permuting cluster labels across cells ($p<0.001$). \textbf{(I-K)} Summary statistics for experimental data. \textbf{(I)} Distribution of the fraction of cells in a session that belong to significant clusters. \textbf{(J)} Distribution of cluster sizes (significant clusters only). \textbf{(K)} Test statistic for cluster-based tuning similarity in each session. Colored diamonds indicate observed values and violin plots show distributions obtained by permuting cluster labels across cells. Green indicates a significant result relative to permuted data ($p < 0.05$) and magenta otherwise. See \nameref{s:Methods} for details.}
    \label{f:cluster_analysis_main}
\end{figure*}

\subsection*{Evidence of functionally-organized correlation-based clusters in ACtx}

A distinctive feature of the clustered model is that noise correlations -- covariations in activity across repeated presentations of the same stimulus \cite{cohen2011measuring} -- are larger between cells in the same \textit{versus} different clusters (Fig.~\ref{f:cluster_analysis_main}B). To better motivate the clustered model, we thus tested for the presence of functional clusters in the data. To identify putative clusters, we estimated pairwise noise correlations and performed hierarchical clustering on the resulting correlation matrix. The statistical significance of detected clusters was assessed via comparison against a trial-shuffled null model that contained no true clustering (\nameref{s:Methods}). 

In the clustered model, significant functional assemblies were clearly detected with hierarchical clustering (Fig.~\ref{f:cluster_analysis_main}A Left), and neurons were accurately assigned to their ground-truth clusters (Fig.~\ref{f:cluster_analysis_main}B). By contrast, significant clusters were rarely identified in noise-correlation matrices from the uniform model (Fig.~\ref{f:cluster_analysis_main}A Right). Correlations in the experimental data were typically weaker and more diffuse relative to the clustered model (Fig.~\ref{f:cluster_analysis_main}F for two examples). Nonetheless, the clustering analysis revealed significant correlation-based assemblies (Fig.~\ref{f:cluster_analysis_main}E). In most sessions, a majority of neurons belonged to significant clusters (Fig.~\ref{f:cluster_analysis_main}I), and cluster sizes ranged from a few to tens of cells (Fig.~\ref{f:cluster_analysis_main}J).

To assess the functional relevance of inferred clusters, we tested whether cells in the same cluster had higher-than-chance tuning similarity (Pearson correlation between trial-averaged stimulus responses \cite{cohen2011measuring}). For each cell, we computed the difference between its average within- and between-cluster tuning similarity, and defined a test statistic (``cluster-based tuning similarity") as the average across cells in significant clusters. The observed value of the test statistic was then compared to a null distribution obtained by randomly permuting cluster labels across neurons (\nameref{s:Methods}).

In the clustered model, cluster-based tuning similarity was significantly greater than chance (Figs.~\ref{f:cluster_analysis_main}C,D). Because neurons in the same cluster exhibit coordinated dynamics, the presence (absence) of stimulus-related drive biases activation (inactivation) of the entire assembly, leading to similar stimulus responses for its constituent neurons. The relationship between correlation-based clusters and tuning similarity was less straightforward in the data. However, we still observed a certain degree of overlap, which was visually apparent in some sessions (Fig.~\ref{f:cluster_analysis_main}G) and verified with the statistical test described above (Fig.~\ref{f:cluster_analysis_main}H). In total, cluster-based tuning similarity was significant in nearly all sessions (Fig.~\ref{f:cluster_analysis_main}K), suggesting that the detected correlation-based clusters exhibit some functional organization. These results provide additional motivation for the clustered model.

\begin{figure*}
    \centering
    \includegraphics[width=\textwidth]{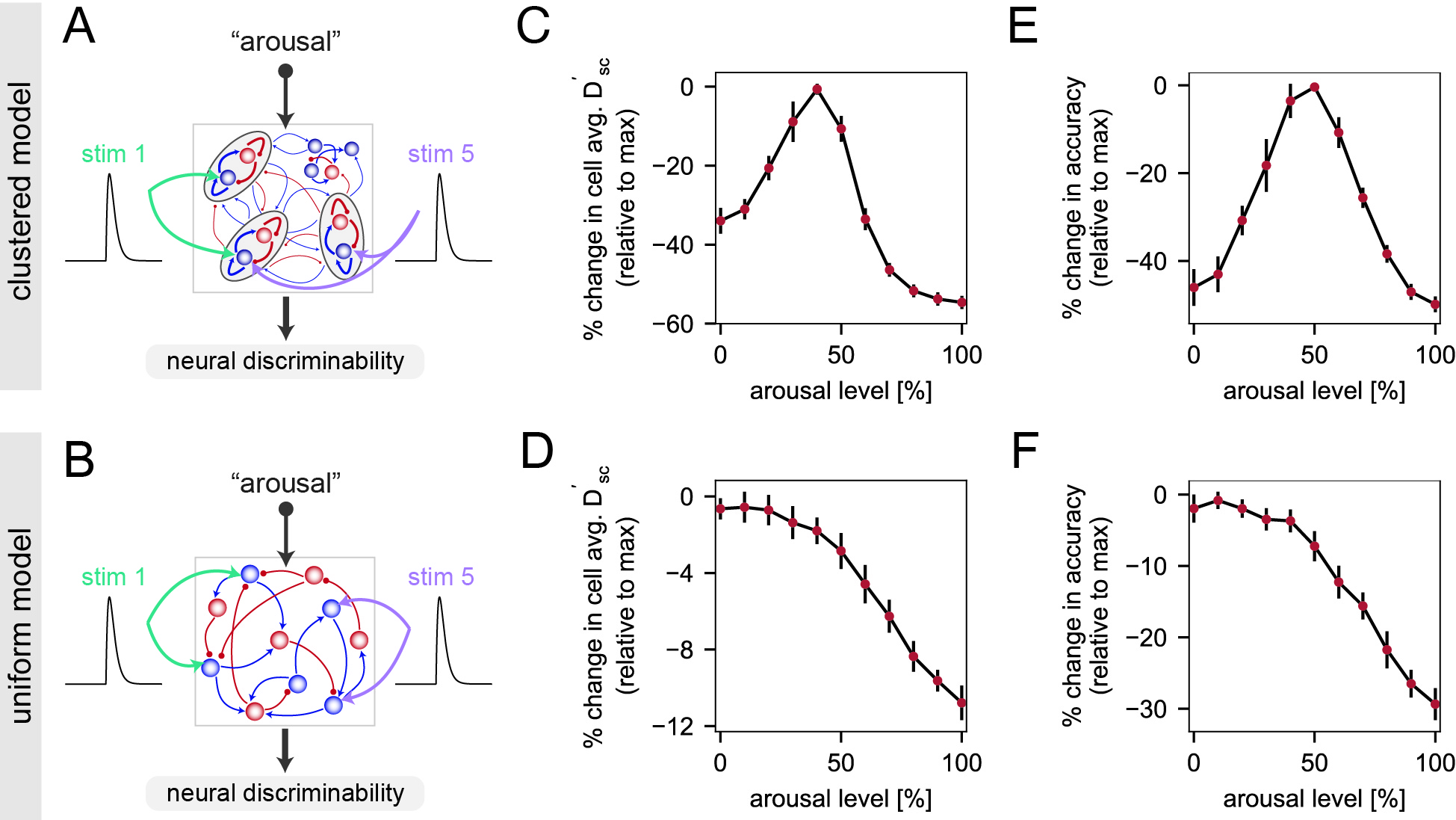}
    \caption[The clustered model captures the inverted-U relationship between stimulus discriminability and arousal.]{\textbf{The clustered model captures the inverted-U relationship between stimulus discriminability and arousal.} \textbf{(A,B)} Five different stimuli were presented several times to the clustered \textbf{(A)} and uniform \textbf{(B)} networks, and measures of neural stimulus discriminability were computed from the responses. \textbf{(C, D)} Percent change in cell-averaged $D'_{\mathrm{sc}}$ \textit{vs.} arousal in the clustered (\textbf{C}) or uniform (\textbf{D}) models (percent change  was computed relative to the maximum across arousal levels; \nameref{s:Methods}). \textbf{(E, F)} Same as \textbf{(C, D)} but for cross-validated population decoding accuracy (\nameref{s:Methods}). For this analysis, linear classification was performed using ensemble activity from $10\%$ of the excitatory cells (in the clustered model, an approximately equal number of cells were used from each cluster). In panels \textbf{C-F}, data points and error bars indicate the mean $\pm$ 1 SD across network realizations. See Fig.~\ref{f:decoding_model_supp} for results with different population sizes.}
    \label{f:decoding_model}
\end{figure*}

\subsection*{The clustered model captures the inverted-U relationship between stimulus discriminability and arousal}

We next examined whether the inverted-U relationships between stimulus discriminability and arousal (Fig.~\ref{f:decoding_allTrials}) could be reproduced in either the uniform or clustered circuit models. We began by investigating how the single-cell discriminability index ($D'_{\mathrm{sc}}$) varied with the arousal modulation (\nameref{s:Methods}). In the clustered model, the cell-averaged $D'_{\mathrm{sc}}$ peaked at moderate arousal (Fig.~\ref{f:decoding_model}C), whereas in the uniform model, it decreased with arousal (Fig.~\ref{f:decoding_model}D). Thus, at the level of single-cell discriminability, only the clustered model captured the inverted-U relationship with arousal observed in the data.

We also examined how the arousal modulation impacted the read-out of stimulus identity from population activity. To this end, we trained ideal-observer linear decoders to discriminate stimuli given responses from an ensemble of neurons sampled from the full network (\nameref{s:Methods}). Similar to single-cell discriminability, cross-validated decoding performance followed an inverted-U relationship in the clustered networks, but decreased with arousal in the uniform networks (Fig.~\ref{f:decoding_model}E,F). Though decoding performance at moderate and high arousal did increase with population size in the clustered networks, the overall inverted-U shape of the decoding curve was relatively robust to variations in the number of sampled neurons (Fig.~\ref{f:decoding_model_supp}).

In sum, the clustered network can explain the inverted-U relationships between stimulus discriminability and arousal observed in the data, while the uniform network fails to do. In what follows, we examine the network mechanisms underlying the inverted-U behavior in the clustered model.

\begin{figure*}
    \centering
    \includegraphics[width=\textwidth]{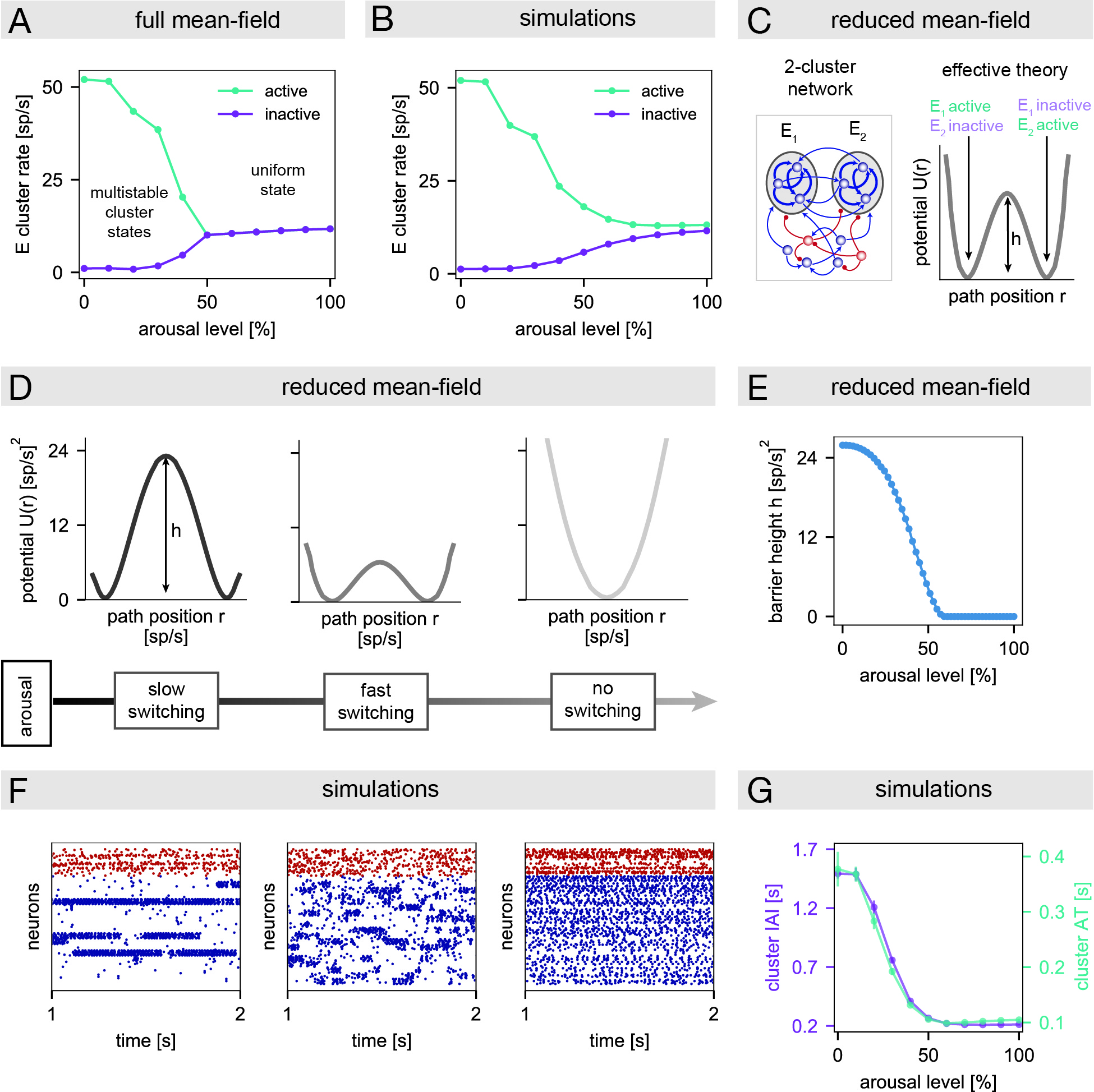}
    \caption[The arousal modulation controls the dynamical regime of the clustered model.]{\textbf{The arousal modulation controls the dynamical regime of the clustered model.} \textbf{(A)} Mean-field firing rates of active and inactive excitatory clusters \textit{vs.} arousal. In these analyses, the mean-field calculations used a larger intracluster coupling than the simulations, so the comparison to panel \textbf{(B)} is only qualitative (\nameref{s:Methods}; see also Fig.~\ref{f:mft_supp}A-C). \textbf{(B)} Average firing rate of active and inactive excitatory clusters from simulations \textit{vs.} arousal. \textbf{(C)} Schematic of the reduced mean-field analysis using a simplified network of two excitatory clusters. The behavior of the two clusters can be described via an effective potential energy, where the two wells correspond to the network's two attractors (\nameref{s:Methods}; see also Fig.~\ref{f:mft_supp}D-H, Table~\ref{t:model_params_2clusterNet}). \textbf{(D)} The effective potential of the 2-cluster network at three increasing levels of arousal. \textbf{(E)} The barrier height $h$ of the effective potential \textit{vs.} arousal. \textbf{(F)} Example raster plots from simulations of the full clustered networks at three increasing levels of arousal. \textbf{(G)} The average cluster inter-activation interval (IAI, left axis) and average cluster activation time (AT, right axis) \textit{vs.} arousal in simulations of the full clustered networks (\nameref{s:Methods}). In panels \textbf{(B)} and \textbf{(G)}, circular markers and error bars indicate the mean $\pm$ 1 SD across network realizations.}
    \label{f:mean_field_vs_sims}
\end{figure*}

\subsection*{The arousal modulation controls the dynamical regime of the clustered model}

Because the inverted-U relationship emerged only in the clustered networks, it must rely on an arousal-induced modulation of the ongoing metastable dynamics unique to that model. To understand the mechanism through which arousal modulates stimulus discriminability, we first used mean-field theory (MFT) to elucidate the effects of arousal on the clustered network's attractor landscape under spontaneous conditions (\nameref{s:Methods}). Though the MFT does not quantitatively describe the simulations, it provides useful qualitative insights.

At low arousal, MFT reveals the presence of multiple attractors in which different subsets of clusters are highly active (multistable cluster states). Beyond a certain arousal, however, the MFT indicates a transition to a regime with only a single-attractor phase (uniform state), wherein all clusters have the same moderate firing rate (Fig.~\ref{f:mean_field_vs_sims}A). The MFT thus predicts that increasing arousal will reduce the contrast between the firing rate of active and inactive firing modes, an intuition that was qualitatively confirmed in network simulations (Fig.~\ref{f:mean_field_vs_sims}B). These modulations of the collective dynamics are driven by the suppression of recurrent synaptic excitation onto pyramidal neurons, which hinders the ability of clusters to achieve and maintain a high activity state.

The arousal modulation could also impact the timescale of cluster switching dynamics. To theoretically elucidate the effects, we analyzed a reduced network composed of two excitatory clusters (Fig.~\ref{f:mean_field_vs_sims}C Left; \nameref{s:Methods}). This network also displays cluster states, but has a simplified landscape with two attractors in which either cluster is active (Fig.~\ref{f:mft_supp}D-H). Using effective mean field theory \cite{mascaro1999effective,mattia2013heterogeneous,mazzucato2019expectation,wyrick2021state}, the attractors can be represented by two potential wells separated by a barrier (Fig.~\ref{f:mean_field_vs_sims}C Right). The height $h$ of this barrier controls the rate of stochastic transitions between attractors \cite{litwin2012slow,hanggi1990reaction,mazzucato2019expectation}.

The barrier height changed with arousal level. At low arousal, the two attractors were separated by a relatively large barrier, indicating inflexible dynamics with slow switching between attractors (Fig.~\ref{f:mean_field_vs_sims}D Left). At intermediate arousal, the two wells were preserved but the barrier height decreased (Fig.~\ref{f:mean_field_vs_sims}D Middle), implying more flexible cluster dynamics with faster switching between configurations. For yet larger arousal, there was a transition from the 2-attractor phase to a single-attractor phase (Fig.~\ref{f:mean_field_vs_sims}D Right); this transition indicates the loss of metastable cluster states. The theoretical insights from the reduced circuit were verified in simulations of the full clustered model. Specifically, we observed decreases in the average cluster inter-activation interval and activation time with increasing arousal (Fig.~\ref{f:mean_field_vs_sims}F Left, Middle; Fig.~\ref{f:mean_field_vs_sims}G), in accordance with the shrinking barrier in the reduced network (Fig.~\ref{f:mean_field_vs_sims}E). Visual inspection of network activity also revealed a degradation of metastable cluster states with increasing arousal (Fig.~\ref{f:mean_field_vs_sims}F Right), consistent with a transition to a uniform phase.

\begin{figure*}
    \centering
    \includegraphics[width=\textwidth]{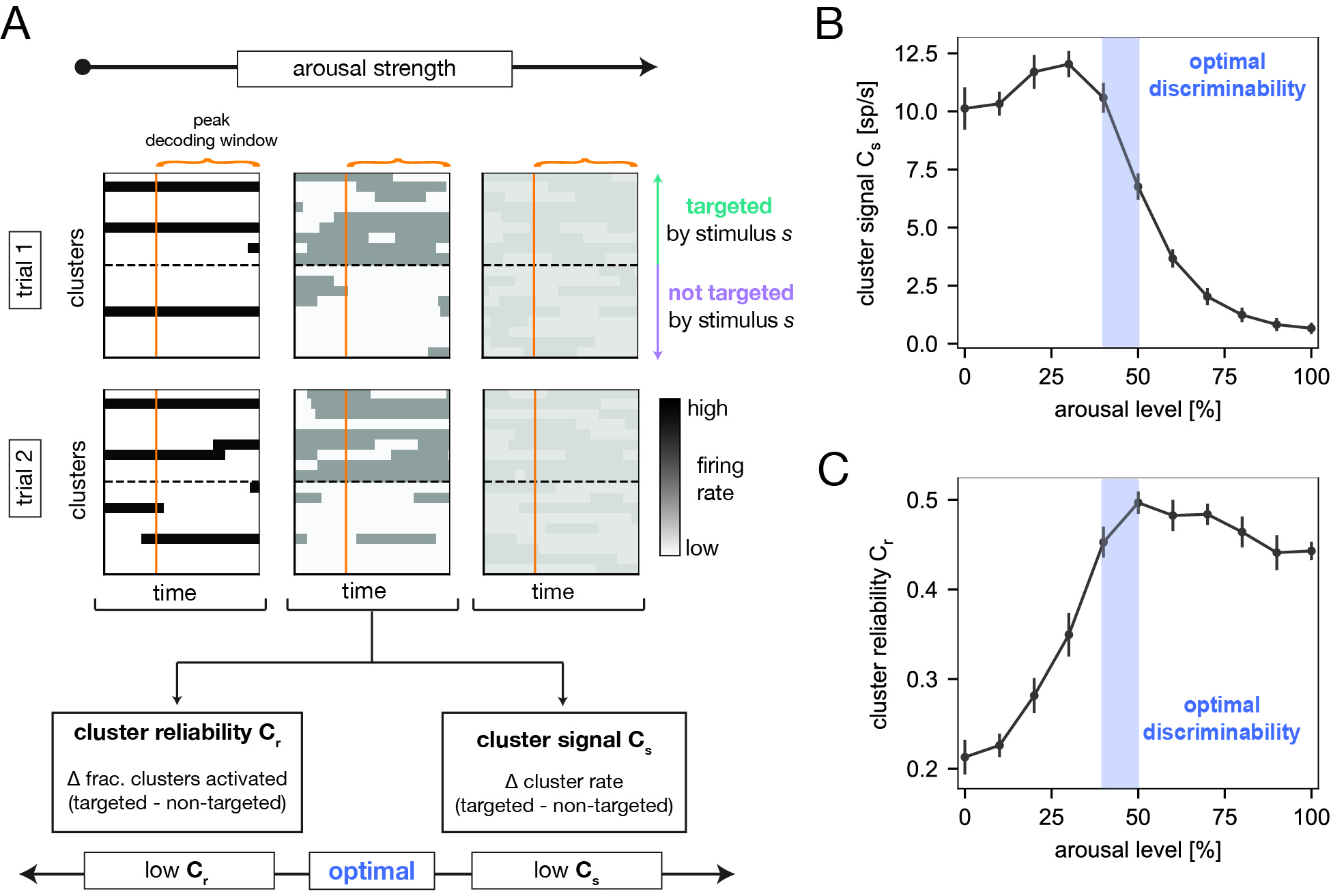}
    \caption[Modulations of cluster dynamics underlie the inverted-U relationship.]{\textbf{Modulations of cluster dynamics underlie the inverted-U relationship.} \textbf{(A)} Schematics demonstrating variations in single-trial evoked responses as a function of arousal in the clustered model. Each panel depicts single-trial cluster firing rates in response to a stimulus $s$ (plotted relative to the time of peak decoding accuracy). At a given arousal level, we computed two quantities to characterize the cluster activity pattern: the ``cluster signal" $C_{s}$ and the ``cluster reliability" $C_{r}$ (\nameref{s:Methods}). \textbf{(B)} The cluster signal \textit{vs.} arousal. \textbf{(C)} The cluster reliability \textit{vs.} arousal. Data points and error bars indicate the mean $\pm$ 1 SD across network realizations. The purple area indicates the region of optimal stimulus discriminability (Fig.~\ref{f:decoding_model}).}
    \label{f:decoding_intuition_model}
\end{figure*}

\subsection*{Modulations of cluster dynamics underlie the inverted-U relationship}

We used the insights of the previous section to develop intuition for the inverted-U nature of stimulus discriminability. To begin, we note that stimulus identity would be perfectly read-out from population activity if each stimulus could strongly activate all of its targeted clusters on every trial and strongly suppress all non-targeted clusters. To quantify the extent to which the ideal scenario occurs, we examined two properties of network activity following stimulus presentation (Fig.~\ref{f:decoding_intuition_model}A; \nameref{s:Methods}): (i) the difference between the average firing rates of targeted and non-targeted clusters (``cluster signal"), and (ii) the difference between the fractions of targeted and non-targeted clusters that are in an activated state (``cluster reliability").

The cluster signal increased slightly and then strongly decreased as a function of arousal strength (Fig.~\ref{f:decoding_intuition_model}B). At low arousal, there is a large separation in the spontaneous firing rates of active and inactive clusters; because stimulus presentation biases the activation of targeted clusters, the cluster signal is high in this regime (Fig.~\ref{f:decoding_intuition_model}A Left). When arousal increases to moderate levels, the contrast between active and inactive clusters decreases but remains substantial; at the same time, transitions between cluster states are induced more easily. These effects enable an increase in the relative amount of targeted cluster activation in response to a stimulus, and the cluster signal remains relatively high (Fig.~\ref{f:decoding_intuition_model}A Middle). As arousal increases further, the spontaneous firing rates of active and inactive clusters further converge. In consequence, the contrast between the average evoked responses of targeted and non-targeted clusters also decreases, and the cluster signal eventually drops to near zero (Fig.~\ref{f:decoding_intuition_model}A Right).

The cluster reliability, in contrast, increased from low to moderate arousal, and then slightly decreased at high arousal (Fig.~\ref{f:decoding_intuition_model}C). At low arousal, spontaneous cluster dynamics are slow and inflexible and only a fraction of all clusters activate in a fixed time window. Because stimuli are not strong enough to completely override the ongoing dynamics, only a fraction of all targeted clusters become activated in response to stimulation, and sometimes non-targeted clusters fail to deactivate. This results in inconsistent activation of targeted clusters and low cluster reliability at low arousal (Fig.~\ref{f:decoding_intuition_model}A Left). As arousal increases, cluster dynamics are more malleable and a larger fraction of clusters activate in a fixed time window. These effects drive the increase in cluster reliability at moderate arousal (Fig.~\ref{f:decoding_intuition_model}A Middle). At very high arousal, non-targeted clusters are not as strongly suppressed (relative to the amount that targeted clusters are activated), which leads to the slight decrease in cluster reliability (Fig.~\ref{f:decoding_intuition_model}A Right); however, it is difficult to estimate reliability at high arousal because the distinction between active and inactive clusters is not well-defined.

The variations in cluster signal and reliability provide intuition for the inverted-U behavior of neural stimulus discriminability. At intermediate arousal, both the signal and reliability are relatively high (Fig.~\ref{f:decoding_intuition_model}B,C). In this optimal regime, stimulus discriminability is maximal. For both lower and higher arousal, either the reliability or signal drop substantially, and discriminability is worse. The key insight is that the arousal modulation affects both the overall strength and consistency of cluster activation patterns, which combine to determine how well the responses to different stimuli can be distinguished.

\begin{figure*}
    \centering
    \includegraphics[width=\textwidth]{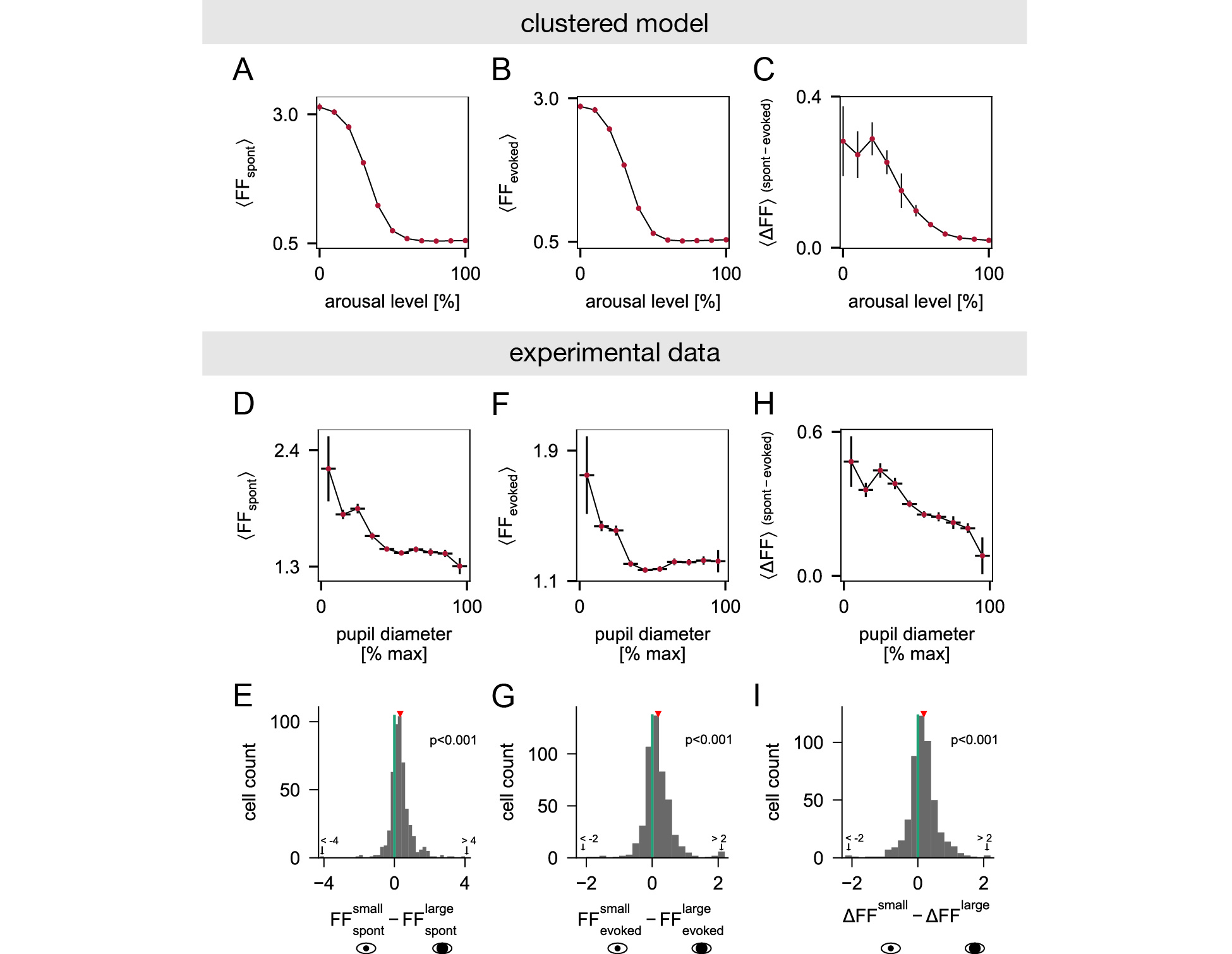}
    \caption[The clustered model qualitatively captures arousal-related modulations of neural variability.]{\textbf{The clustered model qualitatively captures arousal-related modulations of neural variability.}
    \textbf{(A-C)} Results from the clustered model. \textbf{(A)} Population-averaged spontaneous FF ($\mathrm{FF_{spont}}$) \textit{vs.} arousal. Data points and error bars indicate the mean $\pm$ SD of the population-averaged $\mathrm{FF_{spont}}$ across network realizations. \textbf{(B)} Same as \textbf{(A)} but for evoked FF ($\mathrm{FF_{evoked}}$). \textbf{(C)} Same as \textbf{(A)} but for spontaneous minus evoked FF ($\Delta \mathrm{FF}$). \textbf{(D-I)} Results from the experimental data. \textbf{(D)} Population-averaged $\mathrm{FF_{spont}}$ \textit{vs.} pupil diameter (units pooled over sessions). Horizontal error bars indicate pupil diameter bins; data points and vertical error bars indicate mean $\pm$ SEM of $\mathrm{FF_{spont}}$ across cells from all sessions that contributed to the corresponding pupil bin. \textbf{(E)} Distribution of the difference in $\mathrm{FF_{spont}}$ between small and large pupil diameters [$n=487$ units pooled over 9 sessions with average pupil diameter of smallest (largest) decile bin $\leq 33\%$ ($\geq 67\%$) max dilation; red triangle indicates mean difference; $p< 0.001$, Wilcoxon signed-rank test]. \textbf{(F, G)} Same as \textbf{(D,E)}, but for $\mathrm{FF_{evoked}}$ ($p< 0.001$, Wilcoxon signed-rank test). \textbf{(H,I)} Same as \textbf{(D,E)}, but for $\Delta \mathrm{FF}$ ($p< 0.001$, Wilcoxon signed-rank test). See also Figs.~\ref{f:cvISI_spectra_supp}, \ref{f:spectra_fano_supp}. Methodological details are provided in \nameref{s:Methods}.}
    \label{f:fano_factor_main}
\end{figure*}

\subsection*{The clustered model qualitatively captures arousal-related modulations of neural variability}

In the clustered model, the transition from the metastable attractor phase to the uniform phase also predicts certain modulations of neural variability. To delineate these effects, we examined the Fano factor (FF), which measures trial-to-trial variability of single-neuron spike-trains (\nameref{s:Methods}). Increasing arousal was associated with a strong decrease in the FF during spontaneous activity ($\mathrm{FF}_{\mathrm{spont}}$; Fig.~\ref{f:fano_factor_main}A). At low arousal, clusters slowly transition between highly active and inactive states (Fig.~\ref{f:mean_field_vs_sims}F,G); as a result, single-neuron firing rates strongly fluctuate across time and/or trials, leading to high spike-count variability. As arousal increases, cluster switching dynamics accelerate and the distinction between active and inactive rates decreases; these changes lead to a decrease in spontaneous FF as the model transitions towards the uniform phase. The evoked FF ($\mathrm{FF}_{\mathrm{evoked}}$) also decreased with arousal (Fig.~\ref{f:fano_factor_main}B), which is a consequence of stimulus-evoked activity being constrained by the ongoing dynamics. That is, while stimulus presentation does bias activation of targeted clusters, stimuli are not so strong as to be able to activate all targeted clusters simultaneously on every trial (Fig.~\ref{f:decoding_intuition_model}A). The evoked activity thus inherits much of the intrinsic variability present in spontaneous activity. The above-mentioned modulations of cluster dynamics are also reflected in arousal-induced reductions of the coefficient of variation of interspike intervals and low-frequency spike-train power (Fig.~\ref{f:cvISI_spectra_supp}).

Observed trends in the empirical data qualitatively agreed with predictions from the clustered model. At the population level, the spontaneous FF decreased with pupil diameter (Fig.~\ref{f:fano_factor_main}D), and was significantly smaller in states of high pupil-indexed arousal compared to low pupil-indexed arousal (Fig.~\ref{f:fano_factor_main}E; Fig.~\ref{f:spectra_fano_supp}A). The population-averaged evoked FF plateaued at moderate-to-large pupil diameters (Fig.~\ref{f:fano_factor_main}F), but was still significantly smaller in high arousal compared to low arousal conditions (Fig.~\ref{f:fano_factor_main}G; Fig.~\ref{f:spectra_fano_supp}B). Similar trends were also observed for the coefficient of variation of interspike intervals and low-frequency spike-train power (Fig.~\ref{f:cvISI_spectra_supp}). In sum, the model captures qualitative trends in spike-train variability with arousal, but does not quantitatively match the data in terms of the absolute values of the variability measures.

Past studies indicate that stimulus presentation quenches neural variability \cite{churchland2010stimulus}. To quantify this effect in our dataset, we estimated the difference between the spontaneous and evoked FF ($\Delta \mathrm{FF} = \mathrm{FF}_{\mathrm{spont}} - \mathrm{FF}_{\mathrm{evoked}}$), marginalized across pupil diameter. Consistent with past reports, we observed a reduction in the FF during evoked conditions (Fig.~\ref{f:spectra_fano_supp}D). Clustered networks were previously proposed to explain this phenomenon \cite{deco2012neural,litwin2012slow}, and in our model, we also observed some stimulus-induced quenching at low arousal (Fig.~\ref{f:fano_factor_main}C). However, the effect is relatively subtle, since our model operates in a regime where stimuli cannot entirely override ongoing activity. The clustered model also exhibited an interaction between arousal and stimulus-induced variability quenching, wherein $\Delta \mathrm{FF}$ decreased with arousal (Fig.~\ref{f:fano_factor_main}C). In the data, we similarly found a decreasing trend in the population-averaged $\Delta \mathrm{FF}$ with pupil diameter (Fig.~\ref{f:fano_factor_main}H), and a significant reduction in $\Delta \mathrm{FF}$ between highly-constricted and highly-dilated pupil conditions (Fig.~\ref{f:fano_factor_main}I; Fig.~\ref{f:spectra_fano_supp}C).

\section*{Discussion}

We recorded activity from mouse ACtx during passive tone presentation and found that neural stimulus discriminability followed an inverted-U relationship with pupil-linked arousal. We then showed that this inverted-U relationship could be explained via modulations of metastable attractor dynamics in a clustered network model, with optimal stimulus discriminability achieved near a transition in the dynamical regime of the network. The clustered model further predicted a reduction of neural variability with arousal, which was confirmed in the empirical data. This study thus unifies two different phenomena -- arousal-dependent changes in neural discriminability and variability -- under the same computational mechanism: modulations of metastable attractor dynamics.

\subsection*{Neural correlates of an inverted-U relationship between performance and arousal in ACtx}

Several studies have reported inverted-U relationships between pupil-linked arousal and behavioral performance on auditory tasks \cite{waschke2019local, murphy2011pupillometry, beerendonk2024disinhibitory,mcginley2015cortical,hulsey2023decision}, raising questions about the neural origins of optimal performance at intermediate arousal. Previous work found neural correlates of optimal sound detection in mouse ACtx and medial geniculate nucleus at moderate arousal, as evidenced by reduced variability of spontaneous membrane potential dynamics and increased magnitude and reliability of evoked responses \cite{mcginley2015cortical}. Here, we found that intermediate arousal was also associated with optimal neural discriminability of tones in mouse ACtx, even in the absence of a task. 

Our results suggest that arousal-related modulations of ACtx activity might contribute to the inverted-U relationship between arousal and behavioral performance observed during auditory discrimination tasks \cite{hulsey2023decision, beerendonk2024disinhibitory}. That said, the inverted-U relationship for task performance could also receive contributions from arousal-induced modulations of areas up- or down-stream of ACtx. Important future directions include analyzing task-engaged settings and different stages of the auditory and decision-making pathways \cite{khoury2023arousal,sorensen2022arousal,saderi2021dissociation}. It is also important to note that the decoding approach used here assumes that a downstream readout unit acts as an ideal observer implementing linear classification \cite{rigotti2013importance}. While a useful benchmark, neural decoding performance can depend on population size, and information readout and perceptual discrimination performance may differ from predictions based on ideal-observer linear decoding frameworks \cite{panzeri2022structures}. Finally, we note that one previous study reported a monotonic increase of tone decoding accuracy in mouse ACtx \cite{lin2019arousal}. Several factors could contribute to across-study discrepancies, including differences in recording technique, experimental design, or analysis methods (e.g., type of classifier used). Further examining the conditions under which non-monotonic relationships emerge is an important avenue for future work. 

\subsection*{Motivation for the clustered model}

We proposed a mechanism for the inverted-U relationship between arousal and neural discriminability using a network model in which neurons were organized into strongly-coupled clusters representing functional neural assemblies (see \cite{amit1997model,deco2012neural,litwin2012slow, mazzucato2019expectation,mazzucato2015dynamics,wyrick2021state}.) Our empirical dataset exhibited some evidence of functionally-organized cell ensembles, where we found putative correlation-based clusters with larger tuning similarity than expected by chance alone. Such a relationship between noise correlations and stimulus response similarity is reminiscent of prior work showing positive associations between noise and signal correlations in ACtx \cite{ rothschild2010functional,funamizu2023stable,luczak2009spontaneous,winkowski2013laminar}. More broadly, the clustered model is motivated by evidence of strongly-connected groups of cells in sensory cortices \cite{song2005highly,ko2011functional,yoshimura2005excitatory,cossell2015functional,perin2011synaptic,lee2016anatomy}. As in the clustered model, neural data indicates that strongly-coupled neurons display similar stimulus responses \cite{lee2016anatomy,ko2011functional,cossell2015functional} and that spontaneously-activated cell ensembles are also triggered by stimulation \cite{miller2014visual,maclean2005internal}, suggesting that cell assemblies may act as basic cortical processing units. The functional architecture of population activity in ACtx, specifically, also appears consistent with the presence of partially-overlapping and strongly-connected neuronal subnetworks \cite{rothschild2010functional}. 

Metastable attractor dynamics are a key feature of the clustered network \cite{deco2012neural,litwin2012slow, mazzucato2019expectation,mazzucato2015dynamics,wyrick2021state}, and some analyses have suggested activity patterns reminiscent of those in the model. In particular, \textcite{bathellier2012discrete} found that evoked activity patterns in ACtx populations were organized into a small set of discrete ``response modes", and that transitions between modes were abrupt, indicative of attractor-like dynamics. Their study also suggested that sounds are represented by the combined activation pattern of several response modes spread across ACtx \cite{harris2012cell}, somewhat akin to the encoding of stimuli via global cluster activation patterns in our circuit model. Other studies in ACtx have observed transient ``packets" of elevated spiking activity that occur sporadically during spontaneous periods and that constrain stimulus responses \cite{luczak2009spontaneous,luczak2013gating}, as well as evidence of locally-clustered activity in superficial layers \cite{sakata2009laminar}. Although these findings are suggestive of the activity patterns in our network model, more spatially-distributed recordings and targeted perturbation studies are necessary to directly test for the presence of metastable cluster dynamics in ACtx. 

\subsection*{Emergence of arousal-induced modulations of neural discriminability and variability in the clustered model}

Clustered network models with metastable activity have been used to explain several features of cortical activity and computation \cite{la2019cortical,brinkman2022metastable,mazzucato2022neural}, including stimulus-induced quenching of variability \cite{litwin2012slow,deco2012neural, mazzucato2016stimuli}, the emergence of state-sequences during taste processing \cite{ mazzucato2015dynamics,lang2023temporal}, and context-dependent sensory processing and decision-making dynamics \cite{mazzucato2019expectation, wyrick2021state, rostami2024spiking}. Previous work examined the response of clustered networks to relatively small parameter perturbations, leading to monotonic variations in stimulus processing efficacy \cite{mazzucato2019expectation,wyrick2021state}. Here, we explored a broader range of parameter variations, which led to the inverted-U modulation of stimulus discriminability required to explain the data. 

In our model, arousal was implemented as a global decrease in synaptic efficacy between excitatory cells and an increase in external excitatory drive. When the clustered network was subjected to this arousal modulation, the inverted-U relationship emerged via a transition from a dynamical phase with slow switching between multiple metastable attractors to a phase with uniform network activity. Stimulus discriminability was maximized between these two extremes, where stimulus responses were both relatively strong and reliable. Because the transition from the metastable attractor phase to the uniform phase is accompanied by a suppression of slow rate fluctuations, the clustered model also qualitatively captured observed reductions of spiking variability and stimulus-induced variability quenching at high arousal (though see \cite{schwartz2020pupil} for a study in ferrets where variability quenching was independent of pupil size).

\subsection*{Alternative mechanisms and approaches}

We modeled one implementation of arousal that was motivated by empirical evidence and that reproduced the inverted-U relationship. That said, arousal-related neuromodulators affect many biophysical processes to induce diverse alterations of cortical circuit dynamics \cite{picciotto2012acetylcholine,o2012norepinephrine,berridge2003locus,slater2022neuromodulatory,nadim2014neuromodulation,mccormick1997sleep,mccormick1989cholinergic}. For example, neuromodulators can directly excite cortical neurons \cite{fu2014cortical,gasselin2021cell}, alter excitability \cite{mccormick1993neurotransmitter}, synaptic transmission \cite{yang2021cholinergic, salgado2016layer}, and cellular firing modes \cite{mccormick1986acetylcholine}, and regulate circuit function via modulation of distinct interneuron classes \cite{wester2014behavioral}. The modeling framework presented here could be extended to explore additional mechanisms. Importantly, similar modulations of collective network dynamics can be produced by different circuit perturbations \cite{gutierrez2013multiple}. Even here, we identified an alternative arousal implementation that induced similar functional outcomes to the one studied in the main text (though lacked the same degree of biological plausibility; Fig.~\ref{f:sd_nu_ext_e_arousalModel}). Further experimental work is needed to isolate the precise neurophysiological mechanisms of arousal responsible for the inverted-U relationship.

Any circuit mechanism for the inverted-U relationship must incorporate a means of non-linearly modulating the efficacy of stimulus responses. The mechanism we presented here relied critically on the modulation of metastable dynamics in networks with clustering. Importantly, though, our study does not rule out alternative mechanisms that do not require those features (see, e.g., \cite{beerendonk2024disinhibitory,sorensen2022arousal}). Along these lines, one particularly relevant study proposed a rate-based decision-making circuit with two excitatory populations, each selective for a different stimulus, and two interneuron classes (VIP and SST), which were both modulated by arousal \cite{beerendonk2024disinhibitory}. In that model, an increase in arousal first improved discrimination performance via VIP-SST-mediated disinhibition of the excitatory pools; but a further increase in arousal saturated the VIP population, leading to a degradation in performance due to SST-mediated inhibition. This study indicates that another way of achieving non-linear performance modulations is to incorporate other types of complexity, such as multiple inhibitory cell types. Other work has shown that cortical sensory processing is impacted by the presence of spontaneously-generated ``Up" and ``Down" states \cite{pachitariu2015state,curto2009simple}, which are themselves dependent on behavioral state \cite{harris2011cortical} (also discussed further below). Arousal-induced modulation of these global activity patterns could thus also contribute to the inverted-U relationship. Moreover, the right-hand side of the inverted-U relationship might be mediated, at least in part, by motor-related signals, which can suppress tone-evoked activity in ACtx via postsynaptic inhibition of excitatory cells \cite{schneider2014synaptic}. More generally, a large body of literature demonstrates that information processing capabilities in neural systems are often optimal near criticality \cite{beggs2008criticality,o2022critical,shew2013functional,munoz2018colloquium,bertschinger2004real,toyoizumi2011beyond}, and some studies have demonstrated enhanced stimulus discriminability at the transition between asynchronous and synchronous dynamics in cortical network models \cite{tomen2014marginally,yang2025critical}. These studies suggest that other types of dynamical regime transitions -- besides the one studied here -- may also be able to explain the inverted-U relationship. Future work could seek to further test alternative models and mechanisms and determine which are most consistent with experimental observations.

Mechanisms other than cluster-based switching dynamics could also explain arousal-related decreases of neural variability. In particular, modulations of spontaneous ``Up-Down" dynamics -- alternating periods of global silence and global activity observed during anesthesia, sleep, and quiet wakefulness \cite{harris2011cortical,steriade1993novel, steriade2001natural,luczak2007sequential, luczak2009spontaneous,petersen2003interaction} -- could produce similar state-dependent adjustments of single-neuron spiking variability \cite{mochol2015stochastic,shi2022cortical,marguet2011state}. While an advantage of the clustered model is its ability to explain arousal-induced modulations of neural discriminability and variability with a single mechanism, it is possible that both cell assembly dynamics and more distributed Up-Down dynamics contribute to our findings. Indeed, signatures of both have been observed in rodent auditory cortex \cite{harris2012cell,bathellier2012discrete,see2018coordinated,luczak2009spontaneous,luczak2007sequential,luczak2013gating,rothschild2010functional,mochol2015stochastic,curto2009simple}, with prior work suggesting that locally-clustered activity may be more prevalent in superficial layers \cite{sakata2009laminar,harris2012cell}. In our data, we found evidence of correlated subgroups of neurons suggestive of functionally-meaningful neural clusters. However, we also observed more dispersed correlation structure in the data compared to the model, which could be indicative of more globally-coordinated activity fluctuations. Future experiments with more targeted and spatially-expansive electrode placement (i.e., spanning multiple layers and columns) are needed to isolate and refine the network mechanisms contributing to arousal-related modulations of neural variability and stimulus processing. Building network models that combine global Up-Down fluctuations with clustered neural assemblies \cite{setareh2017cortical} may also allow for a more complete description of state-dependent dynamics.

\subsection*{Modality-dependent effects of arousal on sensory-evoked activity}

During perceptual decision-making, inverted-U relationships between arousal and performance arise in both auditory \cite{mcginley2015cortical, hulsey2023decision} and visual \cite{neske2019distinct, hulsey2023decision} tasks. However, relationships between arousal and sensory-evoked activity strongly differ between auditory and visual cortex. In auditory cortex, previous investigations \cite{mcginley2015cortical} (and this study) report inverted-U relationships between pupil-indexed arousal and measures of stimulus response efficacy, as well as suppressed evoked responses during locomotion \cite{zhou2014scaling,schneider2014synaptic,bigelow2019movement,yavorska2021effects}. In contrast, prior work in visual cortex indicates that sensory-evoked responses are enhanced during high arousal \cite{neske2019distinct, reimer2014pupil} and locomotion \cite{niell2010modulation, dadarlat2017locomotion, bennett2013subthreshold, polack2013cellular}. The origin of this divergence remains unclear. Although both auditory and visual cortex appear to share some common architectural features (e.g. functional neural assemblies), there may be important variations in their recurrent circuitry or intrinsic neuron properties that lead to differences in ongoing activity, sensory responses, and interplays with arousal. Another possibility is that differences between auditory and visual cortex reflect an underlying distinction in arousal signaling pathways in the two areas. Indeed, the impacts of locomotion in visual cortex are thought to be mediated by a disinhibitory pathway involving VIP interneurons \cite{fu2014cortical}, but this is not the case in auditory cortex \cite{bigelow2019movement,yavorska2021effects}. Further elucidating the neuromodulatory pathways regulating arousal may help resolve outstanding questions regarding modality-dependent phenomena. 

\section*{Resource Availability}

\subsection*{Lead Contact}

Requests for further information and resources should be directed to the lead contact, Luca Mazzucato (lmazzuca@uoregon.edu).

\subsection*{Materials Availability}

This study did not generate novel reagents.

\subsection*{Data and Code Availability}
~\\
$\bullet$ The electrophysiological and behavioral data analyzed in this study has been uploaded to the DANDI Archive as Neurodata Without Borders (.nwb) files \cite{nwb}. The data is publicly available at:
~\\
https://doi.org/10.48324/dandi.000986/0.251031.1939. 
~\\
~\\
$\bullet$ Original code used to run and analyze model simulations, analyze experimental data, and generate manuscript figures has been deposited on Zenodo. The code is publicly available at: https://doi.org/10.5281/zenodo.17497134.
~\\
~\\
$\bullet$ Any additional information required to reanalyze the data reported in this paper is available from the lead contact upon request.

\section*{Acknowledgements}

This work was funded by National Institutes of Health grants R01NS118461 (D.A. McCormick, L. Mazzucato, S. Jaramillo), R35NS097287 (D.A. McCormick.), R01MH127375 and R01DA055439 (L. Mazzucato), R01AG077681 and R01NS127305 (M. Wehr); and National Science Foundation CAREER Award 2238247 (L. Mazzucato)

\section*{Author contributions}

Conceptualization: L.P., L.M., D.A.M.; methodology: L.P., S.Jo, L.M.; experimental data acquisition: S.Jo., K.Z.; data curation: L.P., S.Ja; investigation: L.P., S.Jo, L.M.; formal analysis, visualization, and software: L.P.; writing – original draft: L.P., S.Jo, K.Z., L.M.; writing – editing: L.P., K.Z., M.W., S.Ja, D.A.M., L.M.; supervision: S.Ja, D.A.M., L.M.; funding acquisition: M.W., S.Ja, D.A.M., L.M.

\section*{Declaration of Interests}

The authors declare no competing interests.

\clearpage
\newpage

\section*{STAR Methods}
\label{s:Methods}

\section*{Experimental model and study participant details}
 
All procedures were carried out with approval from the University of Oregon Institutional Animal Care and Use Committee. Wild-type animals (2 female mice and 3 male mice ranging between approximately 11-20 weeks at the time of surgery) were of C57BL/6J background purchased from Jackson Laboratory and were bred in-house. Mice were kept on a reverse light cycle and had ad libitum access to food and water. For all analyses, experimental sessions were pooled across both sexes. We did not examine potential influences of sex on our results, which is a limitation of this study.

\section*{Method details}

\subsection*{Experimental data}

\subsubsection*{Surgical procedures} 
 
All surgical procedures were performed in an aseptic environment with mice under 1-2$\%$ isoflurane anesthesia, maintaining an oxygen flow rate of 1.5 L/min, and homeothermic maintenance at 36.5 degrees Celsius. Mice were administered systemic analgesia (Meloxicam SR: 4 mg/kg \& Buprenorphine SR: 0.5 mg/kg, Wildlife Pharmaceuticals) and a fluid supplement (1 ml lactated ringer’s solution) subcutaneously. Fur was removed from the skull, and the skin was disinfected. To access auditory areas, the skin, connective tissue, and part of the right temporalis muscle were resected, and cleaned as necessary. A custom-designed headplate was affixed to the skull using dental cement (RelyX Unicem Aplicap, 3M) and covered with silicone elastomer (Kwik-sil, World Precision Instruments), and skin was affixed to the outside edge of the headpost as necessary (Vetbond, 3M). Mice were allowed to recover for three days in an incubator recovery chamber. 

Mice were habituated to handling and head fixation for 2-3 days with increasing duration prior to craniotomy. This was a necessary step for well-being and also helped increase the likelihood that mice entered a broad range of arousal states across the wakefulness spectrum. Habituation to head-fixation atop a treadmill allowed mice to choose to locomote or remain still and quiescent. Craniotomy followed the same aseptic and analgesic procedures as mentioned above. Mice were anesthetized with isoflurane and affixed to the stereotax where a $<$1 mm circular craniotomy was drilled over the right auditory cortex (AP: -2.9 mm, ML: 4.4 mm relative to bregma) with dura left intact; the above-mentioned coordinates were chosen in an attempt to mainly target primary auditory cortex (A1). A small well was created surrounding the craniotomy with flowable composite (Flow-it, Pentron), and a piece of plastic was secured lateral to the well to act as a shield for the probe. The craniotomy was filled with silicone elastomer (Kwik-sil, World Precision Instruments) until the start of the recording session. Mice were allowed to recover overnight, and recovery was monitored.  

\subsubsection*{Neuropixels recordings}
\label{s:neuropixels_recordings}

On the day of a recording, a mouse was affixed to a treadmill and the Kwik-sil was removed. The craniotomy was immediately filled with saline, and a high-density silicon probe (Neuropixels 1.0, imec) \cite{jun2017fully} was inserted perpendicular to the brain surface using a motorized micromanipulator (MP225A, Sutter Instruments) at low speed ($\sim$ 2-4 $\mu$m/second) until all layers of the auditory cortex were covered (1.5-2.5 mm). After the Neuropixels probe reached a desired depth, the remaining saline was removed and the craniotomy was filled with 1\% agarose mixture in saline and covered with mineral oil to keep the brain surface moist. A recording was started at least 20 minutes after the completion of probe insertion to ensure the stability of the probe and the brain. Recordings were made in up to 5 sessions from one mouse depending on the status of the brain surface. For the last recording session, the Neuropixels probe was covered with DiI (Vybrant solution, Thermofisher Scientific) for histology. 

Neurophysiology data was acquired using the PXIe acquisition module (imec) in a NI PXIe-1071 chassis (National Instruments) and OpenEphys software \cite{siegle2017open} at gain of 250 (LFP), and 500 (APs). An output pulse from the OpenEphys software was manually toggled between 1 Hz and 10 Hz to give an accurate and discrete timestamp to the Power 1401 digitizer, which allowed for accurate alignment and further synchronization of the behavioral data. Neuropixels data was sampled at a rate of 30 kHz. The recorded data was pre-processed with common-average referencing \cite{steinmetz2019distributed}, \cite{ludwig2009using} sorted with Kilosort2 \cite{pachitariu2016fast,pachitariu2024spike}, and then manually curated with the phy GUI (https://github.com/cortex-lab/phy). For manual curation, each cluster was compared with other clusters based on the spike waveforms and cross-correlation. Clusters with high similarity were mainly inspected to determine whether they should be merged. Then, the cluster was labeled as a good single unit, multi-units, or noise depending on the quality of the cluster assessed by waveform consistency, amplitude, cross-correlation, and inter-spike intervals. To determine if the good single units were within the auditory cortex, the depth from phy was referenced. In four out of the five mice analyzed, we also verified the recording site of the final session with DiI track spanning after histology. Sessions for which the brain condition was poor, auditory responses were weak, or timestamps could not be aligned were discarded.

\subsubsection*{Histological Analysis} 

Following the last recording session, a mouse was anesthetized and perfused using phosphate buffer and 4\% paraformaldehyde. Then, the brain was kept in 4\% paraformaldehyde, cryo-sectioned (CM3050S, Leica) at 100 $\mu$m thickness, and DAPI-stained. Slides were imaged and DiI tracks were manually registered with the Franklin-Paxinos atlas \cite{franklin1997mouse}. 

\subsubsection*{Spontaneous data and auditory stimulation} 

Each experimental session consisted of alternating spontaneous and auditory stimulation blocks, repeated for up to 2 hours. During spontaneous blocks, neural activity was recorded in the absence of stimulus presentation; each block lasted for five minutes. A spontaneous block was followed by 25 minutes of auditory stimulation. This design enabled us to record substantial amounts of both spontaneous activity ($\sim$ 20-25 minutes/session) and evoked activity ($\sim$ 75-100 minutes/session). The stimulus set consisted of five pure tones (2, 4, 8, 16, or 32 kHz), which were randomly interleaved and sampled from a uniform distribution. Each tone lasted for 25 ms (cosine ramp-up) followed by a 775 ms inter-stimulus interval (ISI). Auditory stimuli were delivered using custom LabView (National Instruments) scripts. Tones were calibrated to 60 dB SPL and waveforms were generated (NI PXI-4461, National Instruments) at 200 kHz sampling rate, conditioned (ED1, Tucker Davis Technologies), and transduced by electrostatic speakers (ES1, Tucker Davis Technologies). 

\subsubsection*{Behavioral measures}

All data collection was conducted using custom LabView scripts. Mice were headfixed atop a cylindrical treadmill (15 cm diameter, 20 cm width) and allowed to freely locomote. Locomotion speed was calculated via a rotary encoder (Encoder Products CO.; 15T-01SF-2500NV1RPP-F03-S1) attached to the axle of the treadmill. Signals from the rotary encoder were continuously converted into cm/s in real-time using LabView software at a rate of 100 Hz, and data was recorded using a Power 1401 digitizer. The running trace was upsampled to match the Neuropixels acquisition rate, and post-hoc analysis was performed using custom python scripts.

The face was lit using an infrared LED (Digi-Key TSHG8200, 830 nm) adjusted to achieve uniform illumination of the face and eye. Additionally, a white LED (RadioShack 5 mm 276-0017) was manually titrated to achieve a wide dynamic range of the pupil, ensuring it remained visible during full dilation. Pupil videos were collected from a camera (Grasshopper 3, FLIR) with a lens (Telecentric TEC-55, Computar) and near-IR Bandpass filter (BN810-43, MidOpt) with FlyCapture software (FLIR). Frames were triggered at 30 Hz through a Power 1401 Digitizer (Cambridge Electronic Design), and online pupillometry was performed using LabView software according to previously described methods \cite{hulsey2023decision}. The pupil diameter trace was upsampled to match the Neuropixels acquisition rate, and post-hoc analysis was performed using custom python scripts.

Raw pupil diameter traces were subject to three processing steps: \textit{(1)} artifact removal, \textit{(2)} smoothing, and \textit{(3)} normalization. The pupil-tracking procedure is imperfect, which can lead to artifacts in the pupil diameter traces such as abrupt drops or spikes. To mitigate the effect of these artifacts, we performed both automated and manual cleaning of the pupil traces in each session. Automated artifact removal consisted of finding and discarding periods of time associated with unnaturally-sharp jumps in pupil diameter values between nearby time points. To find these artifacts, we first normalized the pupil diameter trace in a given session by its maximum value. At each time point $t_{n}$, we then compared the difference in normalized pupil diameter between $t_{n}$ and $t_{n} + 0.5$ ms. If the absolute difference in the normalized pupil diameter between those times exceeded a threshold of 0.08, then we removed the pupil data within a time window starting 250 ms before $t_{n}$ and ending 500 ms after $t_{n}$. This automated procedure removed a large majority of pupil artifacts, but pupil traces were still manually inspected afterwards for outstanding abnormalities. Remaining problematic time windows were tabulated, and the corresponding pupil data was removed from those periods. Pupil traces were also smoothed after artifact removal for easier manipulation. Smoothing was achieved by taking a moving average of the pupil diameter timecourses using windows of length $1/30^{\mathrm{th}}$ of a second sliding forward in $1$ ms steps. After artifact removal and smoothing, the resulting pupil diameter trace for a given session was re-normalized by its maximum value; in all sessions, maximum dilation was associated with movement bouts. This normalization procedure, which has been utilized in several prior studies \cite{mcginley2015cortical,lin2019arousal,hulsey2023decision,neske2019distinct,nestvogel2022visual,collins2023cholinergic}, facilitates combining data across sessions. All pupil-based analyses in the main text were performed using the within-session normalized pupil diameter traces. We also verified that the population decoding results (Fig.~\ref{f:decoding_allTrials}H,I) were robust to an alternative normalization procedure, wherein the pupil traces for each session were all normalized by the \textit{same} value (the largest pupil diameter attained across all sessions combined; Fig.~\ref{f:decoding_dprime_supp_data}F,G). Throughout the text, pupil diameters are reported as a percentage of the maximum value (denoted as ``$\%$ max dilation").

Periods of time corresponding to artifacts in the pupil diameter trace were also removed from the running trace. The running trace was then smoothed using the same process as for the pupil diameter.

\subsubsection*{Additional unit selection criteria}
\label{s:unit_selection}

After following the procedures described in ``\textit{\nameref{s:neuropixels_recordings}}" to identify putative single units, we implemented some additional criteria for the final unit selection process. First, we discarded all clusters whose average firing rate across the duration of the recording was less than 0.25 spikes/second. The remaining criteria mainly involved further analysis of the spike template amplitudes of each cluster that was identified as ``good" after performing the spike sorting and manual curation steps detailed above. Examining the behavior of the template amplitudes (output by Kilosort) for a given cluster across time can reveal potential issues with electrode drift and the general quality of the cluster. Our analysis was designed to search for two potential issues in the spike template amplitudes. First, we considered the shape of the amplitude distribution in a sliding time window, and in each window, we looked for signatures of multiple peaks occurring in the corresponding distribution. The presence of multiple peaks in the amplitude distribution computed from a short block of time is an indication that the particular cluster should not be marked as a well-isolated single unit. Second, we looked for cases when the amplitude appeared to drift towards or away from very low values (i.e., towards or away from the ``noise floor") over time. This scenario could suggest that the cluster was not stably-tracked across the recording.

To determine if the distribution of template amplitudes in a short time segment was composed of two or more separate peaks, we examined the amplitude data in non-overlapping, 5-minute windows over the entire dataset. For each window, we used the `scipy.stats.gaussian\_kde' function from SciPy to estimate the probability density function (pdf) of the amplitude data via kernel density estimation with a Gaussian kernel. For each window, we then determined the locations (i.e., amplitude values) and heights of all peaks in the corresponding pdf. If the pdf from a given window had more than one peak, we computed two additional quantities. First, we computed the ratio of the height of the tallest peak to the height of the second tallest peak in the window; we refer to this quantity as the ``peak height ratio". Smaller peak height ratios tend to correspond to more even splits of the data between the two groups. Second, we computed the percent difference between the locations of the two highest peaks in a window. Larger percent differences between the peak locations correspond to more well-separated groups. After computing these quantities, we found the set of time windows for which the peak height ratio was less than or equal to ten and for which the percent difference between peak locations was greater than or equal to forty. These cut values were selected so as to find time windows for which there were two (or more) well-separated template amplitude ranges that each contributed substantially to the total amount of data in the window. If at least 10\% of all time windows satisfied the above criteria, then the corresponding cluster was not used in subsequent analyses. 

To determine if the template amplitude for a given cluster appeared to drift into or out of the ``noise floor" over time, we first estimated the noise floor as the smallest template amplitude of the cluster across the whole recording. As above, we then considered the pdf of the amplitudes in 5-minute bins. First, we computed the percent difference between the location of the tallest peak in the pdf of a given window and the location of the noise floor. If this percent difference was less than or equal to fifteen, then the corresponding window was marked as having template amplitudes that were concentrated near the noise floor. 

For each window, we also determined the location (i.e., amplitude) of the tallest peak in the pdf. We then computed the smallest and largest of those amplitudes across all time windows, and computed the percent difference between the resulting two values. This quantity, which we refer to as the maximum peak location difference, provides information about the range of template amplitudes sampled across the recording. We removed a cluster from subsequent analyses if the following criteria were met: \textit{(i)} more than 10\% (but not all) of time windows either had template amplitudes concentrated near the noise floor or in the bulk, and \textit({ii)} the maximum peak location difference was greater than or equal to twenty-five. These cut values were chosen so as to try and isolate clusters with significant drift towards or away from low amplitude values. All analyses in the main text were performed after applying the unit selection procedures described in this section. The number of simultaneously recorded single-units that passed all selection criteria ranged from 30 - 235 per session. 

\subsubsection*{Robustness to cell selection criteria}

Because our recordings could have included some cells outside the ACtx, we also tested a more conservative cell selection method that incorporated strict criteria for sound responsiveness. After spike sorting and manual curation (see ``\textit{\nameref{s:neuropixels_recordings}}"), we analyzed the tone-evoked responses of all ``good" single units/cells. To begin, evoked data was aligned to stimulus onset ($t = 0$), and a time-dependent firing rate was computed in each trial by counting spikes in a 100 ms window sliding forward in 1 ms steps. The time $t$ of each window was defined as the location of its right edge, and the first time window (baseline activity) was located at $t = 0$ and the last time window was located at $t = 150$ ms. For a given tone, we then compared the firing rate measurements in each evoked time window (i.e., windows with $t > 0$) to the distribution of baseline firing rates ($t = 0$ window) using the Wilcoxon signed-rank test. The p-value for each evoked window was Bonferroni-corrected for multiple comparisons (i.e., using a correction factor equal to the number of evoked windows), and a cell was considered ``responsive" to the given tone if at least one of the evoked time windows had a corrected p-value $< 0.01$. Only cells that responded to at least two tones according to this criteria (and that passed the additional cuts described above in ``\textit{\nameref{s:unit_selection}}" were kept for further analysis. We found that the main trends in the neural discriminability, clustering, and neural variability analyses were still evident when using this more conservative cell selection method (Fig.~\ref{f:cellSelection_robustness}).

\subsection*{Network modeling}
\label{s:circuit_model}

We modeled a local cortical circuit representing ACtx as a recurrently-connected network of $N$ spiking neurons, $N^{E}$ of which were excitatory (E) cells and $N^{I}$ of which were inhibitory (I) cells. Further details on the circuit modeling are provided below. All model parameters are shown in Table~\ref{t:model_params}.

\subsubsection*{Neural dynamics}
\label{s:LIF_model}

Neural activity evolved according to the leaky-integrate-and-fire (LIF) model with exponential excitatory and inhibitory synapses. In this model, the dynamics of the membrane potential of the $i^{th}$ neuron in population $\alpha \in \{E,I\}$ are described by

\begin{equation}
\tau_{\mathrm{m}}^{\alpha} \frac{dV_{i}^{\alpha}}{dt} = -V_{i}^{\alpha} + \tau_{\mathrm{m}}^{\alpha} I_{\mathrm{rec},i}^{\alpha} + \tau_{\mathrm{m}}^{\alpha} I_{\mathrm{b},i}^{\alpha} + \tau_{\mathrm{m}}^{\alpha} I_{\mathrm{stim},i}^{\alpha} ,
\label{eq:LIF}
\end{equation}
~\\
where $\tau_{\mathrm{m}}^{\alpha}$ is the membrane time constant of cells in population $\alpha$, $I_{\mathrm{rec},i}^{\alpha}$ is the recurrent input to cell $i$ in population $\alpha$ from other neurons in the network, $I_{\mathrm{b},i}^{\alpha}$ represents background external input, and $I_{\mathrm{stim},i}^{\alpha}$ is an additional external input representing sensory stimulation. When the membrane potential $V_{i}^{\alpha}$ reaches a threshold $V_{\mathrm{thresh}}^{\alpha}$, a spike is emitted by the neuron and its membrane potential is reset to a value $V_{\mathrm{r}}^{\alpha}$. After spike emission, the membrane potential remains clamped at the reset value for a refractory period of length $\tau_{\mathrm{ref}}^{\alpha}$.

The recurrent input is a sum of excitatory and inhibitory synaptic currents, such that $I_{\mathrm{rec},i}^{\alpha} = I_{\mathrm{rec},i}^{\alpha E} + I_{\mathrm{rec},i}^{\alpha I}$. These currents obeyed the following differential equations:

\begin{align} 
\tau_{\mathrm{syn}}^{E} \frac{ dI_{\mathrm{rec},i}^{\alpha E} }{dt} = - I_{\mathrm{rec},i}^{\alpha E} + \sum_{j=1}^{N_{E}} W_{ij}^{\alpha E} \sum_{k} \delta(t-t_{j}^{k,E}) 
\label{eq:LIF_currentsE}
\\
\tau_{\mathrm{syn}}^{I} \frac{ dI_{\mathrm{rec},i}^{\alpha I} }{dt} = - I_{\mathrm{rec},i}^{\alpha I} + \sum_{j=1}^{N_{I}} W_{ij}^{\alpha I} \sum_{k} \delta(t-t_{j}^{k,I}). 
\label{eq:LIF_currentsI}
\end{align}
~\\
In Eqs.~\ref{eq:LIF_currentsE} and \ref{eq:LIF_currentsI}, $\tau_{\mathrm{syn}}^{E}$ and $\tau_{\mathrm{syn}}^{I}$ are the excitatory and inhibitory synaptic time constants, and $W_{ij}^{\alpha\beta}$ represents the strength of the synapse from the $j^{th}$ neuron of population $\beta \in \{E,I\}$ to the $i^{th}$ neuron of population $\alpha$; these weights depend on the network architecture (see the section ``\textit{\nameref{s:net_arch}}" below). Finally, $t_{j}^{k,\beta}$ is the time of the $k^{\mathrm{th}}$ spike emitted by the $j^{th}$ neuron of population $\beta$. 

In addition to the recurrent input, each neuron in population $\alpha$ received $C_{\mathrm{ext}}^{\alpha E}$ connections from other excitatory cells outside of the local network. The background synaptic input at the $i^{th}$ neuron of population $\alpha$ evolved according to

\begin{equation} 
\tau_{\mathrm{syn}}^{E} \frac{ dI_{\mathrm{b},i}^{\alpha} }{dt} = - I_{\mathrm{b},i}^{\alpha} + J_{\mathrm{ext}}^{\alpha E} \sum_{j=1}^{C_{\mathrm{ext}}^{\alpha E}} \sum_{k} \delta(t-t_{ij}^{\alpha,k}),
\label{eq:LIF_external_current}
\end{equation}
~\\
where J$_{\mathrm{ext}}^{\alpha E}$ is the strength of external excitatory synapses to cells in population $\alpha$, and where $t_{ij}^{\alpha,k}$ is the $k^\mathrm{th}$ spike time of the $j^{\mathrm{th}}$ external cell targeting neuron $i$ in population $\alpha$. The spike times $t_{ij}^{\alpha,k}$ were generated from a Poisson process with rate $\nu_{\mathrm{ext},i}^{\alpha}$; spike trains were independent for each external synapse to a given cell, and there was no shared input across different cells. The baseline value of the external input rate for cells in population $\alpha$ is denoted as $\nu_{o}^{\alpha}$.

Finally, sensory stimuli were modeled as smoothly-varying, deterministic external inputs $I_{\mathrm{stim},i}^{\alpha}(t)$ that directly entered the voltage equation of the corresponding neuron. Further details on the stimulus inputs are given in the section ``\textit{\nameref{s:model_stim}}".

\subsubsection*{Recurrent network architectures}
\label{s:net_arch}

In the circuit model, the network architecture was either ``clustered" or ``uniform" (Fig.~\ref{f:model_overview}A,B). For the uniform case, neurons of type $\alpha \in \{E,I\}$ received a synaptic connection from $C^{\alpha\beta} = p^{\alpha\beta}N^{\beta}$ randomly chosen neurons of type $\beta \in \{E,I\}$. Moreover, all existing synapses from presynaptic neurons of type $\beta$ to postsynaptic neurons of type $\alpha$ had the same weight,  $J^{\alpha\beta}_{\mathrm{U}}$, in the uniform network. In the clustered model, excitatory and inhibitory neurons were instead arranged into $p$ non-overlapping clusters. Each cluster contained $f^{\alpha}N^{\alpha}$ randomly chosen neurons of type $\alpha$, and the remaining $(1-pf^{\alpha})N^{\alpha}$ neurons were placed into an unclustered ``background" population. Each neuron in a given cluster of type $\alpha$ received $f^{\beta}C^{\alpha\beta}$ connections from other neurons in the same cluster of type $\beta$, $(p-1)f^{\beta}C^{\alpha\beta}$ connections from neurons in different clusters of type $\beta$, and $(1-pf^{\beta})C^{\alpha\beta}$ connections from neurons in the background population of type $\beta$. Each neuron in the background population of type $\alpha$ received $pf^{\beta}C^{\alpha\beta}$ connections from neurons in clusters of type $\beta$ and $(1-pf^{\beta})C^{\alpha\beta}$ connections from other neurons in the background population of type $\beta$. In this way, the total number of non-zero synaptic connections was the same for the uniform and clustered networks. The weights of non-zero synaptic connections between neurons within the same cluster, $J^{\alpha\beta}_{W}$, were generally stronger in magnitude relative to the uniform case ($J^{\alpha\beta}_{W} = J^{\alpha\beta}_{+} J^{\alpha\beta}_{\mathrm{U}}$, $J^{\alpha\beta}_{+} > 1$). Moreover the weights of non-zero synaptic connections between neurons in different clusters, $J^{\alpha\beta}_{B}$, were generally weaker in magnitude relative to the uniform case ($J^{\alpha\beta}_{B} = J^{\alpha\beta}_{-} J^{\alpha\beta}_{\mathrm{U}}$, $0 < J^{\alpha\beta}_{-} < 1$). Synaptic contacts between cells in the background population and cells in clusters were also weakened relative to the uniform model, and given by $J^{\alpha\beta}_{B}$. Finally, connection weights between background neurons were unchanged relative to the uniform architecture and equal to $J^{\alpha\beta}_{\mathrm{U}}$.

The uniform and clustered networks were constructed such that the sum of all synaptic weights was the same for the two architectures. For a given value of $J^{\alpha\beta}_{\mathrm{U}}$, this was accomplished by fixing the intracluster weight factor $J^{\alpha\beta}_{+}$, and solving for the appropriate intercluster weight factor $J^{\alpha\beta}_{-}$. Following this procedure gives

\begin{equation}
J^{\alpha\beta}_{-} = \frac{ f^{\alpha} + f^{\beta} - pf^{\alpha}f^{\beta} - f^{\alpha}f^{\beta}J^{\alpha\beta}_{+} }{f^{\alpha} + f^{\beta} - pf^{\alpha}f^{\beta} - f^{\alpha}f^{\beta}}.
\end{equation}

\subsubsection*{Sensory stimuli}
\label{s:model_stim}

To model stimulus-evoked activity, sensory signals were incorporated as additional, depolarizing external inputs to the cortical circuit (Eq.~\ref{eq:LIF}). For the clustered networks, 50\% of the assemblies were chosen at random to receive input from a particular stimulus; for each selected cluster, stimulus-related input was then applied to 50\% of its E cells (chosen at random). In this way, two different stimuli in general targeted unique but overlapping sets of clusters. For the uniform networks, a given stimulus was modeled as an external input that was applied to a randomly-selected subset of the E cells; for each stimulus, the total number of stimulated neurons was chosen to be the same as in the clustered model. Throughout the text, we refer to the cells and/or clusters that receive input from a particular stimulus $s$ as ``targeted" by that stimulus, and the cells and/or clusters that do not receive input from stimulus $s$ as ``not-targeted" by that stimulus. We presented each model network with five different stimuli, matching the five tones used in the experiments.

If the $i^{th}$ cell of population $\alpha \in \{E,I\}$ was targeted by a given stimulus, then the stimulus-related input to that cell took the form

\begin{align}
I_{\mathrm{stim},i}^{\alpha}(t) = 
\begin{cases}
0 & \text{if } t < t_{\mathrm{stim}} \\
A_{\mathrm{stim}}^{\alpha} \times\nu_{o}^{\alpha}C_{\mathrm{ext}}^{\alpha E}J^{\alpha E}_{\mathrm{ext}} \times s(t) & \text{if } t \geq t_{\mathrm{stim}};
\end{cases}
\label{eq:stim_model}
\end{align}
~\\
otherwise, $I_{\mathrm{stim},i}^{\alpha}(t) = 0 \: \forall t$. In Eq.~\ref{eq:stim_model}, $t_{\mathrm{stim}}$ is the onset time of the stimulus, $A_{\mathrm{stim}}^{\alpha} > 0$ sets the amplitude of the stimulation signal for cells in population $\alpha$, and $s(t)$ describes the stimulus timecourse. Here, $A_{\mathrm{stim}}^{I} = 0$ since only E cells receive sensory stimulation. For the timecourse $s(t)$, we used a difference of exponentials:

\begin{equation}
s(t) = \gamma [e^{-(t-t_{\mathrm{stim}})/\tau_{d}} - e^{-(t-t_{\mathrm{stim}})/\tau_{r}}],
\end{equation}
~\\
where $\gamma = \big[(\tau_{r}/\tau_{d})^{\frac{\tau_{r}}{\tau_{d}-\tau_{r}}} - (\tau_{r}/\tau_{d})^{\frac{\tau_{d}}{\tau_{d}-\tau_{r}}} \big]^{-1}$, $\tau_{r}$ is the rise time constant, and $\tau_{d}$ is the decay time constant. 

\subsubsection*{Arousal modulation}
\label{s:model_arousal}

In the model, an increase in arousal was implemented as a simultaneous modulation of two parameters: (i) a decrease in the strength of recurrent synapses between excitatory cells ($J^{EE}$), and (ii) an increase in the level of external drive to E and I cells ($\nu_{\mathrm{ext}}^{\alpha}$, $\alpha \in \{E,I\}$). The arousal-related reduction in $J^{EE}$ was applied globally to all existing (i.e., all nonzero) E-E synapses: for a pair $(i,j)$ of synaptically-connected E cells, the reduction was modeled by setting $J^{EE}_{ij} = J^{EE}_{ij,o} - \Delta_{J_{EE}} \times  J^{EE}_{ij,o}$, where $J^{EE}_{ij,o}$ is the baseline value of the synapse and $\Delta_{J_{EE}} > 0$ is a parameter that sets the strength of the modulation. For a given $x \in [0,1]$ (where the upper/lower limits of x correspond to arousal levels of $0\%$/$100\%$), the $J^{EE}$ reduction parameter was given by
\begin{equation}
\Delta_{J_{EE}}(x) = \frac{L}{1 + (x^C - 1)^k },
\label{eq:Jee_arousal}
\end{equation}
where $C =  \mathrm{log}_{2} \{1/ [1 + (\frac{1-x_o}{x_o})^{1/k}]\}$, $k=1.25$, $x_{o} = 0.2$, and $L=0.75$. As arousal increases from 0 to 100$\%$ (i.e., as x increases from 0 to 1), $J^{EE}$ decreases as shown in Fig.~\ref{f:model_overview}C(i). The arousal-related increase in $\nu_{\mathrm{ext}}^{\alpha}$ varied from cell-to-cell. For the $i^{th}$ cell in population $\alpha \in \{E,I\}$, the external input modulation was modeled by setting  $\nu_{\mathrm{ext},i}^{\alpha} = \nu_{o}^{\alpha} + \Delta^{\alpha}_{\nu_i}$ , where $\nu_{o}^{\alpha}$ is the baseline value of the external input and $\Delta^{\alpha}_{\nu_i} > 0$ is a cell-dependent parameter that sets the strength of the modulation. For a given $x \in [0,1]$ (where the upper/lower limits of $x$ correspond to arousal levels of $0\%$/ $100\%$), the modulation parameter was given by
\begin{equation}
\Delta^{\alpha}_{\nu_i}(x) = \frac{z_i^\alpha M}{1 + (x^C - 1)^k },
\label{eq:nuext_arousal}
\end{equation}
where $C = \mathrm{log}_{2} \{1/ [1 + (\frac{1-x_o}{x_o})^{1/k}]\}$, $k=1.25$, $x_{o} = 0.2$, and $M=13.125$. For $\alpha \in \{E,I\}$, $z_{i}^{\alpha} \sim \mathrm{Beta}(a,b)$ is a random variable drawn from a Beta distribution with shape parameters $a = 10$ and $b=10$ (same for both E and I populations). Note that because $z_{i}^{\alpha} \in [0,1]$, the external drive always increases with arousal, but by varying amounts for different cells (see Fig.~\ref{f:model_overview}C(ii) for examples). For the mean-field analysis of the reduced 2-cluster network (``\textit{\nameref{s:effective_mft}}"), we used the same arousal model, but multiplied $L$ and $M$ in Eqs.~\ref{eq:Jee_arousal} and \ref{eq:nuext_arousal}, respectively, by a factor of 0.35 (i.e., we used $L = 0.75 \times 0.35$ and $M = 13.125 \times 0.35$). This adjustment recalibrated the arousal parameters such that they varied over an appropriate range for the 2-cluster model.

\subsubsection*{Numerical simulations}
\label{s:numerical_sims}

The dynamical system defined by Eqs.~\ref{eq:LIF}-\ref{eq:LIF_external_current} was integrated using a discrete time step $\mathrm{dt} = 0.5\times10^{-4}$ seconds. All spike times were forced to the simulation grid, and exact updates were performed between time steps. At a given level of arousal, we performed simulations on several realizations of the clustered and uniform networks; different network instances were also associated with different realizations of the arousal model (i.e., different random draws of the external input modulations; see Eq.~\ref{eq:nuext_arousal} above). For most analyses, we generated 10 realizations of the network architecture, and simulated 30 trials of network activity per stimulus for each network realization. For these simulations, each trial lasted 2.5 seconds and stimulus onset occurred at $t_{\mathrm{stim}} = 1$ second; the pre-stimulus period of each trial was considered ``spontaneous" activity. For some analyses, we ran a separate set of simulations to obtain longer continuous blocks of spontaneous activity. For the cluster interactivation and activation timescale analyses (``\textit{\nameref{s:cluster_timescale}}"; Fig.~\ref{f:mean_field_vs_sims}G), we simulated two network realizations, and for each one, we ran 30, 5.2-second-long trials of spontaneous-only activity (no stimulus presentation). For the interspike-interval (``\textit{\nameref{s:cvISI}}"; Fig.~\ref{f:cvISI_spectra_supp}A) and power spectra (``\textit{\nameref{s:power_spectra}}"; Fig.~\ref{f:cvISI_spectra_supp}B,C) analyses, we simulated two network realizations, and for each one, we ran 30, 2.7-second-long trials of spontaneous-only activity. In all simulations, different trials used different random initial conditions for neurons' membrane potentials.

\subsection*{Single-cell discriminability}
\label{s:dprime}

To examine neural discriminability on a single-cell level, we computed a standard metric for quantifying the separability of two univariate stimulus response distributions. Given the responses of an individual cell to repeated presentations of two stimuli $s_A$ and $s_B$, one measure of single-cell discriminability ($d'$) is:

\begin{equation}
d'(A,B) = \frac{|\mu_{A} - \mu_{B}|}{\sqrt{ \frac{1}{2} ( \sigma_{A}^{2} + \sigma_{B}^{2} ) }},
\label{eq:singlecell_prime}
\end{equation}
where $\mu_A$ and $\mu_B$ denote the average responses to the two stimuli, and where $\sigma_A$ and $\sigma_B$ denote the standard deviations of the two response distributions.

To compute an overall discriminability index in the model or data, we began by computing timecourses of the single-cell discriminability relative to stimulus presentation. First, trials were parsed according to arousal level (see below) and then aligned to stimulus onset. For each trial of a given stimulus, we then computed binned spike counts of every cell in a sliding window (see subsections below for window parameters used in the model and data). In total, we obtained an array of spike counts (i.e., responses) of dimension $N_{\mathrm{cells}} \times N_{\mathrm{stimuli}} \times N_{\mathrm{trials}} \times N_{\mathrm{time\:bins}}$. In each time bin, the across-trial mean and standard deviation of the spike counts were used to compute $d'$ for each cell and pair of stimuli, according to Eq.~\ref{eq:singlecell_prime}. To summarize the discriminability of an individual cell $i$ in time bin $t$, we computed its average $d'$ over all stimulus pairs, denoted as $\overline{d'}_{i,t}$. We then computed the average across all cells in each time bin, denoted as $\langle \overline{d'}_t \rangle$. An overall discriminability index for the population was defined as the maximum of the timecourse $\langle \overline{d'}_t \rangle$; we denote this index as the cell-averaged $D'_{sc}$. We also determined the time point $t*$ at which $\langle \overline{d'}_t \rangle$ was maximized, from which we computed an overall discriminability index for each cell $i$ as $D_{\mathrm{sc},i}' = \overline{d'}_{i,t_{*}}$. In the following two subsections, we provide further details on the single-cell discriminability analyses for the network models and experimental data.

\subsubsection*{Experimental data}

To quantify how arousal level impacted single-cell discriminability in the experimental data, we parsed the trials in a given session according to their pupil diameter. To begin, we computed the average pupil diameter across the pre-stimulus period of each trial (100 ms window preceding tone onset). We then split the trials into ten equally-sized partitions according to the deciles of the pre-stimulus pupil diameter distribution (see Fig.~\ref{f:decoding_allTrials}A for an example); this partitioning procedure allowed us to use the maximum number of trials for the analysis. Within each decile bin, we also randomly subsampled the trials to ensure that each partition contained the same number of trials per tone frequency. 

After parsing the data, we computed the single-cell discriminability separately for each pupil-based partition in a session. For this analysis, spikes from each cell were counted in 100 ms windows incremented in 10 ms steps, and we considered a total time span of 450 ms after stimulus onset. We then computed the cell-averaged and single-cell $D'_{sc}$ values using the procedure described above. 

The arousal-conditioned analysis yielded a single value for the cell-averaged $D'_{sc}$ in each pupil diameter decile of a given session. We now explain the procedure for combining results across sessions in order to understand the aggregate effect of arousal on single-cell discriminability (Fig.~\ref{f:decoding_allTrials}F). First, for each pupil decile in a given session, we computed the percent change in the cell-averaged $D'_{sc}$ relative to the maximum value of that quantity across all decile bins in the session. We then computed the average pupil diameter of the trials in each decile bin, and binned each data point in a session (one per decile) according to that average pupil diameter. For this binning step, we used ten non-overlapping bins, each of width $10\%$ max-normalized pupil diameter. If more than one data point from the same session fell within a single pupil diameter bin, we stored the average percent change in cell-averaged $D'_{sc}$ across all the data in that bin. This process was then repeated for each session, yielding a collection of data points (percent changes in cell-averaged $D'_{sc}$) in each pupil diameter bin (gray dots in Fig.~\ref{f:decoding_allTrials}F). Note that because different sessions explored different pupil dilation ranges, not all sessions contributed to every pupil diameter bin; specifically, there was more data at intermediate diameters relative to very small or large ones. To summarize how single-cell discriminability varied with arousal, we computed the average percent change in cell-averaged $D'_{sc}$ across all sessions in each pupil diameter bin (red curve in Fig.~\ref{f:decoding_allTrials}F); the spread of the data across sessions in each pupil bin was indicated by a boxplot (Fig.~\ref{f:decoding_allTrials}F).

To quantitatively test whether single-cell discriminability was improved at intermediate arousal relative to either low or high arousal, we compared the distribution of single-cell $D_{\mathrm{sc}}'$ values at intermediate pupil diameter to the distributions at small or large diameters using a paired statistical test. For each session, we first determined the pupil decile whose trials had an average pupil diameter closest to 50$\%$ of maximum pupil dilation (``central" decile). We also found the set of sessions for which the trials in the first decile bin had an average pupil diameter $\leq 33\%$ of maximum dilation (``low pupil sessions", LS, 10 sessions in total), and the set of sessions for which the trials in the last pupil decile bin had an average pupil diameter $\geq 67\%$ of maximum dilation (``high pupil sessions", HS, all 15 sessions). To compare $D_{\mathrm{sc}}'$ between low and middle pupil diameters, we pooled the single-cell $D_{\mathrm{sc}}'$ values from the first decile bin and central decile bin of each low pupil session into two groups: $\{D'_{\mathrm{sc,low\:pupil}}\}_{\mathrm{LS}}$ and $\{D'_{\mathrm{sc,mid\:pupil}}\}_{\mathrm{LS}}$. To compare $D_{\mathrm{sc}}'$ between high and middle pupil diameters, we pooled the $D_{\mathrm{sc}}'$ values from the last decile bin and central decile bin of each high pupil session into two sets: $\{D'_{\mathrm{sc,high\:pupil}}\}_{\mathrm{HS}}$ and $\{D'_{\mathrm{sc,mid\:pupil}}\}_{\mathrm{HS}}$. We then tested for a significant difference between $\{D'_{\mathrm{sc,low\:pupil}}\}_{\mathrm{LS}}$ and $\{D'_{\mathrm{sc,mid\:pupil}}\}_{\mathrm{LS}}$ (or $\{D'_{\mathrm{sc,high\:pupil}}\}_{\mathrm{HS}}$ and $\{D'_{\mathrm{sc,mid\:pupil}}\}_{\mathrm{HS}}$) using the Wilcoxon signed-rank test. In Fig.~\ref{f:decoding_allTrials}G, we show the distributions of the differences $\{D'_{\mathrm{sc,mid\:pupil}} - D'_{\mathrm{sc,low\:pupil}}\}_{\mathrm{LS}}$ (top panel) and $\{D'_{\mathrm{sc,mid\:pupil}} - D'_{\mathrm{sc,high\:pupil}}\}_{\mathrm{HS}}$ (bottom panel).

\subsubsection*{Network models}

In the network models, spikes from each cell were counted in 100 ms windows incremented in 20 ms steps along the length of a trial. We then used the previously-described procedure to compute the cell-averaged $D'_{sc}$. For a given network realization and arousal level, results were based off 30 trials per each of 5 stimuli. To summarize how the overall single-cell discriminability varied with arousal strength, we computed the cell-averaged $D'_{sc}$ at each sampled arousal level for a given network realization. At each arousal, we then computed the percent change in the cell-averaged $D'_{sc}$ relative to the maximum value over all arousal levels. Finally, we averaged the results across ten network realizations to obtain the results in Figs.~\ref{f:decoding_model}C,D.

\subsection*{Population decoding}
\label{s:population_decoding}

Population decoding analyses assess the extent to which stimulus identity can be read-out from single-trial responses of a neural ensemble \cite{quian2009extracting}. Here, we used cross-validated linear classification methods to examine how well stimuli could be discriminated from population activity as a function of arousal. For this analysis, trials were first parsed by arousal and aligned to stimulus onset; the spikes of each cell in the ensemble were then counted in a sliding window moving along the length of a trial. For a given arousal level, this procedure yielded a spike-count array of dimension $N_{\mathrm{cells}} \times N_{\mathrm{trials}} \times N_{\mathrm{windows}}$. The following two subsections below provide details on the data selection and spike-count window parameters for decoding in the experimental sessions and network models.

After obtaining the spike-count array, a linear decoding analysis was performed separately for each time window in a trial. Cross-validated linear classification was implemented using version 0.24.2 of the scikit-learn Python package, and proceeded in several steps. Within a given time window, trials were split into training and testing sets. This was achieved using ten repetitions of stratified, 5-fold cross-validation. By using stratified folds, we ensured that the training and testing sets contained the same proportion of trials per stimulus. For each train-test split (50 in total), the training data was used to fit a multiclass, linear support vector classifier (`sklearn.svm.LinearSVC' with $C=0.1$, $\mathrm{dual}=\mathrm{False}$ and all other parameters set to the defaults). After fitting, the trained model was used to predict the stimulus identity of each trial in the test set. 

To assess decoding performance, we computed the average cross-validated classification accuracy. Within a given time bin, the accuracy of a single train-test split was defined as the fraction of test trials whose stimulus identity was correctly predicted. The average cross-validated accuracy of the time window was then computed as the average accuracy across all train-test splits. Repeating this process for each time bin yielded a time-course of cross-validated decoding accuracy relative to stimulus onset. The maximum of this time-course (which we refer to as ``peak accuracy" or simply ``accuracy") was then computed to summarize the overall decoding performance. Throughout the text, we refer to the time window corresponding to peak decoding accuracy as the ``peak decoding window". In the following two subsections, we provide further details on the arousal-conditioned decoding analyses for the experimental data and network models.

\subsubsection*{Experimental data}

In the experimental data, all cells were used as features for the population decoding. To quantify how arousal level impacted decoding performance, trials were grouped according to the deciles of their pre-stimulus pupil diameter distribution, as described in the section ``\textit{\nameref{s:dprime}}". Within each decile bin, we randomly subsampled the trials to ensure that each partition contained the same number of trials per tone frequency. Subsequent decoding analyses were then performed independently for each pupil-based partition of the data. When examining the relationship between decoding performance and arousal in the absence of locomotion, we excluded trials with an average pre-stimulus treadmill speed exceeding 2 cm/sec. 

After collecting the relevant subset of data, we computed the spike counts of each cell in every trial using 100 ms windows incremented in 10 ms steps, and we considered a total time span of 600 ms after stimulus onset. We then followed the previously-described procedure to compute the peak decoding accuracy in each pupil decile bin of a session. To combine results across sessions (Fig.~\ref{f:decoding_allTrials}H), we used the method described in ``\textit{\nameref{s:dprime}}". 

To test whether moderate arousal was associated with improvements in population-level decoding, we compared the decoding accuracy at moderate pupil diameters to the accuracy at either small or large diameters. For each session, we first determined the decile whose average pupil diameter was closest to 50$\%$ of maximum pupil dilation (``central" decile). We then found the set of ``low pupil sessions" (those for which the average pupil diameter of trials in the first decile bin was $\leq 33\%$ of maximum dilation; 10 sessions total) and ``high pupil sessions" (those for which the average pupil diameter of trials in the last decile bin was $\geq 67\%$ of maximum dilation; all 15 sessions). For all low pupil (high pupil) sessions, we then tested for a significant difference between the accuracy in the central decile and the accuracy in the first decile (last decile) using the Wilcoxon signed-rank test (Fig.~\ref{f:decoding_allTrials}I).

\subsubsection*{Network models}

For stimulus decoding in the network models, we sampled a subset of the excitatory cells to be used as features in the classification analysis. In the main text (Fig.~\ref{f:decoding_model}E,F), decoding was performed using ensembles composed of 10\% of the excitatory cell population. For the uniform networks, these ensembles were generated by drawing a random sample of cells from the entire excitatory population. For the clustered networks, we randomly sampled an equal number of cells from each subpopulation (i.e, from each cluster and the background population) until the correct sample size was reached; since the total number cells to be sampled was not evenly divisible by the number of subpopulations, the number of cells remaining after sampling equally from each subpopulation were drawn from a randomly-chosen set of the clusters and/or background population. We also explored the impact of using different ensemble sizes in the decoding analyses (Fig.~\ref{f:decoding_model_supp}). Specifically, we considered sample sizes of 1, 2, 4, 8, 16, and 32 neurons/subpopulation, corresponding to a total of 19,  38,  76, 152, 304, and 608 features ($\sim$ 1.2\%, 2.4\%, 4.8\%, 9.5\%, 19.0\% and 38.0\% of the excitatory cell population). 

For a given cell ensemble, the cross-validated decoding accuracy was computed according to the procedure described in the previous section. For this analysis, we used 100 ms spike-count windows incremented in 20 ms steps along the length of a trial. For a given network realization and arousal level, results were based off 30 trials per each of 5 stimuli, and the decoding accuracy was averaged over 25 different runs, where each run used a different random subsample of cells. 

The decoding analysis was performed at several values of arousal for each network realization. For a given network realization, we summarized the impact of arousal by computing the percent change in decoding accuracy at each arousal level relative to the maximum accuracy obtained across all arousal levels. At each level of arousal, we then computed the average percent change in accuracy across ten different network realizations (Fig.~\ref{f:decoding_model}E,F).

\subsection*{Relationships between firing rate and arousal}
\label{s:rate_vs_arousal}

\subsubsection*{Experimental data}

To examine how spontaneous firing rates varied with arousal in the data (Fig.~\ref{f:rate_corr_data_model}A-E), we split the spontaneous periods of each session into smaller windows of length 100 ms. For each window, we computed the spike count of every cell and the average pupil diameter over the window duration. Windows from all spontaneous periods were collected into a single dataset, and were then divided into ten groups according to the deciles of their pupil diameter distribution. For each decile bin, we computed \textit{(i)} the average pupil diameter across all windows in the bin, and \textit{(ii)} the average firing rate of each unit across all windows in the bin (see Fig.~\ref{f:rate_corr_data_model}A,B for examples). Finally, we tested for a monotonic relationship between spontaneous firing rate and arousal by computing the Spearman correlation between a unit's average firing rate in each pupil decile bin and the average pupil diameter in each decile bin (Fig.~\ref{f:rate_corr_data_model}C). A correlation with $p < 0.05$ was considered statistically significant, and the sign of the correlation indicated whether the firing rate of the corresponding unit tended to increase (positive modulation) or decrease (negative modulation) with pupil diameter; non-significant correlations indicated the absence of a clear monotonic relationship between spontaneous firing rate and pupil diameter. We note that the above approach is similar to that used in \textcite{christensen2022reduced}. Fig.~\ref{f:rate_corr_data_model}D shows the fraction of units (averaged across sessions), with significant positive or negative correlations computed with this method. Results for individual sessions are shown in Fig.~\ref{f:rate_corr_data_model}E.

\subsubsection*{Network models}

To quantify how spontaneous activity was impacted by arousal in the network models, we computed single-cell firing-rates in the absence of sensory stimuli. For a fixed value of arousal, rates of all cells were computed during the 800 ms window preceding stimulus onset; in total, we used 150 trials (5 stimuli $\times$ 30 trials/stimulus) per network realization. We then averaged the spontaneous rates of each neuron across trials, and computed the Spearman correlation between the trial-averaged rate of each cell and the arousal modulation strength. A significant ($p < 0.05$) positive/negative correlation indicated a cell whose firing rate tended to monotonically increase/decrease with arousal strength. Fig.~\ref{f:rate_corr_data_model}F shows the fraction of all neurons in the clustered networks that exhibited significant positive or negative correlations with arousal strength. Similar results for the uniform networks are shown in Fig.~\ref{f:rate_corr_data_model}G.

\subsection*{Determining stimulus-responsiveness}
\label{s:tone_responsive}

To determine if a cell responded significantly to a particular stimulus, we compared its pre- and post-stimulus activity. Specifically, for each trial of a given stimulus, we computed cell spike counts in the 100 ms window preceding stimulus onset and in the 100 ms window right after stimulus onset. For each cell, the pre- and post-stimulus spike counts were then compared using the Wilcoxon signed-rank test, and the stimulus response was considered significant if the two-sided p-value was $< 0.05$. In the experimental data, the number of trials per tone (before conditioning on arousal state) ranged from 1076-1517, depending on the session. 

\subsection*{Correlation-based clustering analysis}
\label{s:clustering}

We analyzed noise correlation and tuning similarity matrices to look for evidence of functionally-organized neural clusters in patterns of neural activity (Fig.~\ref{f:cluster_analysis_main}). In what follows, we explain how noise correlations and tuning similarity were computed in the model and data. We then describe the clustering procedure used to extract neural clusters from noise correlation matrices, and the statistical methods employed to test whether the detected clusters were meaningfully organized by tuning similarity.

\subsubsection*{Noise correlations in the network models}

We estimated noise correlations for random samples of cells drawn from either clustered or uniform networks. For a given network realization of the clustered (uniform) model, we subsampled $10\%$ of clustered (all) excitatory cells at random. After subsampling, only cells that responded significantly to at least one stimulus were kept for further analysis (see ``\textit{\nameref{s:tone_responsive}}" for details). To estimate noise correlations, we computed the spike-counts of each cell in the 100 ms post-stimulus window of every trial. We then computed the Pearson correlation between the spike-count vectors of each pair of cells. To avoid capturing correlations driven by arousal- or stimulus-related changes in firing rate, neuron-by-neuron correlation matrices were computed separately for each stimulus and arousal level. A single, overall correlation matrix was then obtained by averaging across all conditions (i.e., across all stimuli and arousals). For a given network realization, results were based off 30 trials/stimulus/arousal level.

We also generated a set of trial-shuffled correlation matrices for each network. For a given stimulus and arousal level, we randomly and independently permuted the trial spike-count vector of each neuron prior to computing pairwise correlations; a single, trial-shuffled correlation matrix was then obtained by averaging across all stimuli and arousal levels, as above. This independent trial-shuffling of the spike-count vectors destroys correlated variability, and leaves behind only correlations that are expected by chance. We repeated this process 100 times, yielding a set of 100 trial-shuffled correlation matrices for each network realization.

\subsubsection*{Noise correlations in the experimental data}

To estimate noise correlations in the experimental recordings, we began by computing single-cell spike counts in the 100 ms post-stimulus window in every trial. We also computed the average pupil diameter across the 100 ms window preceding stimulus onset, and binned trials according to the deciles of the resulting pre-stimulus pupil diameter distribution. To mitigate the impact of correlations due to arousal- or stimulus-related firing rate modulations, trials were separated by tone and pupil decile, and the same number of trials was subsampled for each combination. Then, for a given tone and pupil bin, the noise correlation between a neuron pair was defined as the Pearson correlation between their spike-count vectors from that stimulus and arousal condition. Finally, an overall estimate of pairwise correlations was obtained by averaging noise correlation matrices over 100 different trial subsamples and over all stimulus and arousal combinations. Only cells that responded significantly to at least one tone were included in the analysis (see ``\textit{\nameref{s:tone_responsive}}" for details).

For each session, we also generated a set of trial-shuffled correlation matrices. For each tone and pupil bin combination, we randomly and independently permuted each neuron's spike-count vector prior to computing pairwise correlations; a single, trial-shuffled correlation matrix was then obtained by averaging across tones and pupil bins, as above. The shuffling process was repeated 100 times (once for each subsampling of the data), yielding a set of 100 trial-shuffled correlation matrices for each session.  

\subsubsection*{Tuning similarity in the network models}

To quantify the similarity between the stimulus responses of two neurons, we computed the correlation between their trial-averaged evoked spike-counts for different stimuli. To begin, we computed a time-dependent spike-count for each cell in every trial of a given stimulus using a 100 ms window sliding forward in 5 ms increments. The single-trial stimulus response of a cell was then defined as its spike-count in the 100 ms window following stimulus onset minus the time-averaged spike-count across the 800 ms period preceding the stimulus. The trial-averaged response was then computed as the mean response across all trials of a given stimulus, aggregated over all arousal levels. Finally, the tuning similarity between a pair of cells was defined as the Pearson correlation between their trial-averaged responses to the five different stimuli.

\subsubsection*{Tuning similarity in the experimental data}

The stimulus tuning similarity between each pair of (tone-responsive) neurons was determined from their trial-averaged stimulus responses. For each cell, the single-trial stimulus response was defined as the difference between the spike-counts in the 100 ms window following stimulus onset and the 100 ms window preceding stimulus onset. The trial-averaged response was then computed as the mean response over all trials of a given stimulus, regardless of pupil diameter. Finally, the tuning similarity between a pair of cells was defined as the Pearson correlation between their trial-averaged responses to the five different tones.

\subsubsection*{Hierarchical clustering}
\label{s:h_clustering}

After obtaining a noise correlation matrix, we performed a clustering analysis to extract putative neural clusters corresponding to functionally-coordinated groups of cells. Given the noise correlation $r_{ij}$ between cells $i$ and $j$, we defined the distance between them as $d_{ij} = 1 - r_{ij}$. Hierarchical clustering \cite{kaufman2009finding} was then performed on the distance matrix using SciPy routines. In hierarchical clustering, each neuron begins in its own cluster. The pair of clusters that are closest -- according to a certain ``linkage criteria" -- are then merged, and this is repeated until all neurons are in a single cluster. The algorithm thus results in a hierarchical partitioning of cells into clusters, where with $N$ neurons, the lowest level contains $N$ clusters and the highest level contains 1 cluster. Here, hierarchical clustering was executed using `scipy.cluster.hierarchy.linkage' with the `average' linkage method, followed by `scipy.cluster.hierarchy.cut\_tree' to obtain the cluster partition at each level.

The number of clusters, $k$, is a free parameter. To determine the best solution, we defined a measure of partition quality as
\begin{equation}
Q = \langle \overline{r_{i}^{\mathrm{w}}} - \overline{r_{i}^{\mathrm{o}}} \rangle,
\end{equation}
where $\overline{r_{i}^{\mathrm{w}}}$ is the average correlation between cell $i$ and other cells \textit{within} its cluster (with self-correlation set to zero), $\overline{r_{i}^{\mathrm{o}}}$ is the average correlation between cell $i$ and cells \textit{outside} its cluster, and $\langle \cdot \rangle$ indicates an average over cells. Computing the partition quality at each hierarchical level results in a curve $Q(k)$; the optimal number of clusters $k^*$ was then defined as $\underset{k}{\arg\max} \ Q(k)$. The final output of the clustering procedure is a vector that contains the cluster label of each neuron for the partition with $k^*$ clusters. 

The clustering algorithm always yields a partition of neurons into clusters. It is therefore important to determine whether the detected clusters are significant relative to surrogate data that does not contain true clusters. To this end, we applied the same clustering procedure to observed and trial-shuffled correlation matrices, and determined the optimal partitions in each case. We then computed a quality measure for each cluster in the optimal partition of the observed (un-shuffled) data, and compared the observed statistic to the distribution derived from the optimal clustering of the trial-shuffled data. For a given cluster $c$, the cluster quality was defined as
\begin{equation}
Q_c = \langle \overline{r_{i}^{\mathrm{w}}} - \overline{r_{i}^{\mathrm{o}}} \rangle_{i \in c},
\end{equation}
where $\overline{r_{i}^{\mathrm{w}}}$ is the average correlation between cell $i$ and other cells \textit{within} its cluster (with self-correlation set to zero), $\overline{r_{i}^{\mathrm{o}}}$ is the average correlation between cell $i$ and cells \textit{outside} its cluster, and $\langle \cdot \rangle_{ i \in c}$ indicates an average over cells in cluster $c$). 

The null distribution of cluster qualities $\{Q^{\mathrm{null}}_c\}$ was generated by computing the quality of each cluster in a given trial-shuffled partition, and then aggregating the values across all shuffled partitions (trivial clusters containing only 1 cell were excluded). For each cluster in the observed data, we then computed a p-value as $p = (1 + b_{\mathrm{null}})/(1 + m_{\mathrm{null}})$, where $ m_{\mathrm{null}}$ is the total number of clusters in the null distribution, and $b_{\mathrm{null}}$ is the number of clusters in the null distribution with a quality $Q_c^{\mathrm{null}}$ greater than or equal to the observed statistic $Q_c^{\mathrm{obs}}$. Using the Bonferroni correction for multiple comparisons, detected clusters with $p < 0.05/n_{\mathrm{obs}}$ were considered statistically significant, where $n_{\mathrm{obs}}$ is the number of (non-trivial) clusters in the observed data (i.e., the number of comparisons). Fig.~\ref{f:cluster_analysis_main}A,E show examples of observed and trial-shuffled cluster quality distributions for the network models and experimental data.

\subsubsection*{Cluster-based tuning similarity}
\label{s:cluster_tuningSim}

The functional relevance of detected clusters was assessed with a permutation test, which quantified whether cells in the same cluster had larger tuning similarity than expected by chance. To begin, we defined a test-statistic, ``cluster-based tuning similarity", as
\begin{equation}
G = \langle \overline{s_{i}^{\mathrm{w}}} - \overline{s_{i}^{\mathrm{o}}} \rangle_{i \in\{c^*\}},
\end{equation}, 
where $\overline{s_{i}^{\mathrm{w}}}$ is the average tuning similarity between cell $i$ and other cells \textit{within} its cluster (with self-similarity set to zero), $\overline{s_{i}^{\mathrm{o}}}$ is the average tuning similarity between cell $i$ and all cells \textit{outside} its cluster, and $\langle \cdot \rangle_{i \in \{c^*\}}$ indicates an average over cells in significant clusters). In general, $G$ is large when the average tuning similarity between cells in the same cluster is much greater than between cells in different clusters. To determine if the cluster-based tuning similarity was statistically significant, we compared the value of $G$ computed from the optimal cluster partition (``observed value") to a null distribution obtained by randomly permuting cluster labels across neurons. Specifically, we permuted cluster labels $m_\mathrm{perm} = 1000$ times, and computed a p-value as $p = (1 + b_{\mathrm{perm}})/(1 + m_{\mathrm{perm}})$, where $b_{\mathrm{perm}}$ is the number of permutations that yield a test statistic as large as the observed value obtained from the optimal cluster labels. The cluster-based tuning similarity was considered significant if $p < 0.05$. Fig.~\ref{f:cluster_analysis_main}D,H show examples of the observed and permuted cluster-based tuning similarity in the clustered network model and experimental data. In the clustered model, the cluster-based tuning similarity was always statistically significant (based on results from 10 cell subsamples from each of 10 different networks). 

\subsubsection*{Validity of the clustering procedure}

In the clustered model, ground-truth cluster identities are known. To quantify the accuracy of the clustering procedure, we thus applied it to simulations of the clustered model and computed how well the clusters detected by the algorithm agreed with the ground-truth ones. For a given network realization, we computed the optimal clustering partition as described in the section ``\textit{\nameref{s:h_clustering}}". To quantify the similarity between the true cluster labels and those predicted by the clustering procedure, we then computed the ``adjusted rand score" metric as provided by \textit{scikit-learn}. Using that metric, we found that the clustering algorithm yielded partitions that were $>99\%$ accurate (average score across 10 random cell subsamples from each of 10 different network realizations.

We also verified that the clustering procedure gave reasonable results when applied to simulations of the uniform network model, which does not have true clusters. For a given network realization, we employed the clustering algorithm and statistical test described in ``\textit{\nameref{s:h_clustering}}" to determine significant clusters. Using those procedures, we found that only 0.2\% of clusters detected in cell ensembles from the uniform model were statistically significant (average over clustering results from 10 random cell subsamples from each of 10 different networks). These results are consistent with the fact that the uniform networks do not have strong clustering.

\subsection*{Mean-field analyses}

To obtain theoretical insight into the effects of the arousal modulation on network activity, we performed a series of mean-field analyses for the clustered model. Mean-field theory (MFT) is a commonly-applied technique for studying the collective dynamics of large, recurrently-connected networks of integrate-and-fire neurons \cite{renart2004mean}, and has previously been used to study attractor dynamics in networks of LIF neurons with clusters \cite{amit1997model,mazzucato2019expectation,mazzucato2015dynamics,wyrick2021state}. In what follows, we first explain the mean-field analysis carried out for the full clustered networks with both excitatory (E) and inhibitory (I) assemblies (associated with Fig.~\ref{f:mean_field_vs_sims}A of the main text). We then describe the effective MFT performed on the reduced 2-cluster network (associated with Fig.~\ref{f:mean_field_vs_sims}C-E of the main text). Because observed changes in stimulus processing result only from changes in network dynamics induced by the arousal modulation (versus from changes in the stimuli themselves), all mean-field analyses were performed for the ``spontaneous" condition (i.e., in the absence of sensory stimulation).

\subsubsection*{MFT for the full clustered networks}
\label{s:mft_basic}

Consider a network of LIF neurons composed of $p$ E clusters, $p$ I clusters, 1 ``background" (unclustered) E population, and 1 ``background" I population, for a total of $2(p+1)$ populations. We label the populations with a pair of superscripts $(\alpha, \gamma)$. The first superscript $\alpha \in \{E,I\}$ labels populations as excitatory or inhibitory, and the second superscript $\gamma \in \{1,...,p+1\}$ specifies the population number, where the first $p$ indices correspond to the cluster labels and the $p+1$ index corresponds to the background population. All neurons in the same population described by a specific $(\alpha,\gamma)$ pair have the same intrinsic parameters and the same recurrent connectivity properties (i.e., receive the same number and strength of inputs from their own and other populations). If the external input parameters were also homogeneous across population $(\alpha,\gamma)$, then the statistics of the overall input to each cell would be identical and all cells in the population would share the same average steady-state firing rate, $\nu^{\alpha,\gamma}$. The situation is more complex in the simulations, because the level of external input varies from cell-to-cell (due to the quenched randomness in the external drive component of the arousal modulation; see ``\textit{\nameref{s:model_arousal}}"). However, for the mean-field analyses, we neglect the heterogeneity in the external input rates $\nu^{\alpha}_{\mathrm{ext},i}$ and assume that all neurons in population $(\alpha, \gamma)$ are subject to the cell-average external rate $\overline{\nu^{\alpha}_{\mathrm{ext}}} = \frac{1}{N^{\alpha}}\sum_{i}\nu^{\alpha}_{\mathrm{ext},i}$. Thus, in the mean-field, all neurons in the same population $(\alpha, \gamma)$ are statistically identical and are described by the same average firing rate $\nu^{\alpha,\gamma}$. Though here we neglect the quenched variability in the external inputs, we note that it may be possible to incorporate it using an extended mean-field framework \cite{vegue2019firing}.

The goal of the mean-field analysis is to solve for the steady-state firing rates of each population. To proceed, one makes a set of assumptions about the operating regime of the network, namely, that each neuron's spike train is described by a stationary Poisson process, that the spike trains of different neurons are independent, and that individual spikes from a presynaptic neuron induce only a small change in the voltage of a postsynpatic neuron relative to it's firing threshold \cite{renart2004mean}. Under these conditions, one can make the diffusion approximation and replace the presynaptic input to population $(\alpha,\gamma)$ by a Gaussian white noise with mean $\mu^{\alpha,\gamma}$ and standard deviation $\sigma^{\alpha,\gamma}$. Assuming exponentially-decaying synapses with time constant $\tau_s$, the dynamics of a neuron $i$ in population $(\alpha,\gamma)$ becomes

\begin{align} 
\label{eq:diffusion_dyn}
\tau_{m}^{\alpha} \frac{dV_{i}^{\alpha,\gamma}}{dt} & = -V_{i}^{\alpha,\gamma}(t) + \tau_{m}^{\alpha} I_{i}^{\alpha,\gamma}(t)
\\
\label{eq:diffusion_dyn_current}
\tau_{s} \frac{dI_{i}^{\alpha,\gamma}}{dt} & = -I_{i}^{\alpha,\gamma}(t) + \mu^{\alpha,\gamma} + \sigma^{\alpha,\gamma} \eta_{i}(t)
\end{align}
where $\tau_m^{\alpha}$ is the membrane time constant, $V_{i}^{\alpha,\gamma}$ is the membrane potential, $I_{i}^{\alpha,\gamma}(t)$ is the total synaptic input from both external and recurrent sources, and $\eta_{i}(t)$ is a Guassian white noise obeying $\langle \eta_{i}(t) \rangle = 0$ and $\langle \eta_{i}(t)\eta_{i}(t') \rangle = \delta(t-t')$. The mean $\mu^{\alpha,\gamma}$ and variance $(\sigma^{\alpha,\gamma})^{2}$ of the input depend on the network architecture. For the clustered networks studied here, we have

\begin{align}
\mu^{\alpha,\gamma} = 
\begin{cases}
\displaystyle{\sum_{\beta=E,I}C^{\alpha\beta}f^{\beta}J^{\alpha\beta}_{W}\nu^{\beta,\gamma} + 
\sum_{\beta=E,I} C^{\alpha\beta}f^{\beta}J^{\alpha\beta}_{B}\sum_{\substack{\lambda=1 \\ \lambda \neq \gamma}}^{p}\nu^{\beta,\lambda} +
\sum_{\beta=E,I}(1-pf^{\beta})C^{\alpha\beta}J^{\alpha\beta}_{B}\nu^{\beta,p+1}} \\ \displaystyle{+ \; C^{\alpha E}_{\mathrm{ext}}J^{\alpha E}_{\mathrm{ext}}\overline{\nu_{\mathrm{ext}}^{\alpha}}}, \;\;\; \text{if } \gamma = [1,...,p] \\ \\
\displaystyle{\sum_{\beta=E,I}C^{\alpha\beta}f^{\beta}J^{\alpha\beta}_{B}\sum_{\lambda = 1}^{p}\nu^{\beta,\lambda} +
\sum_{\beta=E,I}(1-pf^{\beta})C^{\alpha\beta}J^{\alpha\beta}_{U}\nu^{\beta,p+1} +
C^{\alpha E}_{\mathrm{ext}}J^{\alpha E}_{\mathrm{ext}}\overline{\nu^{\alpha}_{\mathrm{ext}}}}, \;\;\; \text{if } \gamma = p + 1 
\end{cases}
\label{eq:mft_mean_noQuenched}
\end{align}

and 

\begin{align}
(\sigma^{\alpha,\gamma})^{2} = 
\begin{cases}
\displaystyle{\sum_{\beta=E,I}C^{\alpha\beta}f^{\beta}(J^{\alpha\beta}_{W})^{2}\nu^{\beta,\gamma} +
\sum_{\beta=E,I} C^{\alpha\beta}f^{\beta}(J^{\alpha\beta}_{B})^{2}\sum_{\substack{\lambda=1 \\ \lambda \neq \gamma}}^{p}\nu^{\beta,\lambda} +
\sum_{\beta=E,I}(1-pf^{\beta})C^{\alpha\beta}(J^{\alpha\beta}_{B})^{2}\nu^{\beta,p+1}} \\ \displaystyle{+ \; C^{\alpha E}_{\mathrm{ext}}(J^{\alpha E}_{\mathrm{ext}})^{2}\overline{\nu^{\alpha}_{\mathrm{ext}}}}, \;\;\; \text{if } \gamma = [1,...,p] \\ \\
\displaystyle{\sum_{\beta=E,I}C^{\alpha\beta}f^{\beta}(J^{\alpha\beta}_{B})^{2}\sum_{\lambda = 1}^{p}\nu^{\beta,\lambda} +
\sum_{\beta=E,I}(1 - pf^{\beta})C^{\alpha\beta}(J^{\alpha\beta}_{U})^{2}\nu^{\beta,p+1} +
C^{\alpha E}_{\mathrm{ext}}(J^{\alpha E}_{\mathrm{ext}})^{2}\overline{\nu^{\alpha}_{\mathrm{ext}}}}, \;\;\;  
\text{if } \gamma = p + 1
\end{cases}
\label{eq:mft_sd_noQuenched}
\end{align}
~\\
where $\nu^{\beta, \lambda}$ is the firing rate of population $(\beta, \lambda)$ with $\beta \in \{E,I\}$, $\lambda \in \{1,...,p+1\}$; all other parameters in Eqs.~\ref{eq:mft_mean_noQuenched}-\ref{eq:mft_sd_noQuenched} are defined in the section ``\textit{\nameref{s:circuit_model}}". For each population, $\mu$ and $\sigma^2$ contain recurrent contributions from the same population and from the other populations in the network, as well as an external contribution from the background drive. The system defined by Eqs.~\ref{eq:diffusion_dyn}-~\ref{eq:mft_sd_noQuenched}, along with the threshold and reset conditions for the membrane potential, can be analyzed using the Fokker-Planck framework \cite{renart2004mean}. When $\tau_{s} << \tau_{m}^{\alpha}$, the steady-state firing rate of neurons in population $(\alpha,\gamma)$ satisfies the self-consistent relationship

\begin{equation}
\label{eq:mft_rates}
\nu^{\alpha,\gamma} = \Phi^{\alpha,\gamma}[\mu^{\alpha,\gamma}(\bm\nu),\sigma^{\alpha,\gamma}(\bm\nu)].
\end{equation}
In Eq.~\ref{eq:mft_rates}, $\bm \nu = [\nu^{E,1},...,\nu^{E,p+1},\nu^{I,1},...,\nu^{I,p+1}]$ is the vector of firing rates of each population and $\Phi^{\alpha,\gamma}$ is the transfer function for population $(\alpha,\gamma)$, given by
\begin{equation}
\label{eq:transfer_func}
\Phi^{\alpha,\gamma} = {\displaystyle \Bigg[ \tau_{r} + \tau_{m}^{\alpha} \sqrt{\pi} \int_{q_{r}^{\alpha,\gamma}}^{q_t^{\alpha,\gamma}}  e^{x^2}\mathrm{erfc}(-x)dx \Bigg]^{-1} } 
\end{equation}
where
\begin{equation}
q_{r}^{\alpha,\gamma} = \frac{V_{r}^{\alpha} - \tau_{m}^{\alpha}\mu^{\alpha,\gamma}}{\sqrt{\tau_{m}^{\alpha}}\sigma^{\alpha,\gamma}} + a\sqrt{\tau_{s}/\tau_{m}^{\alpha}}
\label{eq:mft_transferFunc_lower}
\end{equation}
\begin{equation}
q_{t}^{\alpha,\gamma} = \frac{V_{t}^{\alpha} - \tau_{m}^{\alpha}\mu^{\alpha,\gamma}}{\sqrt{\tau_{m}^{\alpha}}\sigma^{\alpha,\gamma}} + a\sqrt{\tau_{s}/\tau_{m}^{\alpha}}
\label{eq:mft_transferFunc_upper}
\end{equation}
and with $a = -\zeta(1/2)/\sqrt{2}$ \cite{brunel1998firing}.

To find allowed states of the network, we numerically solved the set of $2(p+1)$ self-consistent equations defined by Eq.~\ref{eq:mft_rates} in conjunction with Eqs.~\ref{eq:mft_mean_noQuenched} and ~\ref{eq:mft_sd_noQuenched}. Importantly, multiple solutions -- corresponding to different numbers of active and inactive clusters -- can exist for the same set of parameters. In such cases, the solution obtained will depend on the initial guess for the firing rate vector. To systematically deal with this fact, we looked for solutions with $n_{A}$ active clusters and $p-n_{A}$ inactive clusters by setting the initial rates for the first $n_{A}$ E and first $n_{A}$ I populations to $\nu_{\mathrm{high}}^{E}$ and $\nu_{\mathrm{high}}^{I}$, respectively, and the initial rates for the remaining E and I populations to $\nu_{\mathrm{low}}^{E}$ and $\nu_{\mathrm{low}}^{I}$, respectively. By choosing $\nu_{\mathrm{high}}^{E} > \nu_{\mathrm{low}}^{E}$ and $\nu_{\mathrm{high}}^{I} > \nu_{\mathrm{low}}^{I}$ we biased the numerical solver to search for solutions with $n_{A}$ active clusters; the solution space was then mapped by varying $n_{A} \in \{0,...,p\}$. 

We denote a self-consistent solution with $n_A$ active clusters as $\bm{\nu}_{n_A}$. The solution in which all clusters have the same firing rate (i.e., $n_{A}=0$) is referred to as the ``uniform state" and solutions with $n_A \geq 1$ active clusters are referred to as ``cluster states". For cluster states, the $n_{A}$ active clusters of type $\alpha \in \{E,I\}$ have steady-state rate $\nu_{n_A,\mathrm{\uparrow}}^{\alpha}$ and the $p-n_A$ inactive clusters of type $\alpha$ have rate $\nu_{n_A,\mathrm{\downarrow}}^{\alpha}$, where $\nu_{n_A,\mathrm{\uparrow}}^{\alpha} > \nu_{n_A,\mathrm{\downarrow}}^{\alpha}$. Depending on the network parameters, cluster states may not be found.

\subsubsection*{Selecting $J^{EE}_{+}$ for the MFT}
\label{s:mft_sim_Jplus_comparison}

To study the impact of the arousal modulation in the MFT, we first examined the effect of the E-to-E intracluster weight factor $J^{EE}_{+}$, which controls the dynamical regime of the network \cite{mazzucato2015dynamics}. We varied $J^{EE}_{+} \in [12, 19.5]$ using steps of size $\Delta J^{EE}_{+} = 0.025$. At each $J^{EE}_{+}$, we searched for self-consistent solutions $\bm \nu_{n_{A}}$ with $n_{A} \in \{0,...,5\}$ active clusters. Whether or not a cluster solution was found for a particular $n_{A} \geq 1$ depended on the value of $J^{EE}_{+}$ (Fig.~\ref{f:mft_supp}A). 

To compare to the MFT, we ran an additional set of network simulations in which $J^{EE}_{+}$ was varied in the range $[12, 19.5]$ in steps of size $\Delta J^{EE}_{+} = 0.75$. For these simulations, no arousal modulations or sensory stimuli were applied, and we ran 20 trials per each of 5 network realizations; all other parameters were as described in Table~\ref{t:model_params} and the ``\textit{\nameref{s:circuit_model}}" section. For each simulated trial at a given $J^{EE}_{+}$, we computed \textit{(i)} the active cluster rate $\nu_{n_A,\mathrm{\uparrow}}^{E}$ conditioned on a given number of active clusters $n_A$ (see ``\textit{\nameref{s:cluster_rates}}"), \textit{(ii)} the probability $P(n_A)$ of finding $n_A$ active clusters (see ``\textit{\nameref{s:cluster_rates}}"), and \textit{(iii)} the population average firing rate of all E neurons. Analyses were based on 2.3 seconds of simulated activity per trial, and all quantities were averaged across trials and network realizations. Results are shown in Fig.~\ref{f:mft_supp}B; note that the active cluster rate $\nu_{n_A,\mathrm{\uparrow}}^{E}$ is only plotted for values of $n_A$ satisfying $P(n_A) \geq 0.2$.

We observed that cluster states emerged at lower values of $J^{EE}_{+}$ in the simulations compared to the mean-field (Fig.~\ref{f:mft_supp}A,B). This is potentially due to the finite-size of the simulated networks and the inexact incorporation of synaptic dynamics in the mean-field. Although the mean-field does not quantitatively capture the behavior of the simulations, it can still provide insight into the effects of the arousal modulation. In order to qualitatively compare the theory and simulations as a function of arousal, we considered a fixed intracluster weight factor for the simulations ($J^{EE}_{+,\ \mathrm{sim}}$). We then ran the mean-field at a larger intracluster weight factor $J^{EE}_{+, \ \mathrm{mft}}$, which was chosen to achieve the best match with simulations run at $J^{EE}_{+, \ \mathrm{sim}}$ in the absence of the arousal modulation (i.e, using the baseline network parameters). More specifically, we fixed $J^{EE}_{+, \ \mathrm{sim}} = 15.75$ (default value used throughout the main text), and computed the active cluster rate $\nu_{n_{A}^{*},\mathrm{\uparrow},\mathrm{sim}}^{E}[J^{EE}_{+} = 15.75]$ conditioned on the most likely number of active clusters $n_{A}^{*} = 3$. In the mean-field, we then determined the value of $J^{EE}_{+}$ for which the active cluster rate $\nu_{n_{A}^{*},\mathrm{\uparrow},\mathrm{mft}}^{E}$ most closely matched the value $\nu_{n_{A}^{*},\mathrm{\uparrow},\mathrm{sim}}^{E}[J^{EE}_{+} = 15.75]$ from the simulations. This procedure yielded a mean-field intracluster weight factor of $J^{EE}_{+, \ \mathrm{mft}} = 16.725$ (Fig.~\ref{f:mft_supp}A), which was then used for the mean-field calculations performed as a function of arousal in the main text (Fig.~\ref{f:mean_field_vs_sims}A). 

\subsubsection*{MFT as a function of arousal strength}
\label{s:mft_sim_comparison}

The mean-field analysis provides the steady-state firing rates of active and inactive clusters, conditioned on a particular number $n_A$ of active clusters. Together, these rates summarize the collective activity patterns of the network. To elucidate how the arousal modulation impacts the dynamics of the clustered networks, we fixed the mean-field E-to-E intracluster weight factor $J^{EE}_{+, \ \mathrm{mft}}$ according to the procedure described in ``\textit{\nameref{s:mft_sim_Jplus_comparison}}". We then ran the mean-field analysis as a function of the arousal strength, sampling the same data points as the simulations. In the mean-field, varying the arousal strength impacts the E-to-E synaptic weights ($J^{EE}_{W}$, $J^{EE}_{B}$, $J^{EE}_{U}$) and the mean external inputs to E and I cells ($\overline{\nu_{\mathrm{ext}}^{E}}$, $\overline{\nu_{\mathrm{ext}}^{I}}$). All other network parameters were set to the values given in Table.~\ref{t:model_params}. 

For a particular choice of $n_A$, we solved for the active and inactive mean-field rates, $\bm{\nu}_{n_{A},\uparrow}$ and $\bm{\nu}_{n_{A},\downarrow}$, as a function of arousal strength. The mean-field rates were obtained using the procedure described previously in ``\textit{\nameref{s:mft_basic}}". This process was then repeated for different numbers of active clusters $n_A$. In general, whether or not a cluster state solution was found for a particular $n_A \geq 1$ depended on the arousal level; beyond a certain arousal strength only the uniform state was found. 

Fig.~\ref{f:mean_field_vs_sims}A of the main text shows the mean-field rates for active and inactive excitatory clusters as a function of arousal. More specifically, for a given arousal level, the plot shows the cluster rates $\nu^{E}_{n^*_{A},\uparrow}$ and $\nu^{E}_{n^*_{A},\downarrow}$, where $n^*_A$ was the most frequently observed number of active clusters in the simulations at that arousal level (see section ``\textit{\nameref{s:cluster_rates}}"; Fig.~\ref{f:mft_supp}C). If the cluster state solution was not found for a particular arousal level, then the mean-field rate corresponding to the uniform solution is shown. Note that because the mean-field analysis used a different intracluster weight factor than the simulations ($J^{EE}_{+,\mathrm{mft}} \neq J^{EE}_{+,\mathrm{sim}}$; see ``\textit{\nameref{s:mft_sim_Jplus_comparison}}"), the comparison between the mean-field and simulations in Fig.~\ref{f:mean_field_vs_sims} is only meant to be qualitative. 

\subsubsection*{Effective MFT for reduced 2-cluster networks}
\label{s:effective_mft}

The MFT described thus far yields the steady-state cluster firing rates, but it cannot make predictions about dynamical transitions between the metastable states. To further understand the switching behavior of the clustered networks (Fig.~\ref{f:mean_field_vs_sims}D,E), we adapted the effective MFT developed in \cite{mascaro1999effective} and later utilized in \cite{mazzucato2019expectation,wyrick2021state}. For these calculations, we analyzed a reduced version of the full LIF clustered networks composed of two excitatory clusters $E_{1}$ and $E_{2}$, one background (unclustered) excitatory population $E_{b}$, and one background inhibitory population $I_{b}$ (Fig.~\ref{f:mft_supp}D). This 2-cluster network was constructed as described in the section ``\textit{\nameref{s:net_arch}}", with the exception that we did not depress inter-cluster weights (see Table~\ref{t:model_params_2clusterNet} for reduced network parameters). With the chosen parameters (and the arousal level set to $0\%$), the standard MFT predicts the presence of a uniform fixed point and two configurations in which one cluster is active and the other inactive (Fig.~\ref{f:mft_supp}E). The effective MFT enables insight into dynamical transitions between the two cluster states via a dimensionality reduction process that results in a description of the cluster states as wells in an effective potential energy landscape. 

Following \textcite{mascaro1999effective}, the analysis proceeds by splitting the network's populations into two groups: \textit{(i)} a set of ``in-focus" populations whose dynamical behaviors are of interest, and \textit{(ii)} a set of ``ambient" populations. Here, the two clusters $E_{1}$ and $E_{2}$ are taken as the in-focus populations, and their rates $\bm{\nu}^{F} = (\nu^{\mathrm{E,1}}, \nu^{\mathrm{E,2}})$ are treated as parameters. The two background populations, $E_{b}$ and $I_{b}$, are considered ambient populations, with rate vector $\bm{\nu}^{A} = (\nu^{E,b}, \nu^{I,b})$. For some frozen combination of the in-focus rates $\bm{\nu}^{F} = \bm{\nu}^{F}_{\mathrm{in}}$, the rates $\bm{\nu}^{A}$ of the ambient populations are allowed to adapt, and are computed self-consistently by solving the coupled system of equations
~\\
\begin{align} 
\nu^{E,b} & = \Phi^{E,b}\Big[ \mu^{E,b}(\bm{\nu}^{F}_{\mathrm{in}},\bm{\nu}^{A}), \sigma^{E,b}(\bm{\nu}^{F}_{\mathrm{in}},\bm{\nu}^{A}) \Big]
\label{eq:out_focus_rate1}
\\
\nu^{I,b} & = \Phi^{I,b}\Big[ \mu^{I,b}(\bm{\nu}^{F}_{\mathrm{in}},\bm{\nu}^{A}), \sigma^{I,b}(\bm{\nu}^{F}_{\mathrm{in}},\bm{\nu}^{A}) \Big].
\label{eq:out_focus_rate2}
\end{align}
~\\
The solution to Eqs.~\ref{eq:out_focus_rate1} and ~\ref{eq:out_focus_rate2} is denoted as $\bm{\nu}^{A}(\bm{\nu}^{F}_{\mathrm{in}})$. Feedback from the ambient populations then induces new output rates $\bm{\nu}^{F}_{\mathrm{out}} = (\nu_{\mathrm{out}}^{E,1}, \nu_{\mathrm{out}}^{E,2})$ for the in-focus populations, which are given by
~\\
\begin{align} 
\label{eq:in_focus_out_rate1}
\nu^{E,1}_{\mathrm{out}} & = \Phi^{E,1}\Big[ \mu^{E,1}\big(\bm{\nu}^{F}_{\mathrm{in}},\bm{\nu}^{A}(\bm{\nu}^{F}_{\mathrm{in}})\big), \sigma^{E,1}\big(\bm{\nu}^{F}_{\mathrm{in}},\bm{\nu}^{A}(\bm{\nu}^{F}_{\mathrm{in}})\big) \Big] = \Phi^{E,1}_{\mathrm{eff}}\Big[\bm{\nu}^{F}_{\mathrm{in}}\Big]
\\
\nu^{E,2}_{\mathrm{out}} & = \Phi^{E,2}\Big[ \mu^{E,2}\big(\bm{\nu}^{F}_{\mathrm{in}},\bm{\nu}^{A}(\bm{\nu}^{F}_{\mathrm{in}})\big), \sigma^{E,2}\big(\bm{\nu}^{F}_{\mathrm{in}},\bm{\nu}^{A}(\bm{\nu}^{F}_{\mathrm{in}})\big) \Big] = \Phi^{E,2}_{\mathrm{eff}}\Big[\bm{\nu}^{F}_{\mathrm{in}}\Big]
\label{eq:in_focus_out_rate2}
\end{align}
~\\
In Eqs.~\ref{eq:out_focus_rate1}-\ref{eq:in_focus_out_rate2}, the $\mu$'s, $\sigma$'s, and $\Phi$'s are computed similarly to Eqs.~\ref{eq:mft_mean_noQuenched}, ~\ref{eq:mft_sd_noQuenched}, and ~\ref{eq:transfer_func}, but adjusted for the 2-cluster system.

The induced rates of the in-focus populations, $\bm{\nu}^{F}_{\mathrm{out}}$, are in general different from the initial rates, $\bm{\nu}^{F}_{\mathrm{in}}$. By varying $\bm{\nu}^{F}_{\mathrm{in}}$ and computing the difference $\bm{\nu}^{F}_{\mathrm{out}} - \bm{\nu}^{F}_{\mathrm{in}}$ at each point, we obtain a flow map in the $(\nu^{E,1}_{\mathrm{in}}, \nu^{E,2}_{\mathrm{in}})$ plane (see Fig.~\ref{f:mft_supp}F). This flow map captures the response of the clusters to a particular set of quenched input rates $(\nu_{\mathrm{in}}^{E,1}, \nu_{\mathrm{in}}^{E,2})$, and contains the effect of feedback from the ambient populations. In this way, the map reveals the system's fixed points and the flow of the cluster rates $\nu^{E,1}$ and $\nu^{E,2}$ away from the stationary points. Examination of this reduced 2D description indicates that the two cluster states are attractors of the system, and are linked by an unstable fixed point corresponding to the uniform state ($\nu^{E,1} = \nu^{E,2}$; Fig.~\ref{f:mft_supp}G).

To understand how the arousal modulation impacts the cluster dynamics, we performed the effective MFT for several values of arousal. For these analyses, we used the same implementation of arousal described previously in the section ``\textit{\nameref{s:mft_basic}}", wherein the quenched randomness in the external inputs is neglected. For each arousal level, we obtained a compact representation of the system by numerically integrating the 2D flow-field along a trajectory connecting the two cluster states via the unstable fixed point (see Fig.~\ref{f:mft_supp}F). This process results in a 1D effective potential with two wells -- corresponding to the two cluster states -- separated by a barrier whose maxima corresponds to the uniform state (Fig.~\ref{f:mft_supp}H; Fig.~\ref{f:mean_field_vs_sims}C,D). The height $h$ of this barrier is related to the rate of stochastic transitions between the two cluster states \cite{mazzucato2019expectation,wyrick2021state,hanggi1990reaction,litwin2012slow}. Computing the barrier height as a function of arousal strength thus provides insight into the effects of the arousal modulation on the cluster dynamics, with lower barriers indicating faster switching and shorter-lived cluster activation periods (Fig.~\ref{f:mean_field_vs_sims}E).

\subsection*{Measures of cluster activity in the model}

\subsubsection*{Cluster firing rates}
\label{s:cluster_rates}

To compute cluster firing rates in the clustered model, we first computed the time-dependent firing rate $r_{i}(t)$ of each neuron $i$ by convolving its spike train with a Gaussian kernel of width $\sigma = 25$ ms, incremented in 1 ms steps. The firing rate $r_{c}(t)$ of a given cluster $c$, was then computed as the average rate of its constituent neurons: $r_c(t) = \langle r_{i}(t) \rangle_{i \; \in \; \mathrm{cluster} \; c}$.

To quantify how cluster activity varied with arousal (Fig.~\ref{f:mean_field_vs_sims}B) or the intracluster weight factor $J_{EE}^{+}$ (Fig.~\ref{f:mft_supp}B), we computed active and inactive cluster firing rates during the pre-stimulus period of each trial (here taken as the window spanning [-0.8, -0.1]s relative to stimulus onset). To determine the active and inactive rates, we first computed the time-dependent cluster firing rate $r_c(t)$ of every excitatory cluster in each trial (see section ``\textit{\nameref{s:cluster_rates}}"). We then computed the average pre-stimulus cluster firing rate across time and trials, which we denote as $\langle r_{c}^{\mathrm{base}} \rangle$. In a given trial, a cluster was considered ``active" at time $t$ if its baseline-subtracted firing rate, $g_{c}(t) = r_{c}(t) - \langle r_{c}^{\mathrm{base}} \rangle$, was greater than zero (i.e., $g_c(t) > 0 $). Given this criteria for cluster activation, we determined the number of active clusters $n_A$ as a function of time during the pre-stimulus period. By pooling across all time points with a particular value of $n_A$, we then calculated the probability of finding $n_A$ clusters active, as well as the average rate of active and inactive clusters as a function of $n_A$. We denote the trial-averaged active and inactive cluster firing rates for a given $n_A$  as $r_{n_A,\uparrow}$ and $r_{n_A,\downarrow}$, respectively, and the trial-averaged probability of finding $n_A$ active clusters as $P(n_A)$. We determined the most likely number of active clusters, $n_A^{*}$, as the value corresponding to the maximum of the probability $P(n_A)$ (after averaging across network realizations). 

At a fixed arousal level, only some values of $n_A$ occur with high likelihood; moreover, the most likely number of active clusters ($n_A^*$) varies with arousal (see Fig.~\ref{f:mft_supp}C for the probability of finding $n_A$ active clusters at different arousal levels). To summarize the behavior of the clustered networks as a function of arousal, we thus computed the active and inactive cluster rates conditioned on the most likely number of active clusters at each level of arousal (results shown in Fig.~\ref{f:mean_field_vs_sims}B). This analysis was based on 150 trials per network realization (5 stimuli $\times$ 30 trials/stimulus), and results were averaged over 10 different networks. See ``\textit{\nameref{s:mft_sim_Jplus_comparison}}" and Fig.~\ref{f:mft_supp} for details on the analysis of active and inactive cluster rates as a function of the intracluster weight factor $J_{EE}^{+}$.

\subsubsection*{Cluster timescales}
\label{s:cluster_timescale}

To calculate the average cluster interactivation and activation timescales, we analyzed long simulations of spontaneous activity (see the section ``\textit{\nameref{s:numerical_sims}}" for details). For each trial, we used the threshold criteria described in ``\textit{\nameref{s:cluster_rates}}" to determine the time points of cluster activation (discarding the first 0.2 seconds and last 0.1 seconds of each simulation). The cluster interactivation interval and activation time were then calculated as the average duration of all cluster interactivation and activation periods, respectively.  Results were then averaged across 30 trials per network realization. Fig.~\ref{f:mean_field_vs_sims}G shows the cluster interactivation and activation timescales as a function of arousal (average across two different network realizations). 

\subsubsection*{Cluster signal}

To calculate the cluster signal ($C_{s}$; Fig.~\ref{f:decoding_intuition_model}B), we first computed the time-dependent firing rate $r_{c}(t)$ of each excitatory cluster in every trial (see section ``\textit{\nameref{s:cluster_rates}}"). From the individual cluster rates, we next computed the average time-dependent rate across all targeted clusters, $r_{\mathrm{T}}(t)$, and the average time-dependent rate across all non-targeted clusters, $r_{\mathrm{N}}(t)$. We then took the difference between the average targeted and non-targeted cluster rates, $\Delta r_{\mathrm{T,N}}(t) = r_{\mathrm{T}}(t) - r_{\mathrm{N}}(t)$, and averaged the difference across the peak decoding window (100 ms window corresponding to peak decoding accuracy; see ``\textit{\nameref{s:population_decoding}}"). This procedure resulted in a single number $\Delta r^{*}_{\mathrm{T,N}}$ for each trial. The cluster signal was then defined as the average of $\Delta r^{*}_{\mathrm{T,N}}$ across trials. For each network realization, the cluster signal was computed using 150 trials (5 stimuli $\times$ 30 trials/stimulus); results were then averaged across 10 different network realizations.

\subsubsection*{Cluster reliability}

To compute the cluster reliability ($C_{r}$; Fig.~\ref{f:decoding_intuition_model}C), we began by computing the time-dependent, baseline-subtracted rate of each cluster, $g_c(t) = r_c(t) - \langle r_c^{\mathrm{base}} \rangle$ (see section ``\textit{\nameref{s:cluster_rates}}"). The baseline-subtracted rate was then used to determine if a given cluster was active during the peak decoding window (see ``\textit{\nameref{s:population_decoding}}") in a given trial. In particular, defining $g_c^{\mathrm{W}}$ as the average of $g_c(t)$ over the window of interest, a cluster was considered active during that window if $g_c^{\mathrm{W}} > 0$. Given this criteria, we computed the fractions of targeted and non-targeted clusters, $f_{\mathrm{T}_\uparrow}$ and $f_{\mathrm{N}_\uparrow}$, that were active during the peak decoding window in each trial. We then took the difference between those two fractions, $\Delta f_{\mathrm{T_\uparrow,N_\uparrow}} = f_{\mathrm{T}_\uparrow} - f_{\mathrm{N}_\uparrow}$, and defined the cluster reliability as the average of $\Delta f_{\mathrm{T_\uparrow,N_\uparrow}}$ across trials. For each network realization, the cluster reliability was computed using 150 trials (5 stimuli $\times$ 30 trials/stimulus); results were then averaged across 10 different network realizations.

\subsection*{Fano factor analyses}
\label{s:fano_factor}

We used the Fano factor to characterize single-cell spiking variability in both the clustered network model and the experimental data. For a given cell, the Fano factor (FF) is defined as

\begin{equation}
\mathrm{FF} = \frac{\mathrm{var}[n_{\mathrm{sp}}]}{\langle n_{\mathrm{sp}} \rangle},
\label{eq:fano_factor}
\end{equation}
~\\
where $n_{\mathrm{sp}}$ indicates the spike count of the cell within a fixed time window, and where $\mathrm{var}[\cdot]$ and $\langle \cdot \rangle$ indicate the variance and mean across repeated trials (or observation windows), respectively. In both the model and the data, we computed the FF during both spontaneous and evoked conditions.

\subsubsection*{Network model}

In the clustered network model, FFs were computed across 30 trials/stimulus and then averaged across 5 stimuli for each network realization at a fixed arousal level. For this analysis, we focused on cells in the stimulated excitatory clusters (and in the counterpart inhibitory clusters). Cells that had a low spontaneous rate of $<$ 1 spike/second at any arousal level were excluded from the analysis. To compute the FF of a given cell for a particular stimulus, we binned the spikes in each trial using a 100 ms window incremented in 20 ms steps. The FF was then computed in each time bin according to Eq.~\ref{eq:fano_factor}, yielding a time course $\mathrm{FF}(t)$. The spontaneous FF ($\mathrm{FF_{spont}}$) was defined as the value of $FF(t)$ in the bin immediately preceding stimulus onset. To summarize the evoked FF, we averaged the single-cell timecourses $\mathrm{FF}(t)$ across the relevant set of cells for each stimulus and then across all stimuli; we then determined the time point $t_{\mathrm{FF_{min}}}$ corresponding to the minimum of the cell- and stimulus-averaged trace. For a given cell and stimulus, the evoked FF, ($\mathrm{FF_{evoked}}$) was then defined as the value of $FF(t)$ at the time $t_{\mathrm{FF_{min}}}$. For each cell and stimulus, we also computed the difference between the spontaneous and evoked FFs: $\Delta \mathrm{FF} = \mathrm{FF_{spont}} - \mathrm{FF_{evoked}}$. To summarize the results, we averaged each quantity across the relevant set of cells for each stimulus and then across all stimuli; we refer to these overall values as $\langle FF_{\mathrm{spont}} \rangle$, $\langle FF_{\mathrm{evoked}} \rangle$, and $\langle \Delta FF \rangle$. Fig.~\ref{f:fano_factor_main}A-C show $\langle \mathrm{FF_{spont}} \rangle$, $\langle \mathrm{FF_{evoked}} \rangle$, and $\langle \Delta \mathrm{FF} \rangle$, respectively, as a function of arousal (where red markers and error bars indicate the mean $+/-$ 1 SD across 10 network realizations). Figs.~\ref{f:spectra_fano_supp}H-J also show the impact of varying the spike-count window length on $\langle \mathrm{FF_{spont}} \rangle$.

\subsubsection*{Experimental data}

To compute evoked FFs as a function of arousal, we computed the average pupil diameter across the 100 ms pre-stimulus period of each trial; trials were then split into ten groups according to the deciles of the pre-stimulus pupil diameter distribution (using the procedure described previously in ``\textit{\nameref{s:dprime}}"). To compute spontaneous FFs, the spontaneous blocks of each session were divided into 100 ms windows, and the average pupil diameter was computed across each one; windows were then grouped by pupil diameter, using the same partitions as for the evoked data. This procedure ensured that spontaneous and evoked Fano factors were evaluated across similar pupil dilation ranges. To account for differing numbers of windows and trials in each pupil decile partition, we subsampled the data such that all pupil partitions contained the same number of windows and trials per tone. 

For the spontaneous data, the spike count of each cell was computed in each window within a given pupil-based partition. The FF of each cell $i$ was then computed via Eq.~\ref{eq:fano_factor}, and a final estimate of the spontaneous Fano factor, $FF_{\mathrm{spont},i}$, was obtained by averaging across 100 random subsamples of the data. For the evoked FF, trials were first aligned to stimulus onset. In each trial, spikes from each cell were binned using 100 ms windows incremented in 1 ms steps. Using the trials for a given tone and pupil partition, FFs were calculated in each time bin (up to 150 ms after stimulus onset) according to Eq.~\ref{eq:fano_factor}, and results were averaged across 100 random subsamples of the data. This process yielded a time course $FF_{i,s}(t)$ for each cell $i$ and tone $s$. To summarize evoked FFs, we averaged the timecourse $FF_{i,s}(t)$ across the tone-responsive cells for a particular stimulus (see ``\textit{\nameref{s:tone_responsive}}") and then across all stimuli. We then determined the time point $t_{\mathrm{FF_{min}}}$ corresponding to the minimum of the cell- and stimulus-averaged trace. For a given cell $i$ and tone $s$, $FF_{\mathrm{evoked},i,s}$ was defined as the value of $FF_{i,s}(t)$ at the time $t_{\mathrm{FF_{min}}}$. Finally, we obtained an overall evoked FF for cell $i$ -- $FF_{\mathrm{evoked},i}$ -- by averaging $FF_{\mathrm{evoked},i,s}$ across all tones that induced a significant response in that cell. In each pupil bin, we also computed the difference $\Delta FF_{i}$ between the spontaneous and evoked FFs of cell $i$: $\Delta FF_{i} = FF_{\mathrm{spont},i} - FF_{\mathrm{evoked},i}$. Only cells that responded to at least one tone and that had an average spontaneous rate $\geq$ 1 spike/second in every pupil bin were included in the analyses.

To examine the overall effect of arousal on $FF_{\mathrm{spont}}$, $FF_{\mathrm{evoked}}$, or $\Delta FF$, we aggregated each quantity across cells and sessions in a pupil-dependent manner. For each session, we first calculated the average pupil diameter of the trials in each pupil decile partition; we then binned the FF data in each pupil decile according to the decile's average pupil diameter. For this discretization, we used ten non-overlapping pupil bins, each of width $10 \%$ max-normalized pupil diameter. If more than one pupil decile from a given session fell into the same pupil diameter bin, we stored the average value of each data point across those deciles. We then repeated this process for each session, yielding a collection of single-cell $FF_{\mathrm{spont}}$, $FF_{\mathrm{evoked}}$, or $\Delta FF$ values in each pupil diameter bin. Note that because different sessions explored different pupil dilation ranges, not all sessions contributed to every pupil diameter bin; specifically, there was more data at intermediate diameters relative to very small or large ones. To summarize how the FF quantities varied with arousal, we computed the average of each quantity across all cells in each pupil diameter bin. We denote these cell-averaged quantities as $\langle FF_{\mathrm{spont}} \rangle$, $ \langle FF_{\mathrm{evoked}} \rangle $, and $\langle \Delta FF \rangle$. The results of this analysis are shown in Fig.~\ref{f:fano_factor_main}D,F,H, where horizontal error bars indicate pupil diameter bins, and where red markers and vertical error bars indicate the cell-average $+/-$ 1 SEM of each FF quantity ($FF_{\mathrm{spont}}$, $FF_{\mathrm{evoked}}$, or $\Delta FF$) over the cells in each pupil bin. Figs.~\ref{f:spectra_fano_supp}E-J also show the effect of varying the spike-count window length on $\mathrm{FF_{spont}}$.

To test for a difference in each FF quantity between low and high arousal states, we found the set of sessions that expressed a broad range of pupil diameters. Specifically, we considered sessions for which the average pupil diameter of trials in the first pupil decile was $\leq 33\%$ of maximum dilation and for which the average pupil diameter of trials in the last pupil decile was $\geq 67\%$ of maximum dilation (9 sessions in total). We then pooled the single-cell FF quantities in the first and last pupil decile partition across all sessions with a broad pupil range. This yielded a ``small pupil" and ``large pupil" dataset for each FF quantity: (i) $\{FF^{\mathrm{small \ pupil}}_{\mathrm{spont}}\}$ and $\{FF^{\mathrm{large \ pupil}}_{\mathrm{spont}}\}$, (ii) $\{FF^{\mathrm{small \ pupil}}_{\mathrm{evoked}}\}$ and $\{FF^{\mathrm{large \ pupil}}_{\mathrm{evoked}}\}$, and (iii) $\{\Delta FF^{\mathrm{small \ pupil}}\}$ and $\{\Delta FF^{\mathrm{large \ pupil}}\}$. For each FF quantity, the small pupil and large pupil groups were compared using the Wilcoxon signed-rank test. The results are summarized in Fig.~\ref{f:fano_factor_main}E,G,I, which show the distributions of the difference in each FF quantity between small and large pupil diameters ($FF^{\mathrm{small \ pupil}}_{\mathrm{spont}} - FF^{\mathrm{large \ pupil}}_{\mathrm{spont}}$, $FF^{\mathrm{small \ pupil}}_{\mathrm{evoked}} - FF^{\mathrm{large \ pupil}}_{\mathrm{evoked}}$, and $\Delta FF^{\mathrm{small \ pupil}} - \Delta FF^{\mathrm{large \ pupil}}$). Fig.~\ref{f:spectra_fano_supp}A-C also show the small and large pupil FF quantities separately (both for single cells and population averages).

To test for overall decreases in neural variability during stimulus presentation relative to spontaneous conditions, we marginalized the data in a session across pupil diameters. Specifically, we combined the evoked trials or spontaneous windows from each pupil decile partition (see above) into two aggregate datasets. Using the aggregate datasets, we then followed the methods described above to compute (i) a pupil-aggregated spontaneous Fano factor $FF_{\mathrm{spont},i}$ of each cell $i$, and (ii) a pupil-aggregated evoked Fano factor $FF_{\mathrm{evoked},i}$ of each cell $i$. Only cells that responded to at least one tone and that had an average spontaneous rate of $\geq$ 1 spk/second were included in the analysis. To test for stimulus-induced variability quenching, we pooled the single-cell $FF_{\mathrm{spont}}$ and $FF_{\mathrm{evoked}}$ values across all sessions to obtain two groups of data: $\{FF_{\mathrm{evoked}}\}$ and $\{FF_{\mathrm{spont}}\}$. We then compared $\{FF_{\mathrm{evoked}}\}$ and $\{FF_{\mathrm{spont}}\}$ using the Wilcoxon signed-rank test (Fig.~\ref{f:spectra_fano_supp}D).

\subsection*{Variability of interspike intervals}
\label{s:cvISI}

Spike train irregularity was quantified by computing the coefficient of variation of interspike interval (cvISI) distributions during spontaneous activity. Given the interspike-interval (ISI) distribution of a single cell, the cvISI is given by

\begin{equation}
\mathrm{cvISI} = \frac{\sigma_{\mathrm{ISI}}}{\mu_{\mathrm{ISI}}},
\label{eq:cvISI}
\end{equation}
where $\sigma_{\mathrm{ISI}}$ and $\mu_{\mathrm{ISI}}$ are, the standard deviation and mean of the ISIs, respectively \cite{nawrot2008measurement}. The cvISI is equal to 0 for regular, clock-like spiking and is equal to 1 for Poisson spike trains.

\subsubsection*{Network model}

In the clustered model, single-neuron cvISI values were estimated from several simulations (trials) of spontaneous activity at a fixed level of arousal. First, we computed the ISIs of each cell in a given trial. We then collated the ISIs across all trials into a single distribution (separately for each cell). The spontaneous cvISI of a given cell ($\mathrm{cvISI_{spont}}$) was then computed from its trial-aggregated ISI distribution. At each arousal level, calculations were based on 30, 2.5 second-long trials of spontaneous activity. Results were summarized as the average $\mathrm{cvISI_{spont}}$ across all clustered cells that had an average firing rate of at least 1 spike/second at each arousal level. We refer to this population averaged quantity as $\langle \mathrm{cvISI_{spont}} \rangle$. Fig.~\ref{f:cvISI_spectra_supp}A shows $\langle \mathrm{cvISI_{spont}} \rangle$ as a function of arousal, where each data point indicates the mean across 2 network realizations.

\subsubsection*{Experimental data}

To compute cvISIs in the neural data, the spontaneous blocks of each session were split into 2.5 second-long windows, and the average pupil diameter was computed across each one. In order to obtain similar arousal-based partitions of the data across different analyses, the spontaneous windows were then grouped -- based on their pupil diameter -- into one of the ten decile partitions of the pre-stimulus pupil diameter distribution (same splits used for the neural discriminabilty analysis; see ``\textit{\nameref{s:dprime}}"). This discretization allowed us to evaluate changes in the spontaneous cvISI ($\mathrm{cvISI_{spont}}$) across a broad range of arousal states while maintaining several trials in each pupil decile split. Since different pupil diameter partitions contained different amounts of data, we subsampled the same number of windows from each partition. For a given pupil-based split of the data, we then computed the ISIs of each cell in every time window. The single-unit ISIs were then combined across all time windows in the given pupil partition, and the $\mathrm{cvISI_{spont}}$ of each cell was computed from its trial-aggregated ISI distribution. This was repeated for 100 different random subsamplings of the data, and a final estimate of $\mathrm{cvISI_{spont}}$ in each pupil partition was computed as the average across subsamples. Only cells that had an average spontaneous firing rate of at least 1 spike/second in all pupil partitions were included. 

To examine the average behavior of $\mathrm{cvISI_{spont}}$ as a function of arousal, we combined the data over cells and sessions in a pupil-dependent manner, using the method described in ``\textit{\nameref{s:fano_factor}}". The result is shown in Fig.~\ref{f:cvISI_spectra_supp}D, where horizontal error bars indicate the pupil diameter bins, and where red markers and vertical error bars indicate the average $\langle \mathrm{cvISI_{spont}} \rangle$ $+/-$ 1 SEM across all cells in each pupil bin. To quantitatively compare $\mathrm{cvISI_{spont}}$ between low and high arousal conditions, we used the procedure described previously for the FF quantities (see ``\textit{\nameref{s:fano_factor}}"). First, we found all sessions that expressed a broad range of pupil diameters. We then pooled the single-cell $\mathrm{cvISI_{spont}}$ values in the first and last pupil decile partition across all of those sessions (9 in total), yielding two groups of data: $\{\mathrm{cvISI_{spont}^{small \ pupil}}\}$ and $\{\mathrm{cvISI_{spont}^{large \ pupil}}\}$. Finally, the small pupil and large pupil groups were compared using the Wilcoxon signed-rank test, and results were visualized by plotting the distribution of the difference $\mathrm{cvISI_{spont}^{small \ pupil}} - \mathrm{cvISI_{spont}^{large \ pupil}}$ (Fig.~\ref{f:cvISI_spectra_supp}E). Fig.~\ref{f:cvISI_spectra_supp}F,G also show $\mathrm{cvISI_{spont}^{small \ pupil}}$ and $\mathrm{cvISI_{spont}^{large \ pupil}}$ separately (for both single cells and population averages).

\subsection*{Spectral analyses}
\label{s:power_spectra}

We utilized spectral analyses to characterize the temporal structure of spike trains during spontaneous periods in both the network model (Fig.~\ref{f:cvISI_spectra_supp}B,C) and the experimental data (Fig.~\ref{f:cvISI_spectra_supp}H-L). To compute the power spectrum of a neuronal spike train from a single trial (time window) of length $T$, we first binned the spike train at a fine temporal resolution of $\Delta t = 1$ ms. The power spectrum of the binned spike train was then estimated using the multitaper method applied to point processes, as described in \cite{mitra2007observed} and numerically-implemented in \cite{bokil2010chronux}. For the multitaper estimates, we used a time-bandwidth product of $TW = 2$ and averaged over $2TW-1 = 3$ tapers. The multitaper estimate of the spectrum from a given trial was then normalized by the average firing rate of the neuron across that trial; this rate-normalization is equivalent to normalizing the spectrum by that of a Poisson process with the same firing rate. Normalized spectra for a given neuron were then averaged across all trials of a particular condition to obtain a final, normalized power spectrum $S_{\mathrm{norm}}(f)$. The low-frequency power was computed as the average of $S_{\mathrm{norm}}(f)$ between 1-4 Hz.

\subsubsection*{Network model}

In the clustered network model, single-neuron spectra were estimated from several simulated trials of spontaneous activity conditioned on a particular value of arousal. Specifically, for a given network realization and arousal level, we used the method described above to compute the normalized spectrum $S_{\mathrm{norm},i}(f)$ and the low-frequency power $P_{\mathrm{spont},i}^{\mathrm{L}}$ of cell $i$; these calculations were based on 30, 2.5 second trials of spontaneous activity. To summarize the overall extent of low-frequency fluctuations, we computed the average low-frequency power across all clustered cells that had a firing rate of at least 1 spike/second for all arousal levels. We refer to this cell-averaged low-frequency power as $\langle P_{\mathrm{L,spont}} \rangle$. Fig.~\ref{f:cvISI_spectra_supp}B shows the population-averaged power spectrum (rate-normalized) and Fig.~\ref{f:cvISI_spectra_supp}C shows $\langle P_{\mathrm{L,spont}} \rangle$ as function of arousal (averages across 2 network realizations).

\subsubsection*{Experimental data}

To compute power spectra in the experimental data, the spontaneous blocks of each session were divided into 2.5 second-long windows. The windows were then grouped according to pupil diameter using the same scheme that was implemented for the cvISI analysis (see ``\textit{\nameref{s:cvISI}}"). Before computing the power spectra in a given session, we subsampled the same number of windows from each pupil decile partition (we used the maximum possible number of windows given the distribution across pupil partitions); results were then averaged across 50 random subsamplings. For these analyses, we only included cells that had an average spontaneous firing rate $\geq$ 1 spike/second in all pupil decile partitions, and that had a non-zero spike count in all sampled time windows. 

For each pupil decile partition in a session, we computed the normalized spectrum $S_{\mathrm{norm}}(f)$ and low-frequency power $P_{\mathrm{L,spont}}$ of each cell. To examine overall trends in $P_{\mathrm{L,spont}}$ as a function of arousal, we aggregated the data over cells and sessions in a pupil-dependent manner, using the method described in ``\textit{\nameref{s:fano_factor}}". The result is shown in Fig.~\ref{f:cvISI_spectra_supp}I, where horizontal error bars indicate the pupil diameter bins, and where red markers and vertical error bars indicate the cell-average low-frequency power $\langle P_{\mathrm{L,spont}} \rangle$ $+/-$ 1 SEM in each pupil bin. To compute the cell- and session-averaged power spectrum as a function of arousal (Fig.~\ref{f:cvISI_spectra_supp}H), we followed the same procedure, but used three large pupil bins for combining data across sessions ($[0-33)\%$, $[33-67)\%$, and $[67-100]\%$ of maximum dilation). 

To test for overall changes in low-frequency power between low and high arousal states, we used the method described previously for the FF quantities (see ``\textit{\nameref{s:fano_factor}}"). First, we found all sessions that expressed a broad range of pupil diameters. We then pooled the single-cell $P_{\mathrm{L,spont}}$ values in the first and last pupil decile partition across all of those sessions (9 in total), yielding two groups of data: $\{P^{\mathrm{small \ pupil}}_{\mathrm{L,spont}}\}$ and $\{P^{\mathrm{large \ pupil}}_{\mathrm{L,spont}}\}$. Finally, the small pupil and large pupil groups were compared using the Wilcoxon signed-rank test, and results were visualized by plotting the distribution of the difference $P^{\mathrm{small \ pupil}}_{\mathrm{L,spont}} - P^{\mathrm{large \ pupil}}_{\mathrm{L,spont}}$ (Fig.~\ref{f:cvISI_spectra_supp}J). Fig.~\ref{f:cvISI_spectra_supp}K,L also show individual values of $P^{\mathrm{small \ pupil}}_{\mathrm{L,spont}}$ and $P^{\mathrm{large \ pupil}}_{\mathrm{L,spont}}$ (for both single cells and population averages). 

\section*{Quantification and statistical analysis}

Data analysis and simulations were carried out in Python, and utilized the NumPy package \cite{harris2020array}, SciPy library \cite{virtanen2020scipy}, scikit-learn library \cite{scikit-learn}, and Nitime library \cite{rokem2009nitime}. Details of the statistical tests used in this study are provided in the figure legends and/or relevant sections of the \nameref{s:Methods}. For all statistical comparisons of two paired samples, we used the non-parametric Wilcoxon signed-rank test (implemented with the `scipy.stats' module from SciPy). For these analyses, ``n" refers to either the number of trials, number of cells/units, or the number of sessions (as indicated for each analysis), and significance was determined based on the two-sided p-value. For the clustering analysis, cluster significance was assessed by comparing the observed cluster quality against the distribution obtained under a trial-shuffled null model, and the significance of the cluster-based tuning similarity was assessed with a permutation test (see ``\textit{\nameref{s:clustering}}" for details). When relevant, significance levels were corrected for multiple comparisons using the Bonferroni correction. In cases where some data was excluded from a statistical test (e.g., based on firing rate thresholds), the criteria is specified in the relevant \nameref{s:Methods} sections. The meaning of error bars (either SEM or SD) is indicated in each figure caption. Boxplots display the median and the first and third quartiles of the data, with the whiskers extending from the quartiles to $\pm$ 1.5 of the interquartile range. Violin plots display probability densities, and tick marks indicate the minimum, maximum, and median of the data.

\clearpage
\newpage

\section*{Supplementary Information}

\newcommand{\beginsupplement}{%
	\setcounter{table}{0}
	\renewcommand{\thetable}{S\arabic{table}}
	\renewcommand{\theHtable}{S\arabic{table}}
	\setcounter{figure}{0}
	\renewcommand{\thefigure}{S\arabic{figure}}
	\renewcommand{\theHfigure}{S\arabic{figure}}
	
}
\beginsupplement

\clearpage

\begin{figure*}[b!]
	\centering
	\includegraphics[width=\textwidth]{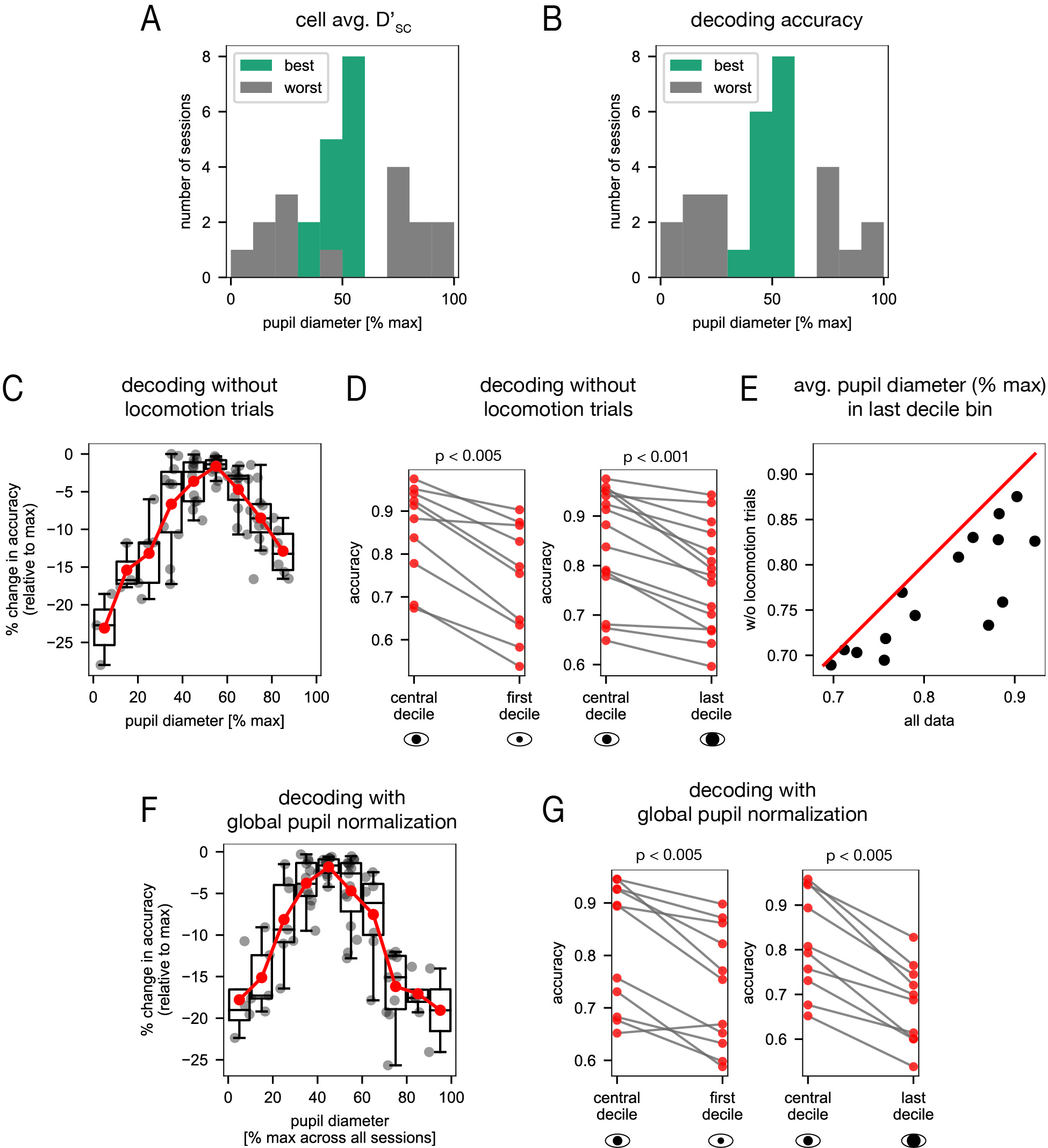}
	\caption[Supplementary analyses of stimulus discriminability in the neural data (related to Fig.~\ref{f:decoding_allTrials}).]{\textbf{Supplementary analyses of stimulus discriminability in the neural data (related to Fig.~\ref{f:decoding_allTrials})}. \textbf{(A,B)} Pupil diameter distributions corresponding to the best and worst cell-averaged $D'_{sc}$ or decoding accuracy. \textbf{(A)} In each session, we determined the pupil decile partition for which the cell-averaged $D'_{sc}$ was largest (best decile) or smallest (worst decile). The histogram shows the distribution of the average pupil diameter of the best decile (teal) and worst decile (gray) across all experimental sessions. \textbf{(B)} Same as \textbf{(A)} but for decoding accuracy.}
\end{figure*}
\addtocounter{figure}{-1} 

\begin{figure*}[t!]
	\centering
	\caption{(Continued from previous page.) \textbf{(C-E)} Session-averaged decoding results when excluding locomotion trials. \textbf{(C)} Percent change in cross-validated decoding accuracy (relative session-maximum) \textit{vs.} pupil diameter (group data from 15 sessions). In each pupil diameter bin, we show single-session data (gray) and the corresponding session-average (red) and boxplot. The session-averaged decoding performance still follows an inverted-U with pupil diameter when locomotion trials are discarded. However, without locomotion trials, large pupil diameters are not as robustly expressed (see panel \textbf{(E)}) and the right hand side of the inverted-U trend is less distinct compared to the case when all data is used (Fig.~\ref{f:decoding_allTrials}H). \textbf{(D)} \textit{Left}: There is a significant decrease in accuracy in the first pupil decile relative to the most central pupil decile of a session (data from n = 10 sessions with average pupil diameter of first decile $\leq 33\%$ max dilation; $p < 0.005$, Wilcoxon signed-rank test). \textit{Right}: There is a significant decrease in accuracy in the last pupil decile relative to the most central pupil decile of a session (data from n = 15 sessions with average pupil diameter of last decile $\geq 67\%$ max dilation; $p < 0.001$, Wilcoxon signed-rank test). \textbf{(E)} The average pupil diameter of trials in the last decile bin of a session without locomotion trials \textit{vs.} when all data is used. The average pupil diameter is noticeably smaller when locomotion trials are excluded. See \nameref{s:Methods} for methodological details. \textbf{(F,G)} Session-averaged decoding results when normalizing the pupil diameter in each session by the global maximum across all sessions, rather than by the maximum within each session separately. \textbf{(F)} Percent change in cross-validated decoding accuracy (relative session-maximum) \textit{vs.} (globally-normalized) pupil diameter (group data from 15 sessions). In each pupil diameter bin, we show single-session data (gray) and the corresponding session-average (red) and boxplot. Results are similar to the case of within-session pupil normalization (Fig.~\ref{f:decoding_allTrials}H). \textbf{(G)} \textit{Left}: There is a significant decrease in accuracy in the first pupil decile relative to the most central pupil decile of a session (data from n = 11 sessions with average pupil diameter of first decile $\leq 33\%$ max dilation; $p < 0.005$, Wilcoxon signed-rank test). \textit{Right}: There is a significant decrease in accuracy in the last pupil decile relative to the most central pupil decile of a session (data from n = 10 sessions with average pupil diameter of last decile $\geq 67\%$ max dilation; $p < 0.005$, Wilcoxon signed-rank test). Results are similar to the case of within-session pupil normalization (Fig.~\ref{f:decoding_allTrials}I). See \nameref{s:Methods} for methodological details.}
	\label{f:decoding_dprime_supp_data}
\end{figure*}

\clearpage
\newpage

\begin{figure*}[b!]
	\centering
	\includegraphics[width=0.95\textwidth]{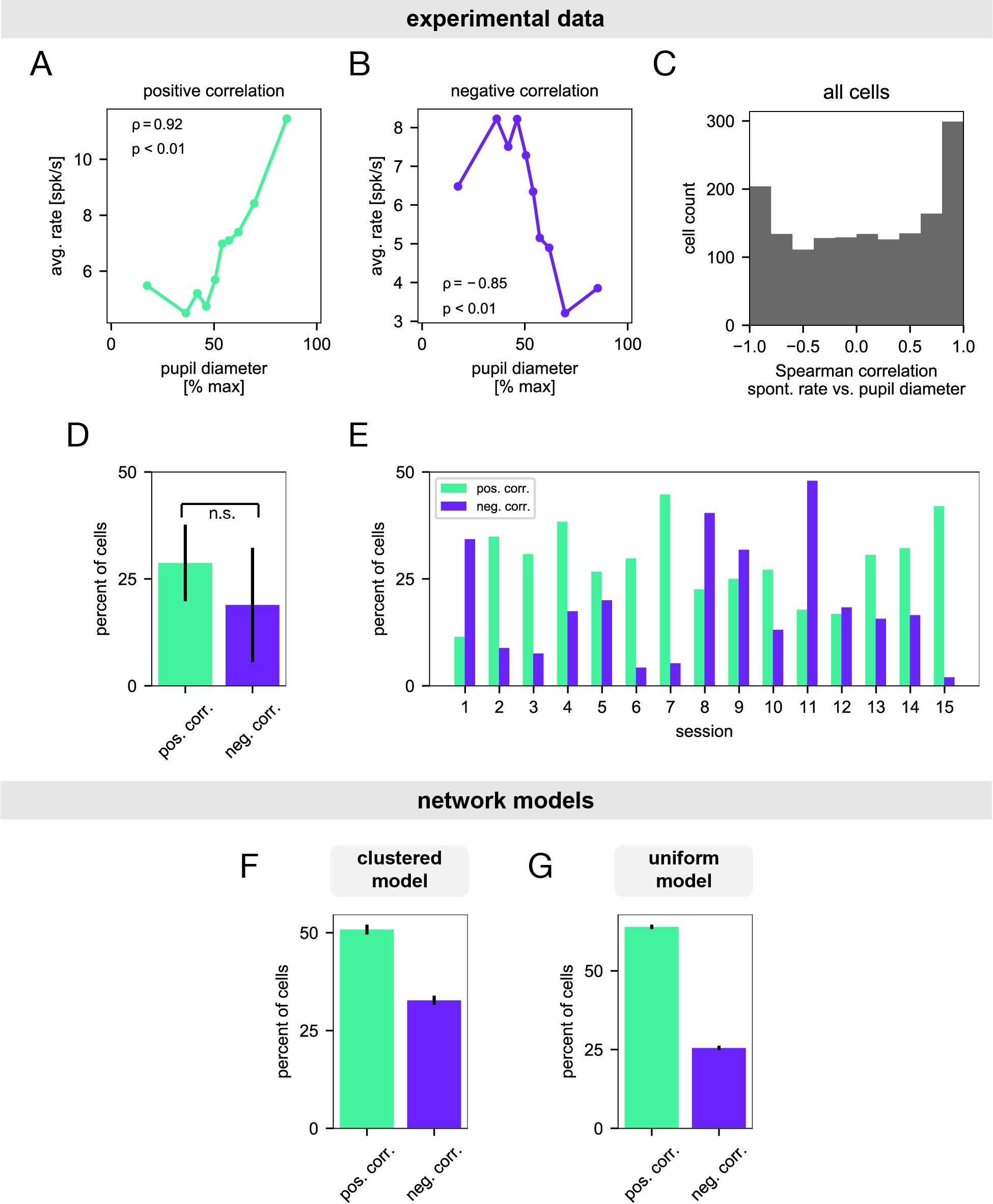}
	\caption[Relationships between spontaneous activity and arousal level (related to Fig.~\ref{f:model_overview}).]{\textbf{Relationships between spontaneous activity and arousal level (related to Fig.~\ref{f:model_overview}).}}
\end{figure*}
\addtocounter{figure}{-1} 

\begin{figure*}[t!]
	\centering
	\caption{(Continued from previous page.) \textbf{(A-E) Experimental data.} \textbf{(A)} A unit whose spontaneous firing rate increased with pupil diameter (Spearman correlation $\rho = 0.92$, $p<0.01$). \textbf{(B)} A unit whose spontaneous firing rate decreased with pupil diameter (Spearman correlation $\rho = -0.85$, $p<0.01$. \textbf{(C)} Histogram of Spearman correlation coefficients between single-cell spontaneous firing rates and pupil diameter. The histogram includes cells from all experimental sessions. \textbf{(D)} Percent of cells whose spontaneous firing rate was significantly positively or negatively correlated with pupil diameter. Bar heights and error bars indicate the mean $\pm$ 1 SD across sessions, and the correlation was considered significant if $p<0.05$. There was no significant difference between the fraction of positively and negatively modulated units ($p=0.135$, $n=15$ sessions, Wilcoxon signed-rank test). \textbf{(E)} Percent of cells in each experimental session whose spontaneous firing rate was significantly positively or negatively correlated with pupil diameter.  \textbf{(F,G) Model networks.} \textbf{(F)} Percent of all neurons whose spontaneous firing rate was positively or negatively correlated (Spearman correlation) with arousal level in the clustered network. Bar heights and error bars indicate the mean $\pm$ 1 SD across 10 network realizations, and a correlation was considered significant if $p<0.05$. \textbf{(G)} Same as \textbf{(F)} but for the uniform network. The data exhibits heterogeneous dependencies between arousal and ongoing activity, as evidenced by a broad distribution of correlation coefficients between spontaneous firing rate and pupil diameter [panels \textbf{(A,B)} for two example units; panel \textbf{(C)} for full distribution of correlation coefficients]. Of those units with a significant correlation, there were comparable fractions with positive and negative relationships between ongoing activity levels and pupil-indexed arousal [panel \textbf{(D)} for session-average results; panel \textbf{(E)} for individual sessions]. Increasing arousal in the network models was also associated with both increases and decreases in spontaneous firing rates [panels \textbf{(F)} and \textbf{(G)} for clustered model and uniform model, respectively]. The arousal implementation thus qualitatively captures the mixed rate modulations observed in the empirical data. In the clustered model, suppression of excitatory synapses onto pyramidal neurons tends to reduce firing rates, while heightened external drive tends to increase network activity. The diversity of rate modulations thus emerges due to the competition between those two effects, along with the cell-to-cell heterogeneity in the external drive (Fig.~\ref{f:model_overview}C). See \nameref{s:Methods} for methodological details.}
	\label{f:rate_corr_data_model}
\end{figure*}

\clearpage
\newpage

\begin{figure*}[b!]
	\centering
	\includegraphics[width=\textwidth]{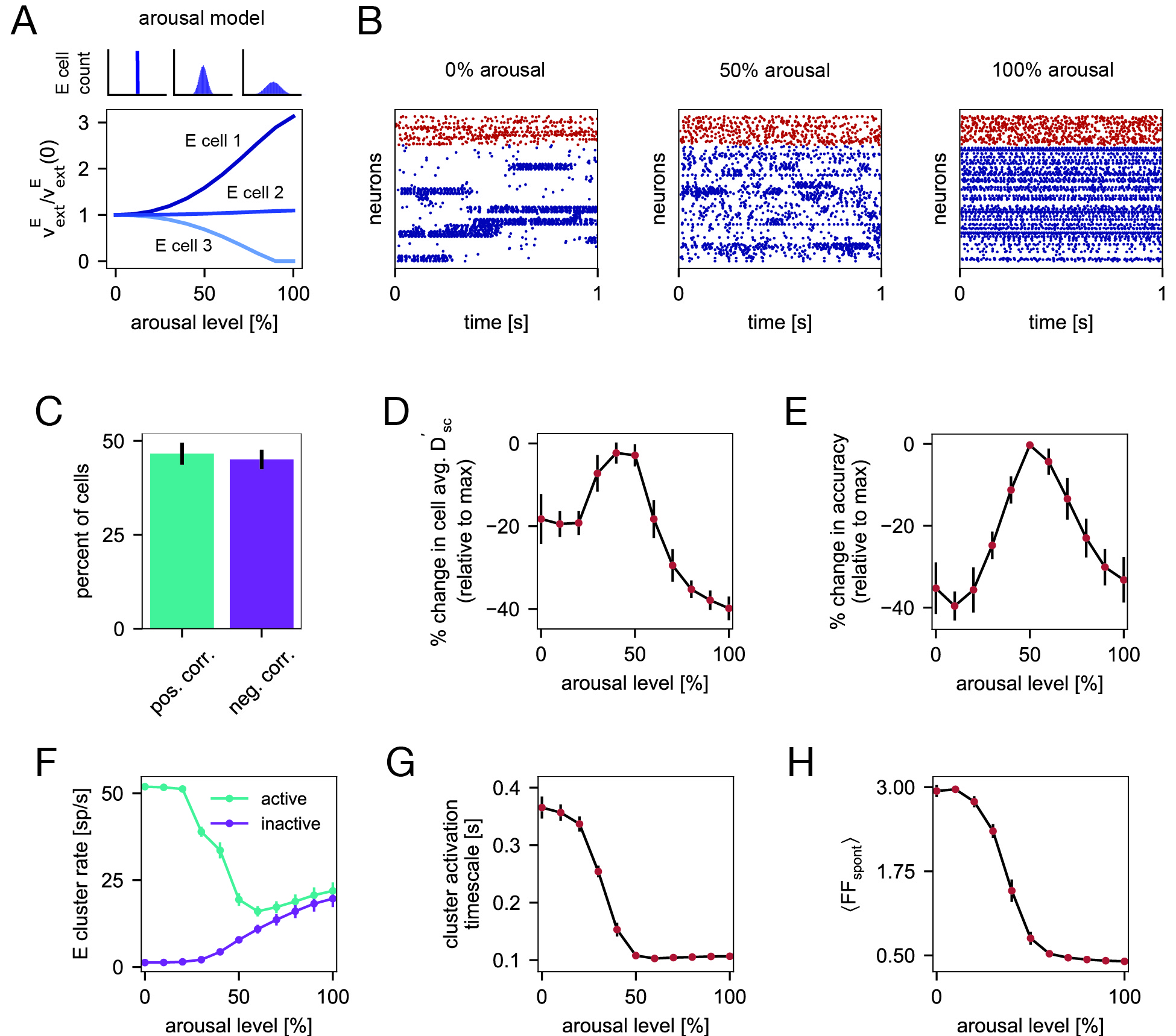}
	\caption[A different implementation of arousal induces similar modulations of network activity and stimulus discriminability (related to Figs.~\ref{f:model_overview}, \ref{f:decoding_model}, \ref{f:mean_field_vs_sims}, and Fig.~\ref{f:fano_factor_main}).]{\textbf{A different implementation of arousal induces similar modulations of network activity and stimulus discriminability (related to Figs.~\ref{f:model_overview}, \ref{f:decoding_model}, \ref{f:mean_field_vs_sims}, and \ref{f:fano_factor_main}).} \textbf{(A)} Schematic of an alternative arousal implementation, where arousal is modeled as a heterogeneous modulation of the background inputs to E cells. The lower plot shows the external input $\nu^{E}_{\mathrm{ext}}$ (relative to its initial value) \textit{vs.} arousal level, where the three different curves correspond to three different excitatory cells. Here, the background input rate to the $i^{th}$ E cell was given by
		$\nu_{\mathrm{ext},i}^{E} = \nu_{o}^{E} + \Delta_{\nu_i}^{E} \nu_{o}^{E}$, where $\nu_{o}^{E}$ is the baseline external input rate to E cells and where $\Delta_{\nu_i}^{E}$ is a cell-dependent parameter that sets the strength of the modulation to cell $i$. Specifically, $\Delta_{\nu_i}^{E}$ was given by Eq.~\ref{eq:nuext_arousal}, with $k=1.25$, $x_o = 0.275$, $M=0.9$, and $z^{E}_i \sim \mathcal{N}(0,1)$ (where $\mathcal{N}(0,1)$ is the standard normal distribution). In this way, increasing arousal increases the variance of the background input rates across cells in the excitatory population, while leaving the spatial average across cells approximately unchanged (inputs were not allowed to go negative). In the clustered model, each assembly was subject to the same realization of the background input distribution, such that all clusters received the same amount of (spatially-averaged) input. \textbf{(B)} Example raster plots from simulations of the clustered network at three increasing levels of arousal. \textbf{(C)} Percent of all neurons whose spontaneous firing rate was positively or negatively correlated with arousal level in the clustered network (\nameref{s:Methods}). A mix of positive and negative modulations are observed. \textbf{(D)} Percent change in cell-averaged $D'_{sc}$ \textit{vs.} arousal in the clustered model (percent change was computed relative to the maximum across all arousal levels; \nameref{s:Methods}). The average single-cell discriminability follows an inverted-U relationship with arousal. \textbf{(E)} Percent change in cross-validated decoding accuracy \textit{vs.} arousal in the clustered model (percent change was computed relative to the maximum across all arousal levels; \nameref{s:Methods}). For this analysis, linear classification was performed using population activity from a random sample of $10\%$ of excitatory cells/cluster. The decoding accuracy follows an inverted-U relationship with arousal.} 
\end{figure*}
\addtocounter{figure}{-1} 

\begin{figure*}[t!]
	\centering
	\caption{(Continued from previous page.) \textbf{(F)} Average firing rate of active and inactive excitatory clusters as a function of arousal (computed during spontaneous activity). At each arousal level, firing rates correspond to the cluster state with $n_{A}^{*}$ active clusters, where $n_{A}^{*}$ is the value that occurred most frequently (\nameref{s:Methods}). The active and inactive cluster rates converge with increasing arousal. \textbf{(G)} The average cluster activation timescale \textit{vs.} arousal in simulations of the clustered network (\nameref{s:Methods}). Cluster activation periods decrease with arousal. \textbf{(H)} Population-averaged spontaneous FF ($\mathrm{FF_{spont}}$) \textit{vs.} arousal (100 ms spike count window; \nameref{s:Methods}). Neural variability decreases with arousal. Results in this figure are based off of simulations from 5 different network realizations (30 simulated trials/stimulus/network, 5 stimuli; see \nameref{s:Methods} for simulation details). Data points (circles) and error bars (vertical bars) indicate the mean $\pm$ 1 SD across networks.}
	\label{f:sd_nu_ext_e_arousalModel}
\end{figure*}

\clearpage
\newpage

\begin{figure*}[h!]
	\centering
	\includegraphics[width=\textwidth]{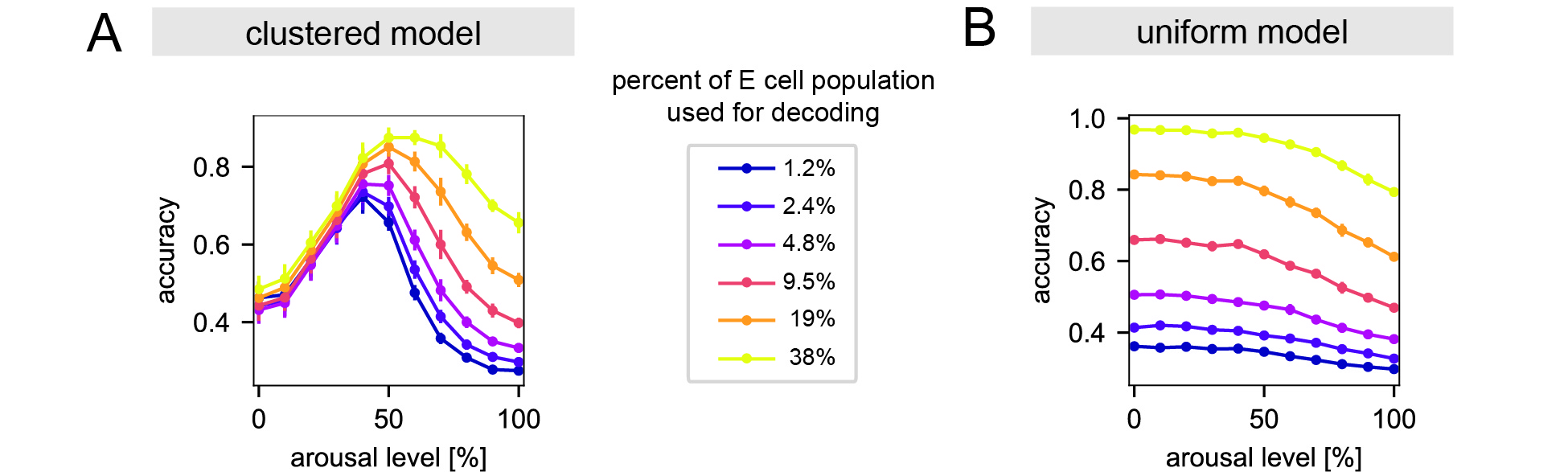}
	\caption[Impact of population size on decoding accuracy in the network models (related to Fig.~\ref{f:decoding_model}).]{\textbf{Impact of population size on decoding accuracy in the network models (related to Fig.~\ref{f:decoding_model}).} \textbf{(A)} Cross-validated decoding accuracy \textit{vs.} arousal in the clustered model. Different curves show results for different population sizes (displayed as a percent of the total excitatory (E) cell population); data points and error bars indicate the mean $\pm$ 1 SD across network realizations. Note that while decoding accuracy increases with sample size for high arousal, the peak always occurs in an intermediate arousal range for the population sizes considered. \textbf{(B)} Same as \textbf{(A)} but for the uniform model. For all sample sizes considered, the decoding accuracy decreases with arousal. See \nameref{s:Methods} for analysis details. In the clustered model, the computational mechanism underlying the inverted-U relationship is a shift in dynamical regime from a metastable attractor phase to a single-attractor uniform phase (Fig.~\ref{f:mean_field_vs_sims}). The transition to the uniform phase explains the increase in decoding accuracy with population size in the high arousal regime of the clustered model. Because neurons become independent in the uniform state, adding more neurons averages out variability and improves performance, even though stimulus responses are weak \cite{moreno2014information}. In terms of population decoding, the inverted-U thus emerges due to competition between response magnitude and response variability. At low arousal, performance is low because variability is high (and pooling has little impact since cells in the same cluster are correlated). At high arousal, a response magnitude-variability tradeoff is present so long as the decoder only samples a subset of neurons from each cluster; then, the decrease in variability obtained by pooling cannot fully compensate for the weak signal, and performance remains low.}
	\label{f:decoding_model_supp}
\end{figure*}

\clearpage
\newpage

\begin{figure*}[b!]
	\centering
	\includegraphics[width=\textwidth]{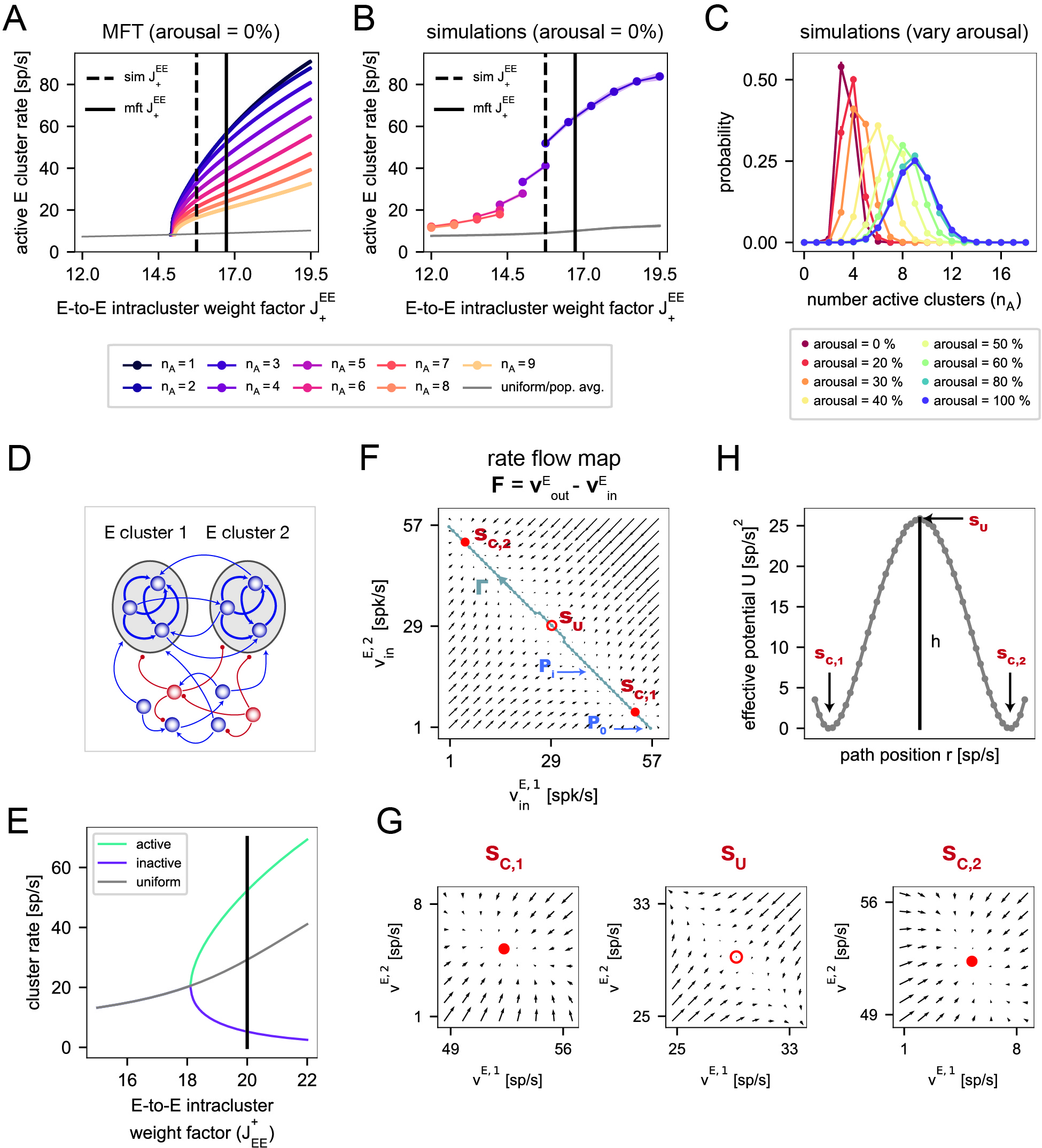}
	\caption[Additional details on the mean-field analysis of the clustered model (related to Fig.~\ref{f:mean_field_vs_sims}).]{\textbf{Additional details on the mean-field analysis of the clustered model (related to Fig.~\ref{f:mean_field_vs_sims}).} \textbf{(A,B)} E-to-E intracluster weight factor controls the onset of cluster states. \textbf{(A)} Effect of the E-to-E intracluster weight factor $J^{EE}_{+}$ on the mean-field solutions of the clustered networks in the absence of arousal. The gray curve shows the rate of the excitatory populations for the solution in which no clusters are active (``uniform" state), and the colored curves show the firing rates of active excitatory clusters for solutions in which $n_A \in \{1,...,9\}$ clusters are active (``cluster" states). When $J^{EE}_{+}$ is below a critical value, the mean-field theory has a single, uniform solution (gray), in which all clusters have the same moderate firing rate. As $J^{EE}_{+}$ is increased above a critical value, additional solutions emerge. These cluster states are characterized by $n_A \geq 1$ active clusters with a rate $\nu_{n_A,\uparrow}$. Note that the stability of the solutions is not indicated.} 
\end{figure*}
\addtocounter{figure}{-1} 

\begin{figure*}[t!]
	\centering
	\caption{(Continued from previous page.) \textbf{(B)} Effect of the E-to-E intracluster weight factor $J^{EE}_{+}$ in the simulations. The gray curve shows the average firing rate of all excitatory neurons and the colored curves show the firing rates of active excitatory clusters conditioned on a particular number $n_A$ of active clusters. At a given $J^{EE}_{+}$, rates are only plotted for values of $n_A$ that occurred with probability $P(n_A) \geq 0.2$. Though there are differences between the theory and simulations (specifically, cluster states emerge at lower $J^{EE}_{+}$ in the simulations), the same qualitative behavior is observed in both cases (the active cluster rate increase with $J^{EE}_{+}$). In both panels, the black dashed line corresponds to the value of the E-to-E weight factor $J_{EE,\mathrm{sim}}^{+}$ that is used in the simulations when studying the impact of arousal, and the black solid line corresponds to the value $J_{EE,\mathrm{mft}}^{+}$ at which the mean-field theory is performed. Note that the arousal-dependent mean-field calculations use a larger $J^{EE}_{+}$ than the simulations ($J_{EE,\mathrm{mft}}^{+} > J_{EE,\mathrm{sim}}^{+}$); the value of $J_{EE,\mathrm{mft}}^{+}$ was chosen to achieve the best match with the simulated firing rates (at $J_{EE,\mathrm{sim}}^{+}$) in the absence of arousal. See \nameref{s:Methods} for details. \textbf{(C)} Probability of observing a certain number of active clusters $n_{A}$ for different arousal levels in the simulations (\nameref{s:Methods}). Circular markers and error bars indicate the mean $\pm$ 1 SD across network realizations. \textbf{(D-H)} Details on the mean-field analysis of the 2-cluster circuit. \textbf{(D)} Schematic of the 2-cluster network, which contains two excitatory clusters and one background excitatory and inhibitory population. \textbf{(E)} Effect of the E-to-E intracluster weight factor $J_{EE}^{+}$ on the mean-field solutions of the reduced 2-cluster network (when the arousal level is zero; \nameref{s:Methods}). When $J_{EE}^{+}$ is below a critical value, the only solution is one in which the two clusters have the same moderate firing rate (``uniform state"). As $J_{EE}^{+}$ is increased above a critical value, an additional solution emerges in which one cluster is active and the other is inactive (``cluster states"), with rates given by the green and purple curves. Note that the stability of the solutions is not indicated. All analyses of the 2-cluster networks in the main text (Fig.~\ref{f:mean_field_vs_sims}D,E) were performed at a fixed E-to-E intracluster weight factor of $J_{EE}^{+} = 20$ (black vertical line). \textbf{(F)} We studied the dynamics of the 2-cluster network using the effective mean-field theory developed in \cite{mascaro1999effective}. To begin, we numerically constructed the rate flow map of the two excitatory clusters, which indicates how the two cluster firing rates will evolve from some initial configuration $\bm{\nu}_{\mathrm{in}}^{E}$. To accomplish this, we tiled the $\nu_{\mathrm{in}}^{E,1}$-$\nu_{\mathrm{in}}^{E,2}$ plane with a grid, and at each grid location, we computed the induced output rates  $\nu_{\mathrm{out}}^{E,1}$ and $\nu_{\mathrm{in}}^{E,2}$ using the effective theory (\nameref{s:Methods}). Here, the rate flow map is visualized by plotting the vector $\mathbf{F} = \bm{\nu}_{\mathrm{out}}^{E} - \bm{\nu}_{\mathrm{in}}^{E}$ at each grid point. From the rate flow diagram, one can identify the three fixed points from the full mean-field theory in \textbf{(E)}, corresponding to the uniform solution ($S_{U}$) and the cluster states in which either the first ($S_{C,1}$) or second ($S_{C,2}$) cluster is active. Moreover, the flow map indicates that the uniform solution is unstable, while the two cluster states are attractors. \textbf{(G)} Close-ups of the three fixed points in \textbf{(F)}. \textbf{(H)} To obtain intuition about transitions between the two attractors, we considered a path $\Gamma$ (gray dotted line in \textbf{(F}) connecting the two cluster states $S_{C,1}$ and $S_{C,2}$ through the unstable fixed point $S_{U}$. For each point $\mathrm{P}_i$ on the path, we computed the line integral $- \int_{\Gamma_{\mathrm{P}_i}} \mathbf{F} \cdot d\bm{\nu}^{E}_{\mathrm{in}}$, where $\Gamma_{\mathrm{P}_i}$ denotes the segment of the path from $\mathrm{P}_{0}$ to $\mathrm{P}_{i}$. This procedure yields a 1-dimensional effective potential $U$, which summarizes the cluster dynamics. Specifically, the potential wells correspond to the two attractors $S_{C,1}$ and $S_{C,2}$, and these configurations are separated by a barrier at the unstable fixed point $S_{U}$ whose height controls the rate of switching between the two cluster states.}
	\label{f:mft_supp}
\end{figure*}

\clearpage
\newpage

\begin{figure*}
	\centering
	\includegraphics[width=\textwidth]{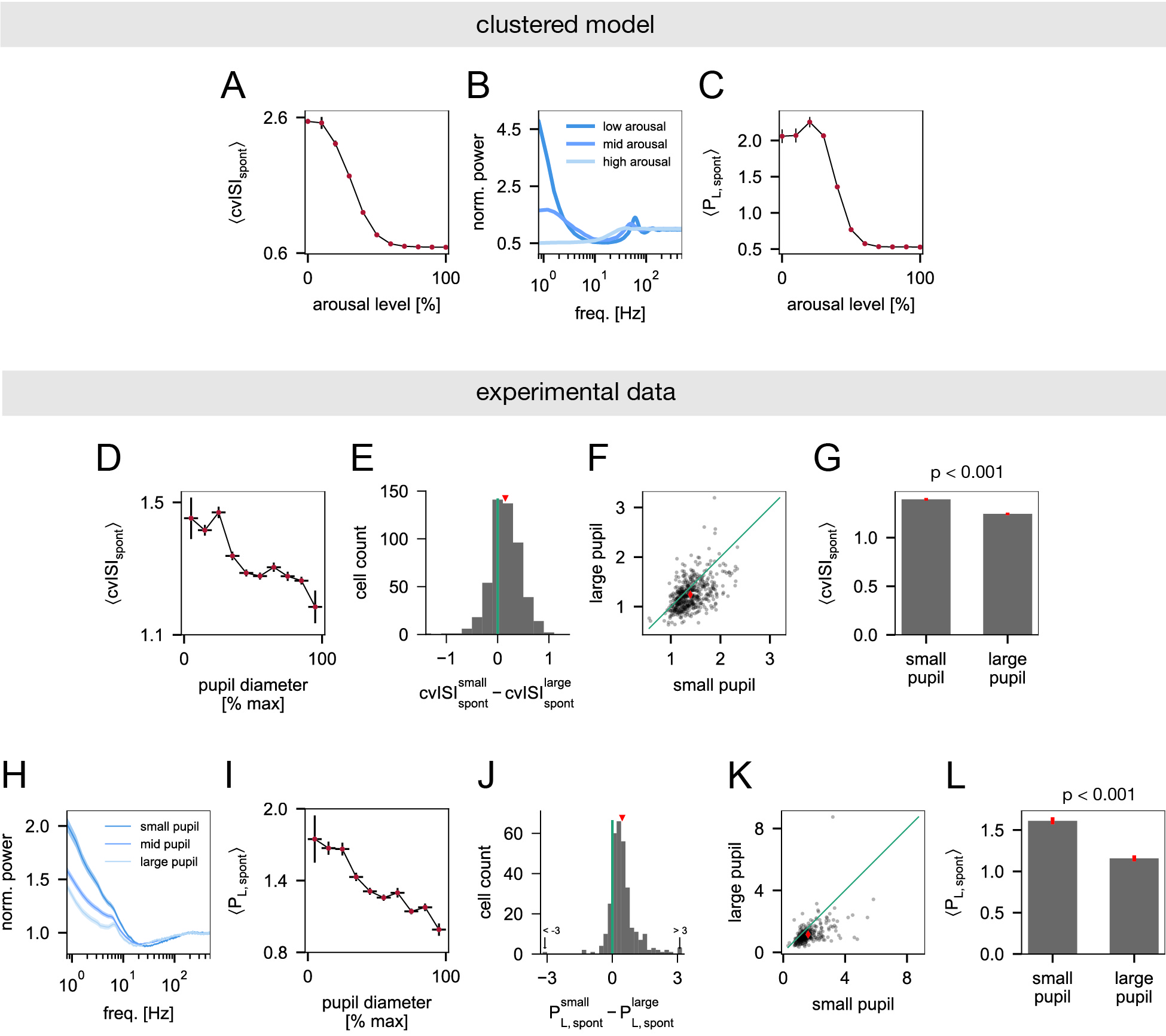}
	\caption[Additional measures of spontaneous neural variability in the clustered model and experimental data (related to Fig.~\ref{f:fano_factor_main}).]{\textbf{Additional measures of spontaneous neural variability in the clustered model and experimental data (related to Fig.~\ref{f:fano_factor_main}).} \textbf{(A-C)} Results from the clustered model. \textbf{(A)} Population-averaged coefficient of variation of interspike intervals during spontaneous activity ($\mathrm{cvISI_{spont}}$) \textit{vs.} arousal. Data points and error bars indicate the mean $\pm$ 1 SD across network realizations. \textbf{(B)} Population-averaged spike-train power spectrum (rate-normalized) of spontaneous activity at low, moderate, and high arousal. \textbf{(C)} Same as \textbf{(A)}, but for spontaneous low-frequency power (average over 1-4 Hz; $\mathrm{P_{L,spont}}$). \textbf{(D-L)} Results from the experimental data. \textbf{(D)} Population-averaged $\mathrm{cvISI_{spont}}$ \textit{vs.} pupil diameter (units pooled over sessions). Horizontal error bars indicate pupil diameter bins; data points and vertical error bars indicate mean $\pm$ SEM across cells from all sessions that contribute to the corresponding pupil bin. \textbf{(E)} Distribution of the difference in $\mathrm{cvISI_{spont}}$ between small and large pupil diameters (red triangle indicates mean difference). \textbf{(F)} $\mathrm{cvISI_{spont}}$ of individual cells in large pupil \textit{vs}. small pupil conditions (red diamond indicates mean values). \textbf{(G)} The mean $\pm$ SEM of $\mathrm{cvISI_{spont}}$ in small pupil and large pupil conditions (small pupil: 1.39 $\pm$ 0.01; large pupil: 1.24 $\pm$ 0.01) $\mathrm{cvISI_{spont}}$ is significantly smaller in states of high pupil-indexed arousal compared to low pupil-indexed arousal [$n = 510$ units pooled over 9 sessions with average pupil diameter of smallest (largest) decile bin $\leq 33\%$ ($\geq 67\%$) max dilation; $p<0.001$, Wilcoxon signed-rank test]. \textbf{(H)} Population-averaged spike-train power spectrum (rate-normalized) of spontaneous activity at small ($0-33\%$ max dilation), moderate ($33-67\%$ max dilation), and large ($67-100\%$ max dilation) pupil diameters. \textbf{(I)} Same as \textbf{(D)} but for population-averaged $\mathrm{P^{L}_{spont}}$. \textbf{(J-L)} Same as \textbf{(E-G)} but for $\mathrm{P_{L,spont}}$; $\mathrm{P^{L}_{spont}}$ is significantly smaller in states of high pupil-indexed arousal compared to low pupil-indexed arousal [small pupil mean $\pm$ SEM: 1.61 $\pm$ 0.04; large pupil mean $\pm$ SEM: 1.16 $\pm$ 0.04; $n = 313$ units pooled over 9 sessions with average pupil diameter of smallest (largest) decile bin $\leq 33\%$ ($\geq 67\%$) max dilation; $p< 0.001$, Wilcoxon signed-rank test]. See \nameref{s:Methods} for methodological details on the cvISI and power spectra analyses.}
	\label{f:cvISI_spectra_supp}
\end{figure*}

\clearpage
\newpage

\begin{figure*}[b!]
	\centering
	\includegraphics[width=\textwidth]{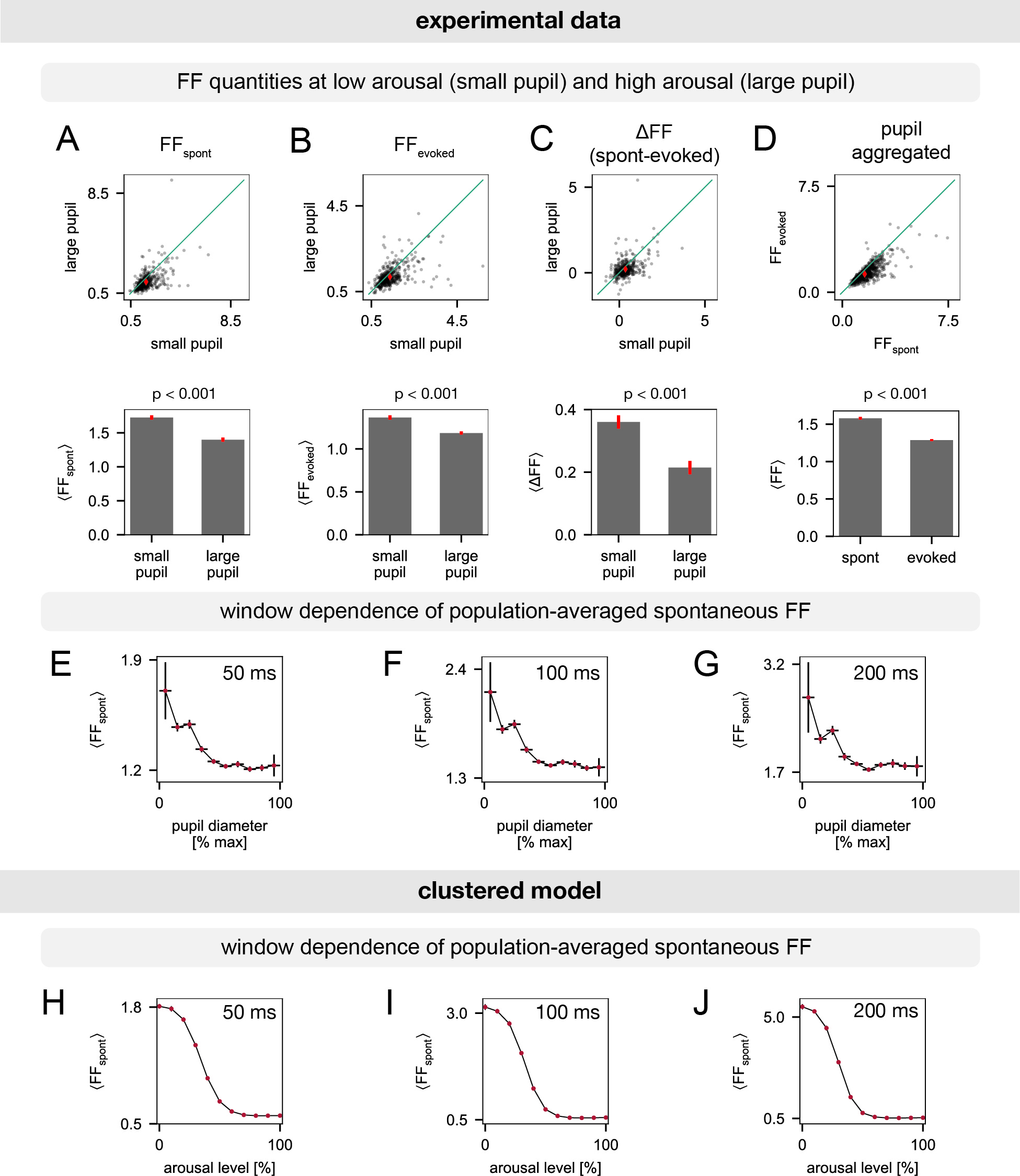}
	\caption[Supplementary analyses of the Fano factor (related to Fig.~\ref{f:fano_factor_main}).]{\textbf{Supplementary analyses of the Fano factor (related to Fig.~\ref{f:fano_factor_main}).} \textbf{(A-D)} FF quantities at low arousal (small pupil) and high arousal (large) pupil in the experimental data. \textbf{(A)} \textit{Top}: Spontaneous Fano factor ($\mathrm{FF_{spont}}$) of individual cells in large pupil \textit{vs}. small pupil conditions. The red diamond indicates the mean values. The distribution of single-cell differences is shown in Fig. 8E. \textit{Bottom:} The mean $\pm$ SEM of $\mathrm{FF_{spont}}$ in small pupil and large pupil conditions (small pupil: 1.72 $\pm$ 0.03; large pupil: 1.40 $\pm$ 0.03). $\mathrm{FF_{spont}}$ is significantly smaller in the large pupil condition [$n=487$ units pooled over 9 sessions with average pupil diameter of smallest (largest) decile bin $\leq 33\%$ ($\geq 67\%$) max dilation; $p< 0.001$, Wilcoxon signed-rank test].}
\end{figure*}
\addtocounter{figure}{-1} 

\begin{figure*}[t!]
	\centering
	\caption{(Continued from previous page.) \textbf{(B)} \textit{Top}: Same as \textbf{(A)}, but for $\mathrm{FF_{evoked}}$. The distribution of single-cell differences is shown in Fig. 8G. \textit{Bottom:} The mean $\pm$ SEM of $\mathrm{FF_{evoked}}$ in small pupil and large pupil conditions (small pupil: 1.36 $\pm$ 0.02; large pupil: 1.18 $\pm$ 0.02). $\mathrm{FF_{evoked}}$ is significantly smaller in the large pupil condition ($n=487$ units, $p< 0.001$, Wilcoxon signed-rank test). \textbf{(C)} \textit{Top}: Same as \textbf{(A)}, but for $\Delta \mathrm{FF}$ (spontaneous - evoked). The distribution of single-cell differences is shown in Fig. 8I. \textit{Bottom:} The mean $\pm$ SEM of $\Delta \mathrm{FF}$ in small pupil and large pupil conditions (small pupil: 0.36 $\pm$ 0.02; large pupil: 0.21 $\pm$ 0.02). $\Delta \mathrm{FF}$ is significantly smaller in the large pupil condition ($n=487$ units, $p< 0.001$, Wilcoxon signed-rank test). \textbf{(D)} \textit{Top:} Pupil-aggregated $\mathrm{FF_{evoked}}$ \textit{vs}. $\mathrm{FF_{spont}}$ of individual cells. The red diamond indicates the mean values. \textit{Bottom:} The mean $\pm$ SEM of the pupil-aggregated FF in spontaneous and evoked conditions (spontaneous: 1.58 $\pm$ 0.02; evoked: 1.29 $\pm$ 0.01). FF is significantly smaller in evoked conditions ($n=1114$ units pooled over 15 sessions, $p< 0.001$, Wilcoxon signed-rank test). \textbf{(E-G)} Population-averaged $\mathrm{FF_{spont}}$ \textit{vs}. pupil diameter for different window sizes in the experimental data. In each panel, horizontal error bars indicate pupil diameter bins, and data points and vertical error bars indicate the mean $\pm$ SEM across cells from all sessions that have sufficient data in the corresponding pupil bin. Spike-count window lengths are indicated in the upper right corner of each plot [panel \textbf{(E)} 50 ms, panel \textbf{(F)} 100 ms, panel \textbf{(G)} 200 ms]. \textbf{(H-J)} Population-averaged $\mathrm{FF_{spont}}$ \textit{vs.} arousal level for different window sizes in the clustered model. In each panel, data points and error bars indicate the mean $\pm$ SD of the population-averaged $\mathrm{FF_{spont}}$ across network realizations. Spike-count window lengths are indicated in the upper right corner of each plot [panel \textbf{(E)} 50 ms, panel \textbf{(F)} 100 ms, panel \textbf{(G)} 200 ms]. In the model, $\mathrm{FF_{spont}}$ increases with the length of the spike-count window in the low arousal regime, where cluster activity is strong. This is consistent with prior work \cite{litwin2012slow}, and indicates the presence of slow rate fluctuations that cause variability to increase over longer integration times. In the data, $\mathrm{FF_{spont}}$ also increases with window length at low arousal. However, we additionally observe increases in $\mathrm{FF_{spont}}$ at moderate and high arousal. These effects suggest that the data may have additional sources of variability that are not present in the model and that impact all arousal levels. The low-arousal variation in $\mathrm{FF_{spont}}$ across window sizes is also less drastic in the data compared to the model, indicating that activity fluctuations are less extreme in the former. Nonetheless, the data exhibits a suppression of neural variability with increasing pupil diameter for all time windows considered. See \nameref{s:Methods} for methodological details on the FF analyses.}
	\label{f:spectra_fano_supp}
\end{figure*}

\clearpage
\newpage

\begin{figure*}[b!]
	\centering
	\includegraphics[width=0.95\textwidth]{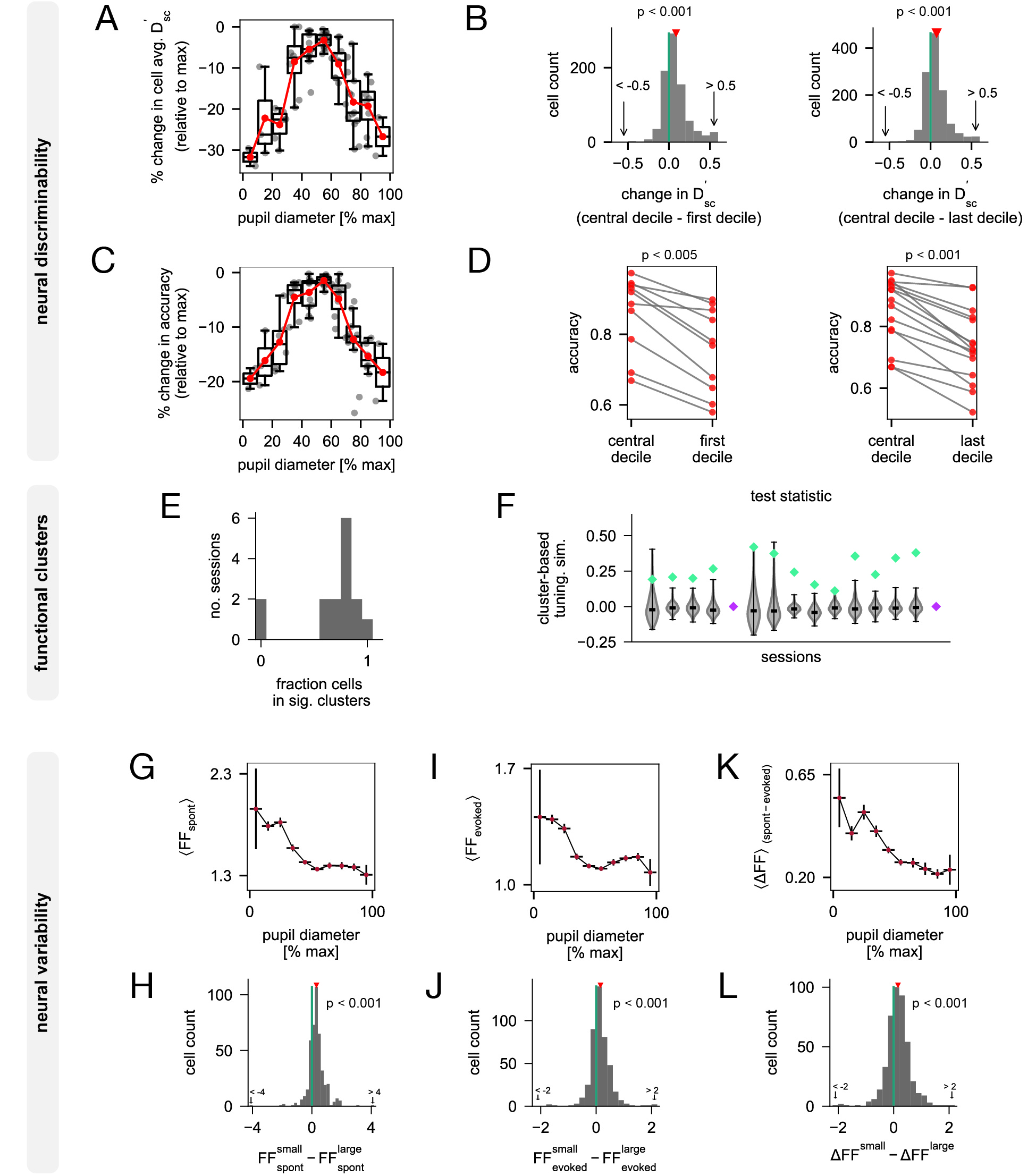}
	\caption[Robustness of main results with stricter cell selection criteria in the experimental data (related to Figs.~\ref{f:decoding_allTrials}, \ref{f:cluster_analysi_main}, and \ref{f:fano_factor_main}).]{\textbf{Robustness of main results with stricter cell section criteria in the experimental data (related to Figs.~\ref{f:decoding_allTrials}, \ref{f:cluster_analysis_main}, and \ref{f:fano_factor_main}).} We examined the robustness of the main results under a more conservative cell selection method (i.e., when implementing strict criteria for sound-responsiveness; see \nameref{s:Methods}). \textbf{(A-D) Neural discriminability analyses} (see \nameref{s:Methods} for methodological details). \textbf{(A)} Percent change in cell-averaged $D'_{\mathrm{sc}}$ (relative to session-maximum) \textit{vs.} pupil diameter (group data from 15 sessions). In each pupil diameter bin, we show single-session data (gray) and the corresponding session-average (red) and boxplot. The population-averaged $D'_{\mathrm{sc}}$ exhibits an inverted-U relationship with pupil diameter.}
\end{figure*}
\addtocounter{figure}{-1} 

\begin{figure*}[t!]
	\centering
	\caption{(Continued from previous page.) \textbf{(B)} \textit{Left:} Difference in $D'_{\mathrm{sc}}$ between the most central and first pupil decile of a session (red triangle indicates mean difference). $D'_{\mathrm{sc}}$ is significantly smaller in the first pupil decile ($n=817$ units pooled over 10 sessions with average pupil diameter of first decile $\leq 33 \%$ max dilation; $p < 0.001$, Wilcoxon signed-rank test). \textit{Right:} Distribution of the difference in $D'_{\mathrm{sc}}$ between the most central and last pupil decile of a session (red triangle indicates mean difference). $D'_{\mathrm{sc}}$ is significantly smaller in the last pupil decile ($n=1211$ units pooled over 15 sessions with average pupil diameter of last decile $\geq 67 \%$ max dilation; $p < 0.001$, Wilcoxon signed-rank test). \textbf{(C)} Same as \textbf{(A)} but for cross-validated decoding accuracy. The average decoding accuracy exhibits an inverted-U relationship with pupil diameter. \textbf{(D)} \textit{Left:} Accuracy in the most central pupil decile and the first pupil decile of a session. The accuracy is significantly smaller in the first pupil decile (data from $n=10$ sessions with average pupil diameter of first decile $\leq 33 \%$ max dilation; $p<0.005$, Wilcoxon signed-rank test). \textit{Right:} Accuracy in the most central pupil decile and the last pupil decile of a session. The accuracy is significantly smaller in the last pupil decile (data from $n=15$ sessions with average pupil diameter of last decile $\geq 67 \%$ max dilation; $p<0.001$, Wilcoxon signed-rank test). \textbf{(E,F) Functional clustering analysis} (see \nameref{s:Methods} for methodological details). \textbf{(E)} Distribution of the fraction of cells in a session that belong to significant correlation-based clusters. Though two sessions do not exhibit significant clusters under the stricter cell selection criteria, the majority of sessions still do. \textbf{(F)} Test statistic for cluster-based tuning similarity in each session. Diamonds indicate observed values and violin plots show distributions obtained by permuting cluster labels across cells. Green diamonds indicate a significant result relative to permuted data ($p < 0.05$) and magenta diamonds otherwise. In most sessions, the observed cluster-based tuning similarity significantly exceeds the distribution obtained under the permutation-based null model, suggesting the presence of neural clusters with some functional organization. \textbf{(G-L) Neural variability analyses} (see \nameref{s:Methods} for methodological details). \textbf{(G)} Population-averaged $\mathrm{FF_{spont}}$ \textit{vs.} pupil diameter (units pooled over sessions). Horizontal error bars indicate pupil diameter bins; data points and vertical error bars indicate mean $\pm$ SEM across cells from all sessions that contribute to the corresponding pupil bin (note that only one session with relatively few cells contributes to the first pupil bin). The spontaneous FF decreases with pupil diameter. \textbf{(H)} Distribution of the difference in $\mathrm{FF_{spont}}$ between small and large pupil diameters (red triangle indicates mean difference). $FF_{\mathrm{spont}}$ is significantly smaller in the large pupil condition [$n=433$ units pooled over 9 sessions with average pupil diameter of smallest (largest) decile bin $\leq 33\%$ ($\geq 67\%$) max dilation; $p < 0.001$, Wilcoxon signed-rank test]. \textbf{(I, J)} Same as \textbf{(G,H)}, but for $\mathrm{FF_{evoked}}$. Although the decreasing trend in the evoked FF is less drastic with the stricter cell selection, $\mathrm{FF_{evoked}}$ is still significantly smaller for large ($\geq 67\%$ max dilation) pupil diameters relative to small ($\leq 33\%$ max dilation) pupil diameters ($n=433$ units; $p<0.001$, Wilcoxon signed-rank test).  \textbf{(K,L)} Same as \textbf{(G,H)}, but for $\Delta \mathrm{FF}$ (difference between $\mathrm{FF_{spont}}$ and $\mathrm{FF_{evoked}}$). $\Delta \mathrm{FF}$ generally declines with pupil diameter, and $\Delta \mathrm{FF}$ is significantly smaller for large ($\geq 67\%$ max dilation) pupil diameters relative to small ($\leq 33\%$ max dilation) pupil diameters ($n=433$ units; $p<0.001$, Wilcoxon signed-rank test).}
	\label{f:cellSelection_robustness}
\end{figure*}

\clearpage
\newpage

\begin{table*}[h!]
	\begin{center}
		\begin{tabular}{m{0.15\linewidth}m{0.4\linewidth}m{0.45\linewidth}}
			\hline \hline
			\\
			Parameter & Description & Value \\ \\
			\hline \hline
			\\
			$N^{E}$ & number of E cells & 1600 \\
			$N^{I}$ & number of I cells & 400 \\
			
			$\tau_{\mathrm{m}}^{E}$ & membrane time constant of E cells & 20 ms \\
			$\tau_{\mathrm{m}}^{I}$ & membrane time constant of I cells & 20 ms \\
			$\tau_{\mathrm{syn}}^{E}$ & E synaptic time constant & 5 ms \\
			$\tau_{\mathrm{syn}}^{I}$ & I synaptic time constant & 5 ms \\
			$\tau_{\mathrm{ref}}^{E}$ & refractory period of E cells & 5 ms \\
			$\tau_{\mathrm{ref}}^{I}$ & refractory period of I cells & 5 ms \\
			
			$V_{\mathrm{thresh}}^{E}$ & threshold potential of E cells & 1.5 mV \\
			$V_{\mathrm{thresh}}^{I}$ & threshold potential of I cells & 0.75 mV \\
			$V_{\mathrm{r}}^{I}$ & reset potential of I cells & 0 mV \\
			$V_{\mathrm{r}}^{I}$ & reset potential of I cells & 0 mV \\
			
			$p^{EE}$ & E-to-E recurrent connectivity fraction & 0.2 \\
			$p^{IE}$ & E-to-I recurrent connectivity fraction & 0.5 \\
			$p^{EI}$ & I-to-E recurrent connectivity fraction & 0.5 \\
			$p^{II}$ & I-to-I recurrent connectivity fraction & 0.5 \\
			
			$J^{EE}_U$ & uniform E-to-E synaptic strength & $0.63/\sqrt{N}$ mV \\
			$J^{IE}_U$ & uniform E-to-I synaptic strength & $0.63/\sqrt{N}$ mV \\
			$J^{EI}_U$ & uniform I-to-E synaptic strength & $-1.9/\sqrt{N}$ mV \\
			$J^{II}_U$ & uniform I-to-I synaptic strength & $-3.8/\sqrt{N}$ mV \\
			
			$p$ & number of E and I clusters & 18 \\
			$f^{E}$ & fraction of E cells/cluster & 0.05 \\
			$f^{I}$ & fraction of I cells/cluster & 0.05 \\
			
			$J^{EE}_{+}$ & E-to-E intracluster weight factor & 15.75 \\
			$J^{IE}_{+}$ & E-to-I intracluster weight factor & 5.45 \\
			$J^{EI}_{+}$ & I-to-E intracluster weight factor & 6.25 \\
			$J^{II}_{+}$ & I-to-I intracluster weight factor & 5.0 \\
			
			$C_{\mathrm{ext}}^{EE}$ & number of external synapses to E cells & 320 \\
			$C_{\mathrm{ext}}^{IE}$ & number of external synapses to I cells & 320 \\
			$J^{EE}_{\mathrm{ext}}$ & external E-to-E synaptic strength & $2.3/\sqrt{N}$ mV \\
			$J^{IE}_{\mathrm{ext}}$ & external E-to-I synaptic strength & $2.3/\sqrt{N}$ mV \\
			$\nu_{o}^{E}$ & baseline external input rate to E cells & 7 spks/s \\
			$\nu_{o}^{I}$ & baseline external input rate to I cells & 7 spks/s \\
			
			$A_{\mathrm{stim}}^{E}$ & relative stimulus amplitude for E cells & 0.05 \\
			$A_{\mathrm{stim}}^{I}$ & relative stimulus amplitude for I cells & 0 \\
			$t_{\mathrm{stim}}$ & stimulus onset time & 1 s \\
			$\tau_{r}$ & stimulus rise time & 75 ms \\
			$\tau_{d}$ & stimulus decay time & 100 ms \\
			----- & sampled arousal levels (clustered network) & $[0, 10, 20, 30, 40, 50, 60, 70, 80, 90, 100]\%$ \\
			----- & sampled arousal levels (uniform network) & $[0, 10, 20, 30, 40, 50, 60, 70, 80, 90, 100]\%$ \\ \\
			* & \multicolumn{2}{l}{additional parameters related to the stimulus and arousal implementation are provided in the} \\
			& \multicolumn{2}{l}{\nameref{s:Methods}} \\
			\\
			\hline \hline
		\end{tabular}
	\end{center}
	\caption[Baseline parameter values for the spiking network model (related to \nameref{s:Methods}).]{\textbf{Baseline parameter values for the spiking network model (related to \nameref{s:Methods}).}} 
	\label{t:model_params}
\end{table*}
\clearpage

\clearpage
\begin{table*}[t!]
	\begin{center}
		\begin{tabular}{m{0.2\linewidth}m{0.4\linewidth}m{0.3\linewidth}}
			\hline \hline
			\\
			Parameter & Description & Value \\ \\
			\hline \hline
			\\
			$N^{E}$ & number of E cells & 640 \\
			$N^{I}$ & number of I cells & 160 \\
			
			$\tau_{\mathrm{m}}^{E}$ & membrane time constant of E cells & 20 ms \\
			$\tau_{\mathrm{m}}^{I}$ & membrane time constant of I cells & 20 ms \\
			$\tau^{E}_{\mathrm{syn}}$ & E synaptic time constant & 5 ms \\
			$\tau^{I}_{\mathrm{syn}}$ & I synaptic time constant & 5 ms \\
			$\tau_{\mathrm{ref}}^{E}$ & refractory period of E cells & 5 ms \\
			$\tau_{\mathrm{ref}}^{I}$ & refractory period of I cells & 5 ms \\
			
			$V_{\mathrm{thresh}}^{E}$ & threshold potential of E cells & 4.86 mV \\
			$V_{\mathrm{thresh}}^{I}$ & threshold potential of I cells & 5.98 mV \\
			$V_{\mathrm{r}}^{I}$ & reset potential of I cells & 0 mV \\
			$V_{\mathrm{r}}^{I}$ & reset potential of I cells & 0 mV \\
			
			$p^{EE}$ & E-to-E recurrent connectivity fraction & 0.2 \\
			$p^{IE}$ & E-to-I recurrent connectivity fraction & 0.5 \\
			$p^{EI}$ & I-to-E recurrent connectivity fraction & 0.5 \\
			$p^{II}$ & I-to-I recurrent connectivity fraction & 0.5 \\
			
			$J^{EE}_U$ & uniform E-to-E synaptic strength & $0.8/\sqrt{N}$ mV \\
			$J^{IE}_U$ & uniform E-to-I synaptic strength & $2.5/\sqrt{N}$ mV \\
			$J^{EI}_U$ & uniform E-to-I synaptic strength & $-10.6/\sqrt{N}$ mV \\
			$J^{II}_U$ & uniform E-to-I synaptic strength & $-9.7/\sqrt{N}$ mV \\
			
			$p$ & number of E and I clusters & 2 \\
			$f^{E}$ & fraction of E cells/cluster & 0.125 \\
			$f^{I}$ & fraction of I cells/cluster & 0 \\
			
			$J^{EE}_{+}$ & E-to-E intracluster weight factor & 20 \\
			$J^{IE}_{+}$ & E-to-I intracluster weight factor & 1 \\
			$J^{EI}_{+}$ & I-to-E intracluster weight factor & 1 \\
			$J^{II}_{+}$ & I-to-I intracluster weight factor & 1 \\
			
			$C_{\mathrm{ext}}^{EE}$ & number of external synapses to E cells & 128 \\
			$C_{\mathrm{ext}}^{IE}$ & number of external synapses to I cells & 128 \\
			$J^{EE}_{\mathrm{ext}}$ & external E-to-E synaptic strength & $14.5/\sqrt{N}$ mV \\
			$J^{IE}_{\mathrm{ext}}$ & external E-to-I synaptic strength & $12.9/\sqrt{N}$ mV \\
			$\nu_{o}^{E}$ & baseline external input rate to E cells & 7 spks/s \\
			$\nu_{o}^{I}$ & baseline external input rate to I cells & 7 spks/s \\
			
			----- & sampled arousal levels for the effective MFT & 50 evenly spaced values in $[0,100]\%$ \\
			\\
			* & \multicolumn{2}{l}{additional parameters related to the arousal implementation are provided in the \nameref{s:Methods}} \\
			\\
			\hline \hline
		\end{tabular}
	\end{center}
	\caption[Baseline parameter values for the reduced 2-cluster network model (related to \nameref{s:Methods}).]{\textbf{Baseline parameter values for the reduced 2-cluster network model (related to \nameref{s:Methods}).}} 
	\label{t:model_params_2clusterNet}
\end{table*}

\bibliographystyle{unsrtnat}
\bibliography{refs.bib}

\end{document}